\documentclass[reprint]{revtex4-2}
\usepackage{graphicx} 
\usepackage[utf8]{inputenc}
\usepackage{amsmath}
\usepackage{amssymb}
\usepackage{mathtools}
\usepackage{subcaption}
\usepackage{float}
\usepackage{leftindex}
\newcommand{\delrho}{\Delta\rho}
\newcommand{\dx}{\Delta x}
\newcommand{\half}{\tfrac{1}{2}}
\newcommand{\Q}{\mathcal{Q}}
\newcommand{\eps}{\varepsilon}
\newcommand{\smallminus}{\scalebox{0.5}[1.0]{\( - \)}}
\newcommand{\Oh}{\text{Oh}}
\newcommand{\We}{\text{We}}

\begin{document}

\title{Nonlinear Stability in the Two-Fluid Model, v1}
\author{Alexander López-de-Bertodano}
\affiliation{School of Exact Sciences, National University of Central Buenos Aires, Tandil, Argentina}
\author{Alejandro Clausse}
\affiliation{School of Exact Sciences, National University of Central Buenos Aires, Tandil, Argentina\\
School of Nuclear Engineering, Purdue University, West Lafayette, Indiana, USA\\
Instituto Tecnológico de Buenos Aires, Argentina}

\date{September 2025}

\maketitle

\section{Abstract}

The 1-D Two-Fluid Model (TFM) promises a powerful and computationally cheap platform for simulating multi-fluid flow phenomena. However, runaway Kelvin-Helmholtz instabilities plagued previous approaches, necessitating aphysical regularizations. We demonstrate a novel physics-based approach, using a simple turbulent viscosity model to provide nonlinear stabilization, compatible with inertial coupling. We develop a set of analytical and numerical tools to investigate the resulting dynamics, including turbulent cascades, chaos, and formation of churn or slug flow. Our approach opens up a wide range of new capabilities for the TFM by capturing the Kelvin-Helmholtz instability physically.

\section{Introduction}

The Two-Fluid Model (TFM) is a continuum approach to fluid dynamics, originally conceived for mixture flows in up to 3 dimensions \cite{Landau1941}. It subsequently found wide use in 1-D interfacial flows \cite{Ishiibook}, with applications in the design of nuclear reactors, geothermal plants, condensers, heat exchangers, rocket engines, and so on.

In problems that may be reduced to one dimension (pipe flows), the TFM has one main advantage, computability. However, dimensional reduction constitutes a substantial approximation and introduces two key weaknesses:

\begin{enumerate}

\item \textbf{Higher-Dimensional Structure.}\ 3-D flow and interfacial structure may be baked into a 1-D TFM on an empirical basis, but cannot be derived from it. The sacrifice of this capability is the fundamental trade-off of simplification to 1-D. Modalities that seek to retain full predictive power by moving to higher dimension (e.g.\ Volume of Fluid or Level Set) incur a corresponding computational cost. 

\item \textbf{Badly Behaved Local Instabilities.} Historically, the TFM has been subject to unruly local wave behavior resulting from the Kelvin-Helmholtz instability. Short wavelength components grow too quickly, meaning solutions become dominated by noise or cease to exist altogether. The model is, in such a case, \textit{ill-posed}. 

\end{enumerate}

These two issues are not really separate. The real culprit of ill-posedness is a failure to appropriately account for higher-dimensional structure throughout the development of the Kelvin-Helmholtz, not the instability itself.

It is important that we state precisely what we mean by ill-posedness. In prior work on the TFM \cite{Gidaspow1994}, three ideas have often been regarded as equivalent and interchangeable: (1) the presence of complex linear eigenvalues, (2) ellipticity, and  (3) ``ill-posedness".

The original work on well-posedness by Hadamard \cite{Hadamard1902} analyzed the wave and Laplace equations. For these simplest cases, the three ideas do perfectly intersect. An initial value (Cauchy) problem is well-posed on the \textit{hyperbolic} wave equation with its constant, real eigenvalues. It is ill-posed on the \textit{elliptic} Laplace equation with its constant, imaginary eigenvalues.

In more complex systems, however, these equivalences break down. Dispersive systems \cite{WhithamBook} need not be cleanly elliptic or hyperbolic. In some TFMs, eigenvalues even shift between complex and real based on wavelength. Signs also matter. Positive imaginary eigenvalues, corresponding to wave growth, are workable if confined to a finite range of wavelengths. Negative imaginary eigenvalues are no issue whatsoever.

Throwing out a system of PDEs because it has complex eigenvalues is like throwing out a system of ODEs because it has any kind of fixed point other than oscillatory centers. Many famous dynamical models (predator-prey, Lorenz, Van der Pol) would need to be classed as ``ill-posed".

For the purposes of the present analysis, we therefore adopt the following definition: a well-posed system has unique solutions (``possible et determiné"), and an ill-posed system lacks them. We are now free to analyze the linear dynamics on their own terms, as well as the nonlinear effects that may cause a TFM to be well- or ill-posed more generally.

The problem of ill-posedness has reared its head repeatedly in the development of the TFM, and to combat it a slough of fixes have emerged \cite{Dinh2003}. These regularizations are universally unsatisfactory, either proving too weak or involving aphysical forces. 

The most convenient and widespread approach in simulation has been to employ ``coarse" (large node) meshes, sidestepping low-wavelength Kelvin-Helmholtz behavior through a combination of numerical viscosity and the exclusion of sub-mesh length scales. Premier codes like RELAP \cite{Pokharna1997} rely on coarse meshes. This methodology makes the correspondence between numerical solutions and experimental or analytical results dependent on the level of mesh refinement used.

We maintain that, to be of predictive value, numerical solutions of a TFM must converge up to arbitrarily fine mesh resolution, via the application of physical forces only.

In pursuit of such a TFM, M.\ López-de-Bertodano et al.\ pioneered the use of surface tension \cite{Bertbook}. 
Surface tension introduces a short-wavelength cutoff to linear perturbation growth, but further work \cite{Fullmer2014} \cite{Chaos2019} showed that this cutoff is not sustained in numerical solutions when transitioning to a nonlinear regime. Surface tension is a beginning but not the end; we explore its role thoroughly in Sections \ref{linstab} and \ref{nonlinear}.

It has been previously held (see e.g. \cite{Pokharna1997}) that numerical approaches to the TFM could not practically handle mesh scales where higher-order terms become strong enough to constrain wave growth. This limitation no longer stands. 

Inertial coupling (IC), which may be used to model conservative aspects of higher-dimensional structure, is a fully effective stabilizer for runaway short waves \cite{Clausse2021} \cite{Clausse2023}. In some regimes, however, it is more appropriate to leave the IC at least partially unstable in order to capture Kelvin-Helmholtz wave growth. In such cases, it is necessary to introduce nonlinear dissipation to bound the amplitude of the fluctuations.

Our novel approach to the ill-posedness problem is inspired by the thesis of W. Fullmer \cite{Fullmerthesis}, who originated the idea of using a simple eddy model to represent turbulent dissipation around a fluid-fluid interface. The power of this approach lies in allowing viscosity to scale with the relative velocity across the interface. 

We present a methodology for the construction, analysis, and numerical solution of well-posed, Kelvin-Helmholtz unstable TFMs. We well-pose these cases through a combination of linearly and nonlinearly stabilizing higher-order forces. We hope that this work will open the door to an even broader class of well-posed, physically predictive models for a variety of applications.

\section{Model Variables and Equations} \label{varseqs}


For simplicity, we assume two-fluid incompressible flow in a square channel of height $H$. The constant fluid densities are labeled $\rho_1$ and $\rho_2$, with $\rho_1 \ge \rho_2$ by convention. The volume fractions of each fluid in a given perpendicular cross section are $\alpha_1$ and $\alpha_2$, with the identity $\alpha_1 + \alpha_2 = 1$; we will mostly denote $\alpha_2$ as just $\alpha$ (the void fraction) and $\alpha_1$ as $1-\alpha$. The average fluid velocities per cross section are likewise labeled $u_1$ and $u_2$. We then define:
\begin{equation}
    \begin{aligned}
        \rho_m \equiv (1-\alpha) \rho_1 + \alpha \rho_2 &: \text{``Mixture" density} \\
        \delrho \equiv \rho_1 - \rho_2 &: \text{Density difference} \\
        j \equiv (1-\alpha) u_1 + \alpha u_2 &: \text{Volumetric flux} \\
        u_r \equiv u_2 - u_1 &: \text{Relative velocity} \\
        J \equiv \alpha (1-\alpha) u_r &: \text{Drift flux}
    \end{aligned}
\end{equation}

We introduce a function $\Gamma(\alpha)$ \cite{Pauchon1992} to describe the inertia of relative motion, including inertial coupling (IC). The mixture kinetic energy (per area) can be written:
\begin{equation} \label{eq:kinetic}
    T = \half \Gamma \! J^2 - \delrho J j + \half \rho_m j^2
\end{equation}
In the case of no IC, the inertial function is:
\begin{equation} \label{eq:gamma0}
    \Gamma = \Gamma_0 \equiv \frac{\rho_1}{1-\alpha} + \frac{\rho_2}{\alpha}
\end{equation}
and the kinetic energy reduces to:
\begin{equation} \label{eq:primkinetic}
    T = \half \left[ \rho_1 (1-\alpha) u_1^2 + \rho_2 \alpha u_2^2\right]
\end{equation}
which is valid for smooth stratified flow.\ Following Clausse and López-de-Bertodano \cite{Clausse2021} \cite{Clausse2023}, we continue in terms of:
\begin{equation} \label{eq:Q}
    \begin{aligned}
        W \equiv \Gamma J &: \text{Relative momentum} \\
        \Q(\alpha) \equiv \Gamma^{-1} &: \text{Reciprocal inertial function} \\
        \Q'(\alpha) \equiv \frac{d\Q}{d\alpha} &: \text{First derivative of } \Q
    \end{aligned}
\end{equation}

We seek to solve for 4 ``basis" variables, chosen as $\alpha$, $j$, $W$, and a pressure field $P$. We therefore require 4 governing equations.

We begin with mass conservation for the lighter fluid:
\begin{equation} \label{eq:voidprop}
    \partial_t \alpha + \partial_x (\alpha u_2) = \boxed{\partial_t \alpha + \partial_x (\alpha j + \Q W) = 0}
\end{equation}
From this and the mass conservation equation for the heavier fluid:
\begin{equation} \label{eq:voidprop1}
    \partial_t \alpha_1 + \partial_x (\alpha_1 u_1) = 0
\end{equation}
we then derive the ``incompressibility" condition:
\begin{equation} \label{eq:incomp}
    \begin{aligned}
        \partial_t \alpha_1 + \partial_x (\alpha_1 &u_1) + \partial_t \alpha + \partial_x (\alpha u_2) = \\
        & \boxed{\partial_x j = 0}
    \end{aligned}
\end{equation}

The variational method on $T$ (Eq.\ \ref{eq:kinetic}) obtains the left-hand side of the relative momentum equation \cite{Clausse2021}:
\begin{equation} \label{eq:relmotion}
    \boxed{
    \begin{aligned}
        \partial_t (W - \delrho j) + \partial_x \mathcal{A}_W = & - g_x \delrho - g_y \delrho H \partial_x \alpha \ \\
        + \sigma \! H \partial_x \! \left( \! \tfrac{\partial_{x_{}}^2 \alpha}{[1 + (H \partial_x \alpha)^2]^{3/2}} \! \right) + & F_{\text{visc},W} - \textstyle \sum_{i} \tfrac{(-1)^i}{\alpha_i} D_i \\
        - \tfrac{C_D \rho_D}{H} | & u_r| u_r
    \end{aligned}
    }
\end{equation}
where $\mathcal{A}_W$ is the relative momentum flux \cite{Clausse2024}:
\begin{equation}
    \mathcal{A}_W = \half \Q' W^2 + W j - \half \delrho j^2
\end{equation}
in agreement with the primitive variable equations derived by Geurst \cite{Geurst1985}. Derivations of the right-hand side terms, representing conservative and dissipative forces acting between the phases, are given in Appendix \ref{appx:momeqs}. $g_x$ is the component of gravity parallel to the channel, $g_y$ is the perpendicular component, $\sigma$ is the surface tension, $F_{\text{visc},W}$ are the viscous forces, $D_i$ is the wall drag on the $i$'th fluid, and $C_D$ is the interfacial drag coefficient.

The viscous term will be of particular interest for the present analysis. In simplified form it reads:
\begin{equation}
    F_{\text{visc},W} = \partial_x \! \left[ \nu \partial_x (W - \delrho j) \right]
\end{equation}
where $\nu$ is the viscosity. If the flow is incompressible (Eq.~\ref{eq:incomp}) there remains only:
\begin{equation} \label{eq:simplevisc}
    F_{\text{visc},W} = \partial_x \! \left( \nu \partial_x W \right)
\end{equation}
with:
\begin{equation}
    \nu = \nu_k + \nu_t
\end{equation}
where $\nu_k$ is the kinematic viscosity and $\nu_t$ is the turbulent eddy viscosity, given by:
\begin{equation} \label{eq:turbvisc}
    \nu_t = l_m |u_r|
\end{equation}
where $l_m$ is the ``mixing length", held constant for simplicity. This formulation provides superior nonlinear stabilization by dissipating preferentially where $u_r$ is large. We may also consider ``local averaging" of the above:
\begin{equation} \label{eq:localavgvisc}
    \nu_t(x) \equiv l_m \int_{x-\delta_m}^{x+\delta_m} |u_r|\, \frac{dx}{2\delta_m}
\end{equation}
where $\delta_m$ is the local averaging length, the range over which flow conditions at any point are taken to affect surrounding eddies. We set $\delta_m = 2l_m$ throughout.

Finally, we have the mixture momentum equation \cite{Clausse2021}  \cite{Clausse2023}: 
\begin{equation} \label{eq:mixmotion}
    \boxed{
    \begin{aligned}
        \partial_t (\rho_m j - \delrho J) + \partial_x \mathcal{A}_m & = g_x \rho_m
        + g_y \rho_m H \partial_x \alpha \\
        + \,\alpha \sigma \! H \partial_x \! \left(\tfrac{\partial_{x_{}}^2 \alpha}{[1 + (H \partial_x \alpha)^2]^{3/2}} \! \right) +&\, F_{\text{visc},m} - \textstyle \sum_{i} D_i - \partial_x P
    \end{aligned}
    }_{_{_{}}}
\end{equation}
where $\mathcal{A}_m$ is the mixture momentum flux \cite{Clausse2021}, $F_{\text{visc},m}$ are the mixture-equation viscous forces, and $P$ is the ``reference" pressure, selected as the heavy-phase interfacial pressure $P_{\text{int},1}$ (see Eq.\ \ref{eq:refpressure}). 

The advantages of stating the momentum equations in this form are several \cite{Clausse2021}. First, the relative momentum equation is independent of the pressure field. 
Second, casting the mixture momentum equation in terms of the divergence-free $j$ simplifies its computation. Additionally, each momentum equation-variable pair governs a distinct dynamic: the mixture momentum equation and $j$ control system oscillations, analyzed in detail by Ishii and collaborators \cite{Ishiithesis} \cite{SIZ1976}, while the relative momentum equation and $W$ control local waves, which are the source of the ill-posedness problem.

\section{Linear Stability} \label{linstab}

The linear stability of local waves in the TFM has been treated many times \cite{Clausse2021} \cite{Clausse2023} \cite{Stuhmiller1977} \cite{Ansari2011}. We follow these earlier studies but compliment linear analysis with detailed numerical testing. See Appendix B for more about the linear dynamics.

\subsection{Pure Advection} \label{pureadvect}

Assuming incompressibility (Eq.\ \ref{eq:incomp}), $j$ cannot sustain perturbations of the form $e^{ikx}$ if $k \neq 0$. The pressure $P$, though it may respond to changes in $\alpha$ and $W$, does not feed back into the mass or relative momentum equations (Eqs.\  \ref{eq:voidprop} and \ref{eq:relmotion}). We may therefore exclude $j$ and $P$ from the perturbed state vector.

We omit system oscillations from the analysis by setting $\partial_t j = 0$. The following two-equation system then represents the local wave dynamics:
\begin{equation} \label{eq:pureadvectsys}
    \begin{aligned}
      \partial_t \alpha + \partial_x (& \Q W + \alpha j) = 0 \\
      \partial_t W + \partial_x (\half & \Q' W^2 + W \! j) = 0
    \end{aligned}
\end{equation}
where we have left out all force terms from Eq.\ \ref{eq:relmotion} for the time being. Casting Eq.\ \ref{eq:pureadvectsys} in matrix form yields:
\begin{equation} \label{eq:pureadvectmatrix}
    \frac{\partial}{\partial t} \!
    \begin{pmatrix}
        \! \alpha \! \\ \! W \!
    \end{pmatrix}
    +
    \begin{pmatrix}
        \Q' W + j & \Q \\ \half \Q'' W^2 & \Q' W + j
    \end{pmatrix}
     \! \frac{\partial}{\partial x} \!
    \begin{pmatrix}
        \! \alpha \! \\ \! W \!
    \end{pmatrix}
    = 0
\end{equation}

We make the standard assumption that the equilibrium state we perturb is ``flat": $\partial_x \alpha_0 = \partial_x W_0 = 0$. An infinitesimal perturbation of $\alpha$ and $W$ of the form $\vec{\eps}  e^{i(k x-\omega t)}$ must then satisfy:
\begin{equation}
    \begin{vmatrix}
        - \omega + (\Q' W + j)k & \Q k \\ \half \Q'' W^2 k & - \omega + (\Q' W + j) k
    \end{vmatrix}
    = 0
\end{equation}
where $W$, $Q$, and its derivatives in $\alpha$ now denote their equilibrium values. We solve for the dispersion frequencies:
\begin{equation} \label{eq:pureadvectdisp}
    \omega = (\Q' W + j)k \pm \sqrt{\half \Q \Q'' W^2 k^2}
\end{equation}
elsewhere referred to as ``eigenvalues".

$\Q' W + j$ is the ``eigenvelocity". The radicand contains $W^2$, $k^2$, and $\Q$, which may never be negative, and $\Q''$, which therefore determines the sign of the radicand. If $\Q'' \ge 0$, the root is real, creating two modes with split velocities. If $\Q'' < 0$, the root is imaginary, so one mode decays and other grows.

The growth rate is given by the imaginary part of $\omega$:
\begin{equation}
    \text{Im}(\omega) =
    \begin{cases}
		\sqrt{-\half \Q \Q''} |W \! k| & \text{if } \Q''<0\\
        \hfil 0 & \text{if } \Q'' \geq 0
	\end{cases}
\end{equation}

As stated in \cite{Clausse2021}, this growth is a manifestation of the Kelvin-Helmholtz instability. Synonymously, we call $\Q'' \! < 0$ the ``inertial instability condition". 

In the purely advective model, this condition results in divergent growth rates as $k \to \infty$, making the linearized problem \textit{ill-posed} (i.e.\ no solution exists) despite well-behaved initial conditions.  

To demonstrate, define any initial perturbation of the interface $\Delta \alpha$ so that at high $k$ its Fourier transform scales as $k^{-p}$ (with p $>$ 1). We denote the Fourier transforms of the perturbations as follows:
\begin{equation}
    \begin{aligned}
        \Delta \widetilde{\alpha}(k,t) \equiv \mathcal{F}[\Delta\alpha] \qquad \ \Delta \widetilde{W}(k,t) \equiv \mathcal{F}[\Delta W]
    \end{aligned}
\end{equation}
Taking Eq.\ \ref{eq:Wlinresponse} and omitting constant factors, we can show that the $L_2$-norm of the response in $\Delta W$:
\begin{equation} \label{eq:linresponsenorm}
   L_2[\Delta W] = \sum_{k}^\infty |\Delta \widetilde{W}(k,t)|^2 \! \sim \! \sum_{k}^\infty k^{-2p} \sinh^2 \frac{Wk}{2} t
\end{equation}
diverges for $t > 0$.


The prototypical case of negative $\Q''$ is found in smooth stratified flow. For the simplest form possible take $\rho_1 = \rho_2 = 1$:
\begin{equation} \label{eq:simpleinertia}
    \Gamma = \tfrac{1}{\alpha} + \tfrac{1}{1-\alpha} \Rightarrow \Q = \alpha(1-\alpha) \Rightarrow \Q'' = -2
\end{equation}
in which case one obtains:
\begin{equation}
    \text{Im}(\omega) = \sqrt{\alpha(1-\alpha)}|W \! k|
\end{equation}

\begin{figure}
    \includegraphics[width=0.45\textwidth]{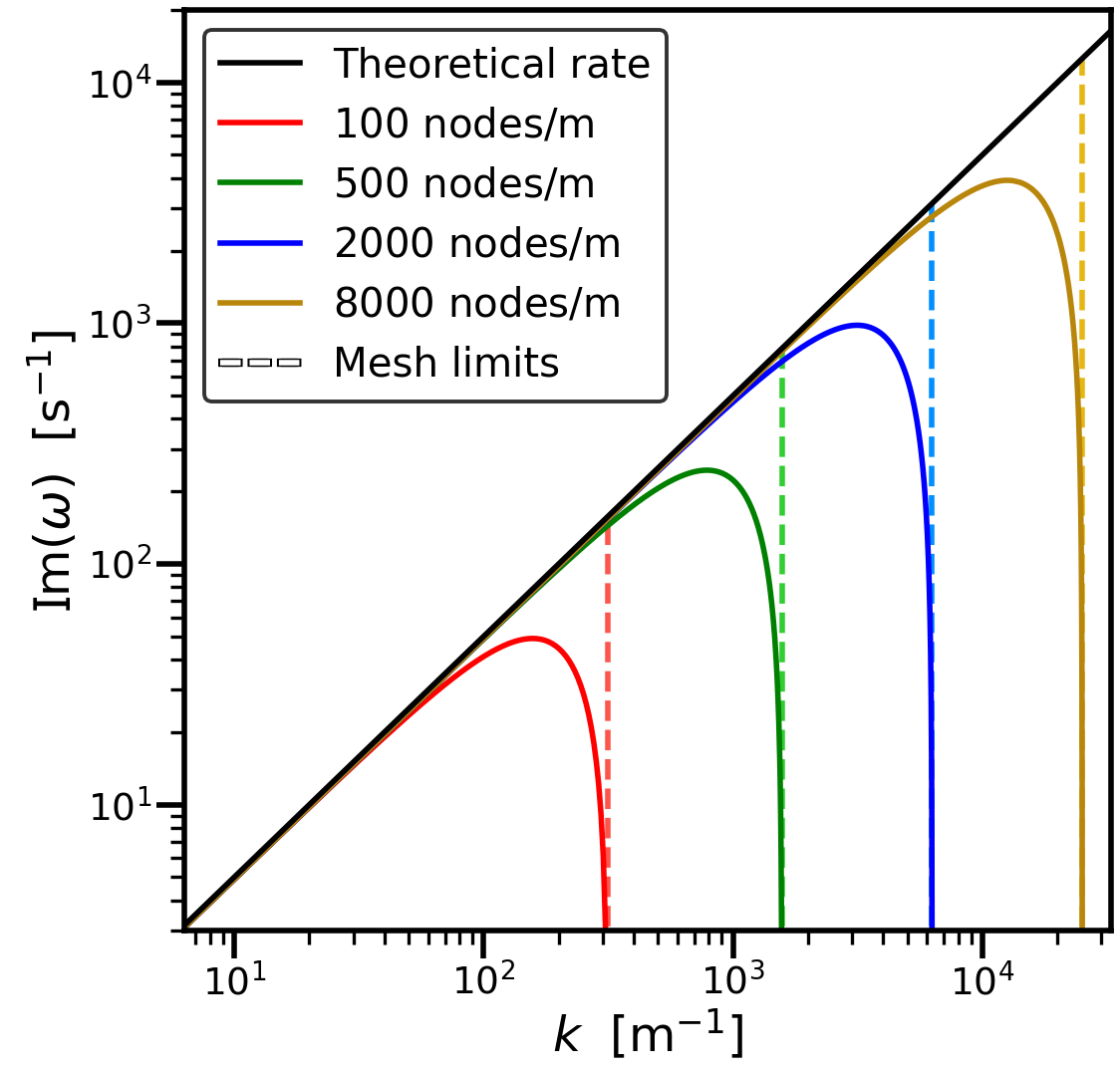}
    \caption{Wave growth rate, $\text{Im}(\omega)$, vs.\ wavenumber, $k$, under pure advection. $\alpha = \half,\, W = 1 \frac{\text{m}}{\text{s}}$. Log scale.}  \label{fig:PureAdvDisp}
\end{figure}

We demonstrate the divergent nature of numerical solutions for this simplest case in Fig.\ \ref{fig:PureAdvDisp}. Values of $\text{Im}(\omega)$ are obtained by solving for the response to each $k$ individually in a 1\,m test section with periodic boundary conditions over very short times. At each resolution, $\text{Im}(\omega)(k)$ tracks the theoretical rate before approaching 0 at the mesh limit $k_{max} = \pi/\dx$. Refining the mesh increases this limit without bound. Details of the numerics are provided in Appendix \ref{appx:numerics}. 

It should be noticed that the divergence of $\text{Im}(\omega)$ is not unique to the 1-D model. The classic dispersion relation for the Kelvin-Helmholtz instability in 2-D \cite{LambBook}:
\begin{equation}
    \omega = \frac{\rho_1 u_1 +\rho_2u_2}{\rho_1+\rho_2}k \pm \frac{\sqrt{-\rho_1\rho_2(u_2-u_1)^2k^2}}{\rho_1+\rho_2}
\end{equation}
also shows a linear divergence proportional to $k$.

All in all, the linear ill-posedness problem is caused by runaway growth of short wavelengths. This result seems counter-intuitive given the fact that the local wave energy (cf. Eq.\ \ref{eq:kinetic}):
\begin{equation} \label{eq:localwaveenergy}
    \boldsymbol{\mathcal{E}} \equiv \int T_{\text{local}} \, dx = \int \half \Q W^2 \, dx
\end{equation}
is conserved by advection. We may in fact write the corresponding energy flux (Eq.\ \ref{eq:localEflux}):
\begin{equation}
    \mathcal{F}_{\text{local}} = (\Q' W + j) \, T_{\text{local}}
\end{equation}
However, the lowest-order energy perturbation term:
\begin{equation}
    \Delta \boldsymbol{\mathcal{E}} = \int \tfrac{1}{4}\Q''W^2(\Delta \alpha)^2 + \half \Q (\Delta W)^2 \, dx
\end{equation}
remains constant in time despite growing perturbations so long as $\Q'' < 0$. 

Negative $\Q''(\alpha)$, namely inertial instability, is the root of ill-posed linear wave growth. One may thus wish to get rid of it. However, it is not desirable, in every case, to construct the inertia so that $\Q'' \geq 0$ at any $\alpha$.

Inertial coupling in the TFM equations results from energy not reflected in the averaged flow variables  \cite{Wallis1990}. This energy may arise from local velocity fluctuations, vortices, or other unresolved flow features. In our formalism, IC is accounted for by $\Gamma(\alpha)$, whose reciprocal $\Q(\alpha)$ is key to understanding the stability of local wave modes.

In some regimes (stratified flow in particular) sufficient stabilizing inertia may not be forthcoming. Furthermore, wherever one sets $\Q''$ to be positive, inertial wave growth does not occur. Over-application of this approach rids the TFM of predictive power over Kelvin-Helmholtz-type phenomena along with ill-posedness. To retain all possible physics, inertial instability should be retained and incorporated.

\subsection{Viscosity} \label{viscosity}

Since using inertia to tame local wave growth is not desirable in every case, it is natural to look toward dissipative mechanisms as well. Dissipation of the relative motion comes from interfacial drag and viscosity. However, the interfacial drag includes no derivatives, so its contribution to $\omega$ does not scale in $k$. It is therefore uncompetitive with the inertial growth term at short wavelengths -- see Eq.\ \ref{eq:fulldisp}. Viscosity, having a second derivative, contributes to the dispersion with a promising $k^2$ scaling.

We thus define a viscous force term in simple form:
\begin{equation}
    \partial_t W + \partial_x (\half \Q' W^2 + W \! j) = \ldots + \tilde{\nu} \, \partial_x^2 W
\end{equation}
where $\tilde{\nu}$ is the effective viscosity, a constant. This form describes the contribution of the real viscous forces to the linearized equations if there is no inertial coupling (cf. Eq.\ \ref{eq:simplevisc}). If inertial coupling is included, it may only be approximate. 

We then introduce the following notation for dispersion frequencies:
\begin{equation}
    \omega = \bar{\omega} \pm \sqrt{\Delta}
\end{equation}
where, with viscosity:
\begin{equation}
    \begin{aligned}
        \bar{\omega} &=  (\Q' W + j)k - \frac{i \tilde{\nu} k^2}{2} \\
        \Delta &= \frac{\Q \Q'' W^2 k^2}{2} - \frac{\tilde{\nu}^2 k^4}{4}
    \end{aligned}
\end{equation}

Despite the decay term in $\bar{\omega}$, the + mode still exhibits finite (but not divergent) high-$k$ growth if $\Q'' < 0$:
\begin{equation}
   \lim_{k \to \infty} \text{Im}(\omega) = \frac{\Q |\Q''| W^2}{2 \tilde{\nu}}
\end{equation}
Repeating the process used to obtain Eq.\ \ref{eq:linresponsenorm} shows that the $L_2$-norm of the linear response to an allowable interfacial perturbation:
\begin{equation}
   L_2[\Delta W]  \! \sim \! \sum_{k}^\infty k^{-2p} \! \sinh^2(t)
\end{equation}
exists but is unbounded in time. 

Though the viscous model is not linearly ill-posed, it allows for unchecked growth of arbitrarily high-$k$ ``noise" waves.\ 
Note that numerical dissipation eliminates this shortcoming. A first-order upwind scheme on Eq.\ \ref{eq:pureadvectsys} would have numerical viscosity:
\begin{equation}
    \tilde{\nu}_\text{num} = \Delta x (\Q' W + j)
\end{equation}
where $\Delta x$ is the mesh size. The key difference is that now the $\alpha$ equation acquires a diffusion term with the same coefficient:
\begin{equation}
    \partial_t \alpha + \partial_x (\Q W + \alpha j) = \tilde{\nu}_\text{num} \partial_x^2 \alpha
\end{equation}
The result of numerical viscosity is to modify the base dispersion frequency:
\begin{equation}
    \bar{\omega} = (\Q' W + j)k - i \, \tilde{\nu}_\text{num} k^2
\end{equation}
without contributing to $\Delta$:
\begin{equation}
    \Delta = \half \Q \Q'' W^2 k^2
\end{equation}
The new, artificial term in $\tilde{\nu}_\text{num}$ can constrain wave growth effectively, though only insofar as $\Delta x$ is kept sufficiently large. Higher-order numerical methods may exhibit almost no numerical dissipation, neutralizing the effect.

\subsection{Gravity and Surface Tension}
\label{surfacetension}

Dissipative forces fail to stabilize the linear model. That leaves us with the conservative forces, gravity and surface tension.

Gravity parallel to the channel ($g_x$) does not directly influence $\omega$. The perpendicular component ($g_y$) contributes a term to the radicand:
\begin{equation} \label{eq:gravradicand}
    \Delta = \Q (\half \Q'' W^2 k^2 + g_y \delrho H k^2)
\end{equation}
This term has the same order in $k$ as the destabilizing $\Q''$ term. It imposes a threshold in $W^2$\,\,under which the model is Lyapunov stable for all wavelengths. Above this threshold, though, the model is again unstable with divergent growth rates, 
indicating instant ill-posedness.

\begin{figure} 
    \begin{subfigure}[b]{0.45\textwidth}
    \includegraphics[width=\textwidth]{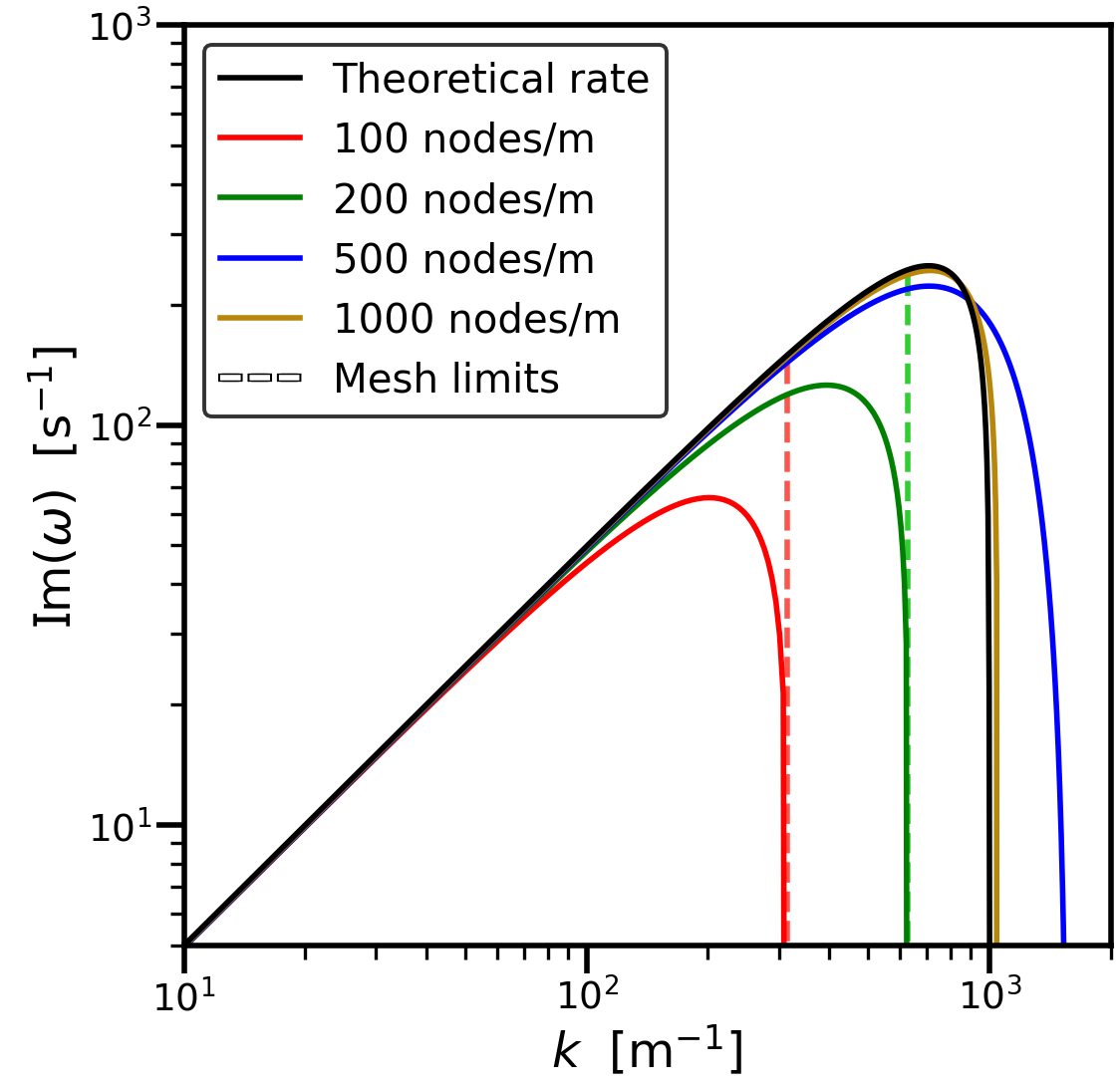}
        \caption{Coarse mesh results. Logarithmic scale. \newline} 
    \end{subfigure}
    \begin{subfigure}[b]{0.45\textwidth}
        \includegraphics[width=\textwidth]{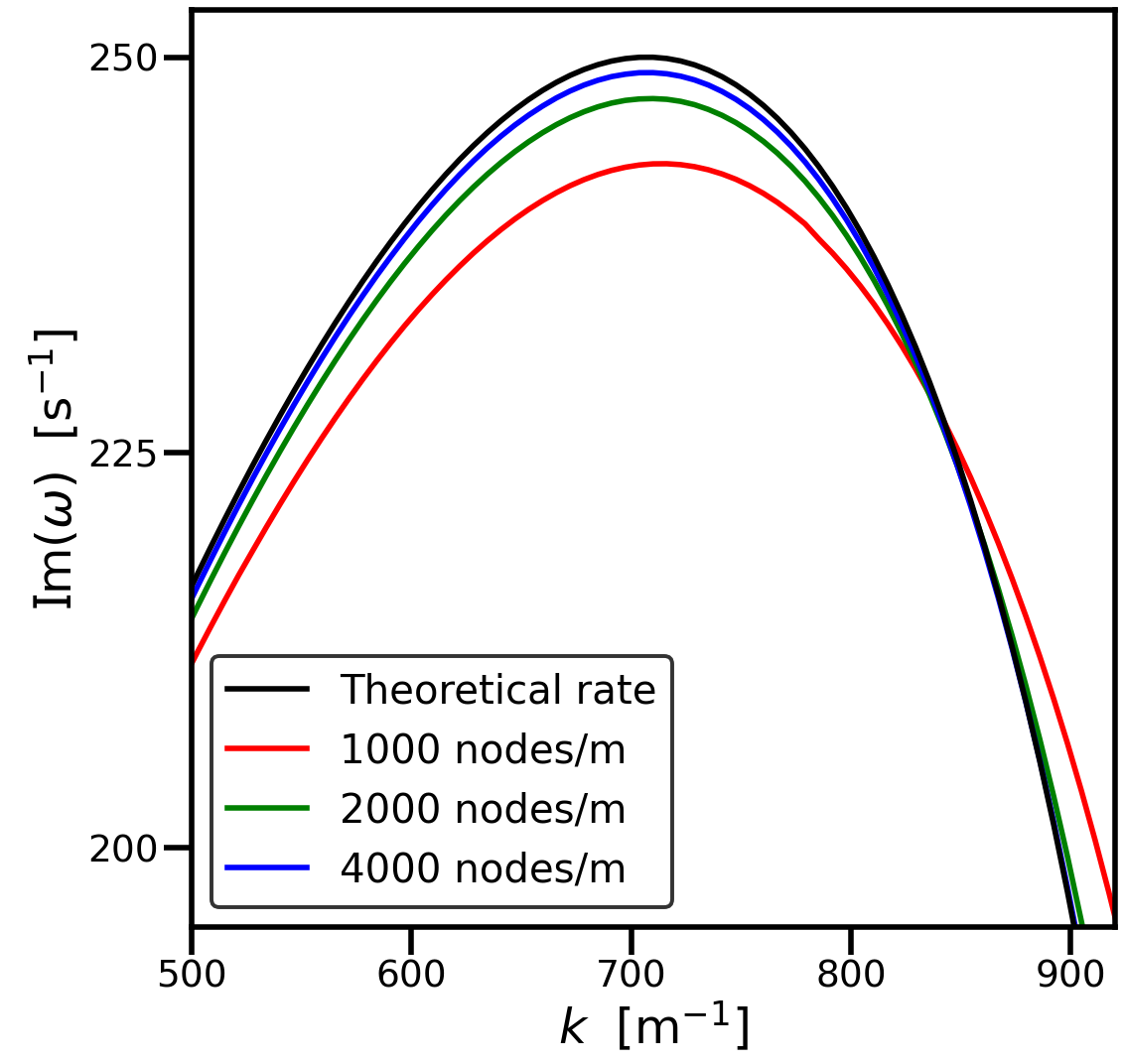}
        \caption{Fine mesh results near $\max(\text{Im}(\omega))$. Linear scale.}
    \end{subfigure}
    \caption{Wave growth rate, $\text{Im}(\omega)$, vs.\ wavenumber, $k$, with surface tension.\ $\alpha=\half,\, W = 1 \frac{\text{m}}{\text{s}},\, \sigma = 10^{-4} \frac{\text{m}^3}{\text{s}^2},\, H = 10^{-2}\, \text{m}$} \label{fig:STDisp}
\end{figure}

For its part, surface tension adds a higher-order term to the radicand:
\begin{equation}
    \Delta = \Q (\half \Q'' W^2 k^2 + \sigma H k^4)
\end{equation}
This new term imposes a cutoff in $k$ above which the linear perturbations do not grow, but instead acquire additional oscillatory frequency. Given $\Q'' < 0$:
\begin{equation}
    \text{Im}(\omega) =
    \begin{cases}
		\sqrt{\Q\left(\half|\Q''| W^2 k^2 - \sigma H k^4\right)} & \text{if } k < k_\text{cut} \\
        \hfil 0 & \text{if } k \geq k_\text{cut}
	\end{cases}
\end{equation}
The cutoff wavenumber is given by:
\begin{equation} \label{eq:kcutoff}
    k_\text{cut} = |W| \sqrt{\frac{|\Q''|}{2 \sigma \! H}} 
\end{equation}

As in the pure advection case, we use the simple inertia of Eq.\ \ref{eq:simpleinertia} for numerical tests, now including surface tension. Fig.\ \ref{fig:STDisp} shows the new \textit{convergence} of short-time solutions, with increasing mesh resolutions tending to the predicted cutoff behavior.

Linear solutions exist, and the response of high-$k$ modes is tamed, but it would be premature to say we have solved the ill-posedness problem here. Modes at $k<k_\text{cut}$ still undergo exponential growth, which would eventually invalidate the linear approximation and, if sustained, cause extremely large amplitudes. This model must be tested in the \textit{nonlinear} regime in order to gauge its true efficacy.

\section{Nonlinear Stability} \label{nonlinear}

We now wish to analyze solutions over longer times, where nonlinear interactions between wavelengths become important and amplitude dependence emerges. We begin with numerical solutions to gain intuition, and gradually elucidate the observed dynamics by analytical means. The resulting insights will allow us to construct a nonlinearly well-posed model.

We construct our initial conditions somewhat differently than in the purely linear test case. Instead of probing individual wavelengths, we perturb the interface with a Gaussian pulse:
\begin{equation} \label{eq:gaussianpert}
    \Delta \alpha(x,t=0) = \eps e^{-x^2/2s^2}
\end{equation}
where $\eps$ is a small finite magnitude and $s$ is the pulse width. We select:
\begin{equation} \label{eq:gaussianpertvalues}
    \eps = 10^{-5} \qquad s^2 = 5 \, \text{mm}^2
\end{equation}
The Gaussian has a smooth Fourier series, which enhances the readability of the nonlinear response. We continue to utilize periodic boundary conditions in order to allow moving waves enough time to develop at minimum computational cost. 

\subsection{Breakdown of the Linear Approximation} \label{LinBreakdown}

We begin by testing our minimal linearly well-behaved model, i.e. the one with uncoupled inertia and surface tension only (Sec.\ \ref{surfacetension}). 
Since this model is conservative no other mechanisms are needed to maintain an energy balance. We may prove this using the surface energy:
\begin{equation} \label{eq:surfenergy}
    U_{\sigma} = \frac{\sigma}{H} \! \left[1 + (H \partial_x \alpha)^2 \right]^{\frac{1}{2}}
\end{equation}
which is conserved by an additional ``local energy" flux:
\begin{equation}
    \begin{aligned}
    \mathcal{F}_{\text{local},\sigma} = \mathcal{S} \, \partial_x (\Q W) - \Q W & \partial_x \mathcal{S} \\
    \mathcal{S} \equiv \sigma H \partial_x \alpha \left[1 + (H \partial_x \alpha)^2 \right]&^{-1/2}
    \end{aligned}
\end{equation}
derived from the underlying PDEs (Eq. \ref{eq:surfEflux}).

\begin{figure} 
    \begin{subfigure}[b]{0.45\textwidth}
    \includegraphics[width=\textwidth]{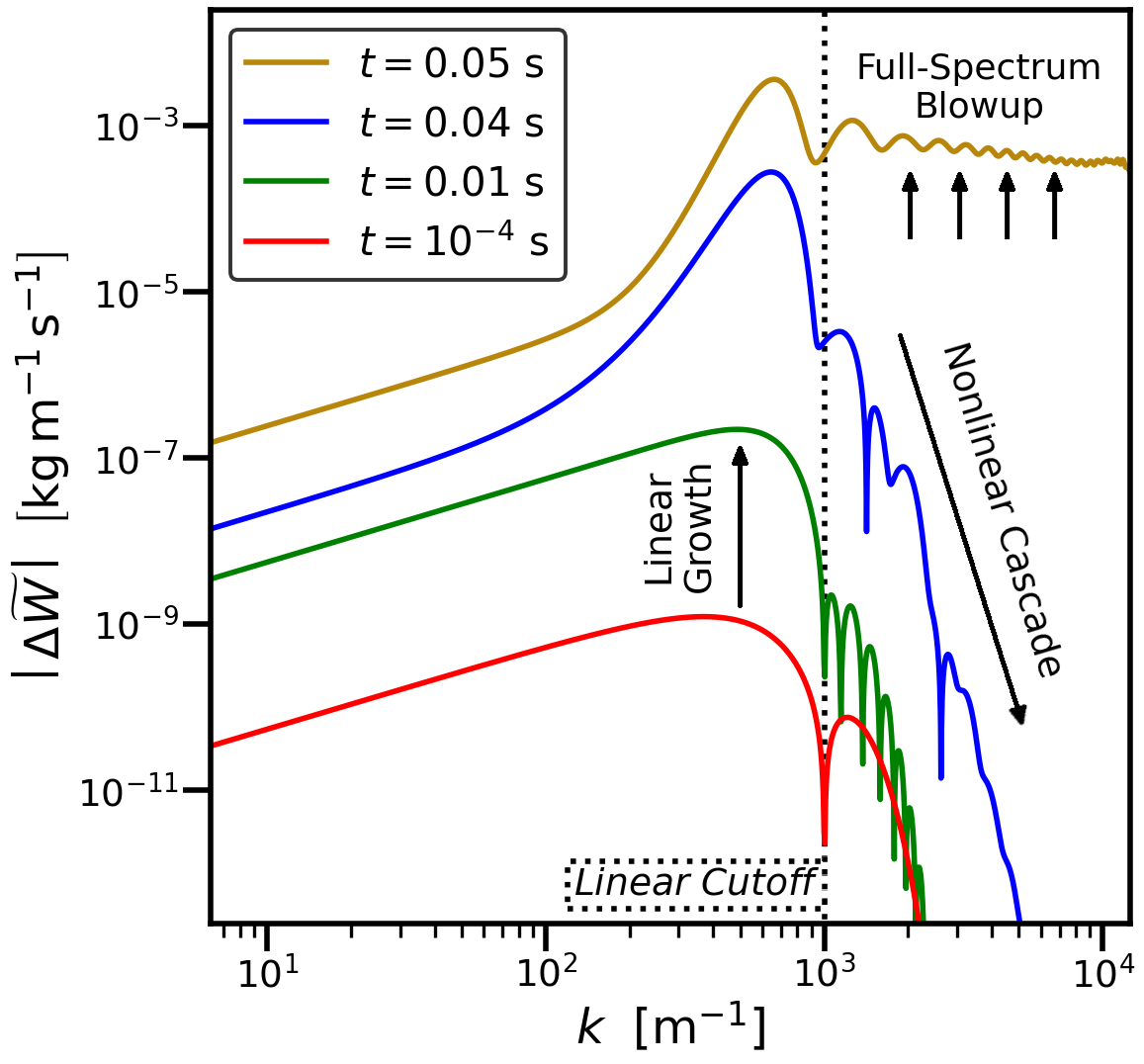}
   \caption{Time evolution of relative momentum Fourier transform. Logarithmic scale. \newline} 
     \end{subfigure}
    \begin{subfigure}[b]{0.45\textwidth}
        \includegraphics[width=\textwidth]{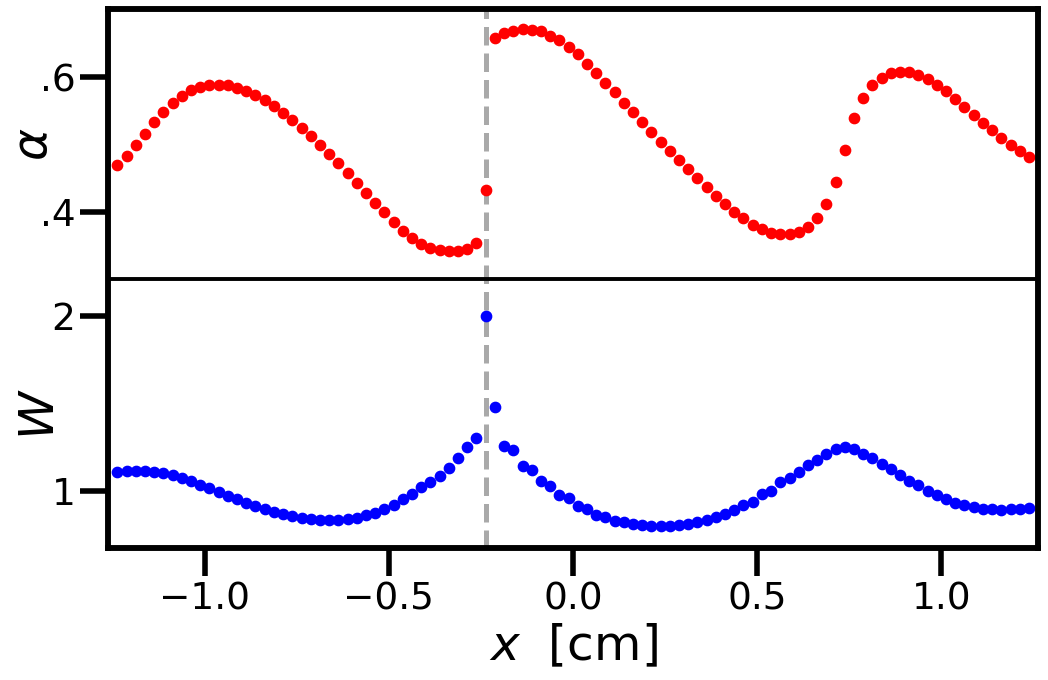}
        \caption{An interface shock at $t = 0.0495$, lined up with a relative momentum spike. \newline}
    \end{subfigure}
    \begin{subfigure}[b]{0.45\textwidth}
        \includegraphics[width=\textwidth]{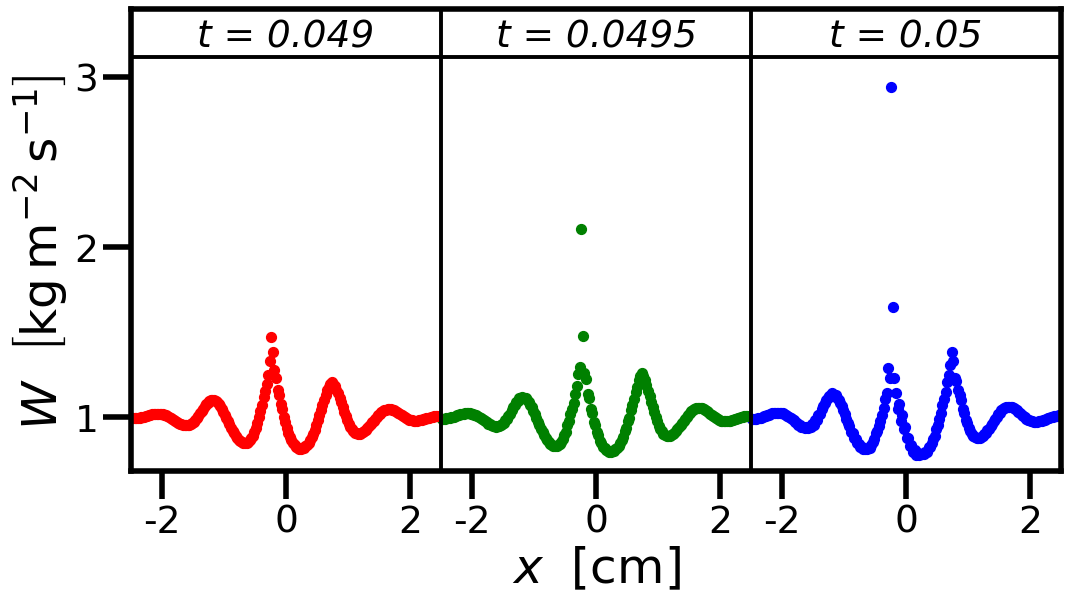}
        \caption{Fast runaway growth of a momentum spike near the initial perturbation location x = 0.}
    \end{subfigure}
    \caption{Development of a Gaussian perturbation ($s =$ 2.24 mm, $\eps = 10^{-5}$) at 4000 nodes/m. Well-posed linear wave growth triggers a nonlinear cascade (a), creating a shock in $\alpha$\,(b) aligned with a runaway spike in $W$\,(c).} \label{fig:FiniteTimeST}
\end{figure}

Figure \ref{fig:FiniteTimeST} shows results from this model over finite times. We use $\Delta x = 1/4$ mm and $\Delta t = \Delta x/100 \, \text{m} \, \text{s}^{-1} $ with periodic boundary conditions. Initially, waves follow a linear growth profile: growth below $k_\text{cut}$, a dip at the cutoff, and sine oscillation above (see Appx.\ \ref{appx:linear}). However, the growing sub-cutoff components quickly begin to ``feed" into the higher-$k$ modes, resulting in a cascade of power down the spectrum.

One may understand the essence of this mode-to-mode transfer by extending the linear approximation of the equations to second order. Take an equilibrium state vector and a perturbation:
\begin{equation}
    \Psi_0 \equiv (\alpha_0, W_0);\, \psi \equiv (\Delta \alpha, \Delta W)
\end{equation}
Writing the perturbation in terms of its Fourier transform:
\begin{equation} \label{eq:psifourier}
    \psi(x,t) = \sum_k \widetilde{\psi}(k,t) \, e^{ikx}
\end{equation}
we may represent the first-order (linear) approximation of the equations in general form:
\begin{equation}
    \partial_t \widetilde{\psi}(k,t) \approx \mathbf{M}_1 (\Psi_0,k) \, \widetilde{\psi}(k,t)
\end{equation}
where $\mathbf{M}_1$ is the 2$\times$2 linearization matrix. See Eq.\ \ref{eq:pureadvectmatrix} for an example. Each mode only interacts with itself and the background equilibrium state.

The extension to 2nd-order, however, involves a convolution between wavelengths:
\begin{equation} \label{eq:2ndorderconvol}
    \partial_t \widetilde{\psi}(k,t) \approx ... + \mathbf{M}_2 \sum_{k'} \widetilde{\psi}(k-k',t) \, \widetilde{\psi}(k',t)
\end{equation}
where $\mathbf{M}_2$ is a 2$\times$2$\times$2 matrix. The convolution in Fourier space results from products of variables in the real-space PDEs, and it governs interactions between modes. 

Suppose for now that higher-order contributions are negligible. The convolutive structure of Eq.\ \ref{eq:2ndorderconvol} naturally leads to the spectrum cascade seen in Fig.\ \ref{fig:FiniteTimeST}a. If significant modes initially exist up to wavenumber $k_\text{cut}$, they convolve and contribute to modes up to $2 k_\text{cut}$, which in turn affect modes up to $4 k_\text{cut}$ more weakly, and so on.

In the present case, nonlinear growth of increasingly short waves creates narrowing spikes in $W$ coupled with shocks in $\alpha$ (Fig.\ \ref{fig:FiniteTimeST}b). We call them ``shock-spike" waves; they turn out to be an emblematic feature of the inertially unstable TFM.

If growth occurs via the convolution of low-$k$ modes, the feedback of high-$k$ modes may in turn limit the spectrum. 
Evidently this limiting does not occur at present. Instead, once a momentum spike's width nears the mesh limit, it blows up violently (Fig.\ \ref{fig:FiniteTimeST}c).

\begin{figure} 
    \includegraphics[width=0.45\textwidth]{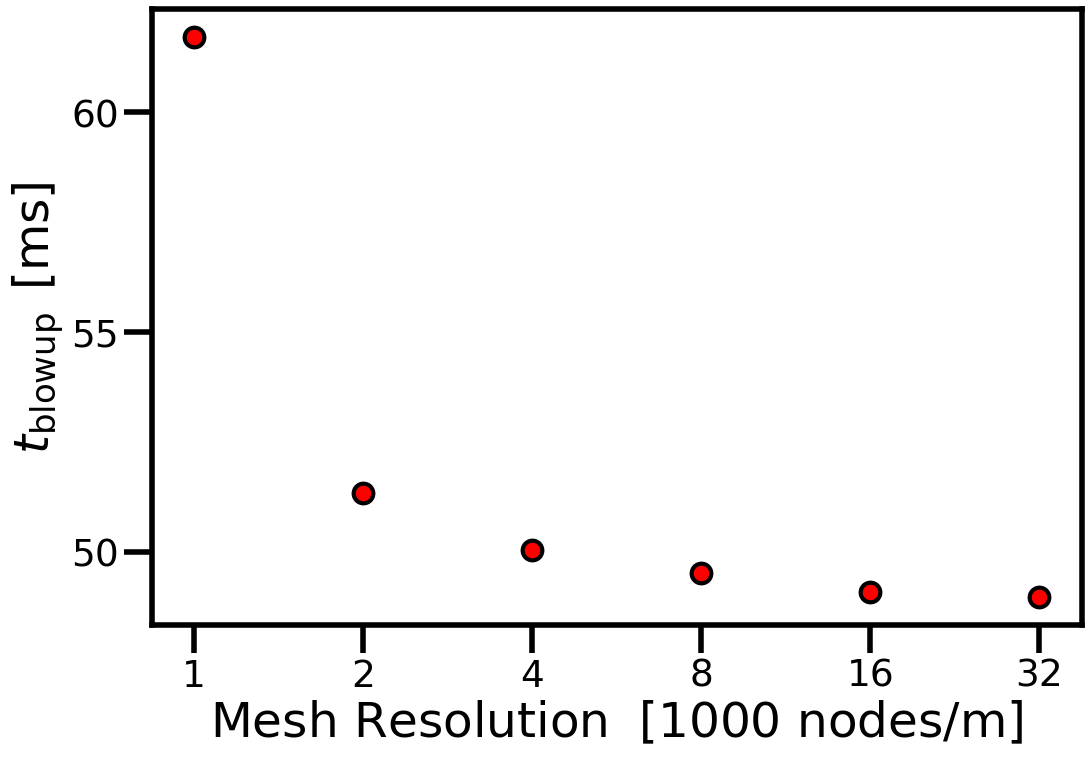}
    \caption{Blowup time on increasingly fine meshes. Measured as the time when the $W$-spike exceeds 3~m~s$^{-1}$.}  \label{fig:BlowupTimeConservative}
\end{figure}

It is worth asking whether the time to this blowup increases significantly as the mesh is refined. Perhaps if the system has more modes to pour energy into, it can hang on for longer times -- perhaps for arbitrarily long. Fortunately from a computing perspective, the trend is the opposite. Blowup times asymptote downward (see Fig.\ \ref{fig:BlowupTimeConservative}) as the nonlinear effects are more accurately represented. The time required to fill additional modes turns out to be negligible. 

This minimal model is ill-posed, not instantly, but after a finite time. Some other mechanism is required to tame this \textit{nonlinear} ill-posedness.

Continuing to analyze the issue via Fourier series is possible but difficult; 
at the end of this Section we will use a perturbative expansion to model the interactions among a finite ladder of modes. First, however, we develop heuristics to understand the effects of high-order forces in the highly nonlinear regime.

\subsection{Characteristics and the Black Box Method} \label{characteristics}

In developing our analytical methods, we look to simpler PDE systems. A family of nonlinear wave equations with analytical solutions descends from the Korteweg-de\,Vries (KdV) equation \cite{KdV1895}:
\begin{equation} \label{eq:KdV}
    \partial_t u + \partial_x (\half u^2) + \sigma \partial_x^3 u = 0
\end{equation}
and the Burgers equation \cite{Burgers1948}:
\begin{equation} \label{eq:Burgers}
    \partial_t u + \partial_x (\half u^2) = \nu \partial_x^2 u
\end{equation}

Many approaches to these problems involve some variation of the characteristic method, meaning the assumption of solutions in terms of one combined variable:
\begin{equation}
    \theta = x - v_c t
\end{equation}
where $v_c$ is the characteristic velocity. The one-soliton solution to the KdV equation assumes a globally constant $v_c$, while the pre-shock solution of the inviscid Burgers problem utilizes $v_c(\theta) = u(x=\theta,t=0)$; i.e.\ constant velocity along characteristics.

The nature of the TFM seriously complicates the application of this methodology. The linearized problem generates candidate characteristic velocities given by $\frac{w(k)}{k}$, but usually two of them. They can be complex as well as $k$-dependent. 

It is unclear how these challenges may be overcome in an analytical approach to the general nonlinear problem. However, inspiration may be found in the principal dispersion term:
\begin{equation}
    \bar{\omega} = (\Q' W + j) k
\end{equation}
which appears in every relation in Section \ref{linstab}. We name the factor multiplying $k$ the ``eigenvelocity":
\begin{equation}
    v_\text{eig} \equiv (\Q' W + j)
\end{equation}
The eigenvelocity is constant in $k$ and controls the kinetic energy flux of local waves (Eq.\ \ref{eq:localEflux}).

\begin{figure} 
    \begin{subfigure}[b]{0.45\textwidth}
    \includegraphics[width=\textwidth]{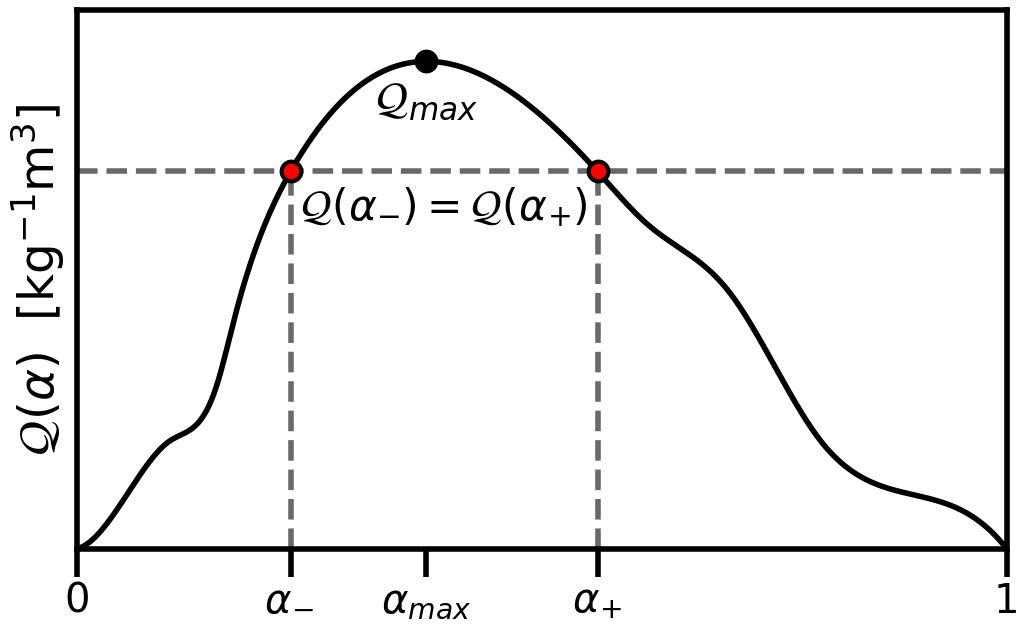}
   \caption{Equal values of the inverse inertial function $\Q(\alpha)$ at two void fractions around a peak, $\alpha_-$ and $\alpha_+$. \newline} 
     \end{subfigure}
    \begin{subfigure}[b]{0.45\textwidth}
        \includegraphics[width=\textwidth]{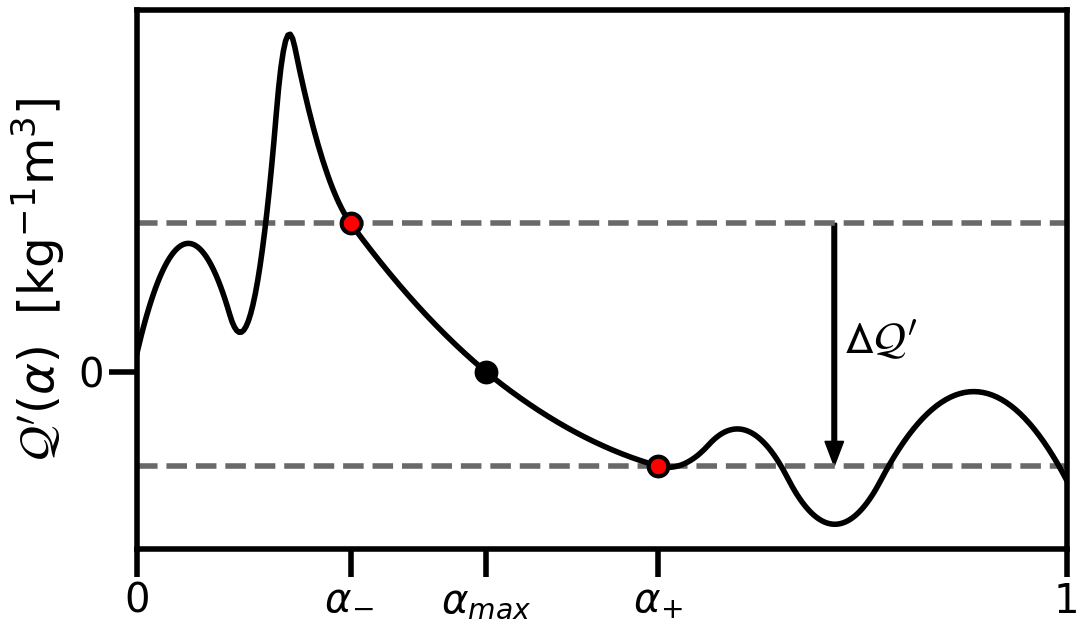}
        \caption{The derivative of the inverse inertial function $\Q'(\alpha) = d\Q / d\alpha$. $\Delta \! \Q'$ between $\alpha_-$ and $\alpha_+$.}
    \end{subfigure}
    \caption{A sample inverse inertial function $\Q(\alpha)$ (a), and its derivative $\Q'$ (b), illustrating the decrease in $\Q'$ around a local maximum in $\Q$.} \label{fig:Qplots}
\end{figure}

To make use of these properties, we construct a thought experiment where, due to piecewise constant flow variables, the eigenvelocity temporarily dominates momentum and energy transport. We start by taking some inverse inertial function $\Q(\alpha)$ with a local maximum $\Q_{max}$ at void fraction $\alpha_{max}$ and select two surrounding values of $\alpha$ yielding equal $Q$:
\begin{equation}
   \alpha_- < \alpha_{max} < \alpha_+ \qquad \Q(\alpha_-) = \Q(\alpha_+) < \Q_{max} 
\end{equation}
We may select $\alpha_-$ and $\alpha_+$ such that:
\begin{equation}
    \Q'(\alpha_-) > 0 > \Q'(\alpha_+)
\end{equation}
We then define:
\begin{equation}
    \Delta \! \Q' \equiv \Q'(\alpha_+) - \Q'(\alpha_-) < 0
\end{equation}
See Fig.\ \ref{fig:Qplots} for an illustration. 

\begin{figure}
    \includegraphics[width=0.45\textwidth]{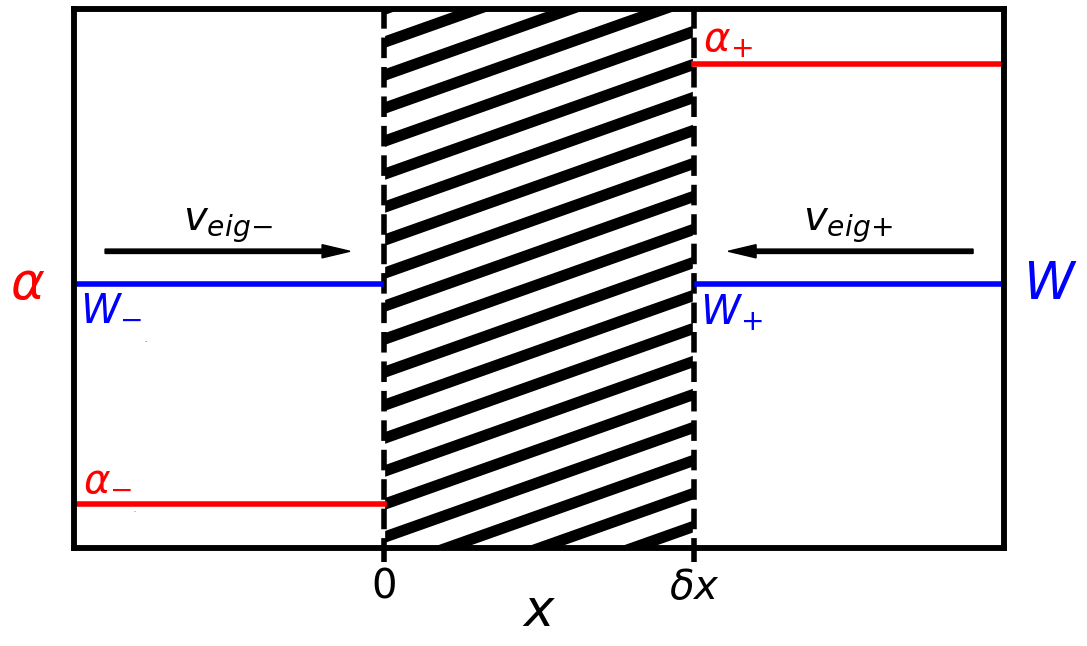}
    \caption{Illustration of flow conditions either side of the ``black-box" region.}
    \label{fig:flowregion}
\end{figure}

Suppose now that at each side of a flow region length $\delta x$ the void fraction is given by $\alpha_-$ and $\alpha_+$ sequentially. Also set the relative momentum density at each side of the region to be equal, $W_- = W_+$, as shown in Figure \ref{fig:flowregion}. We do not specify the conditions inside the region, making it a ``black box". 

We construct the case so that no gradients exist outside the box:
\begin{equation} \label{eq:nogradients}
    \begin{aligned}
        \partial_x \alpha (x<0) &= \partial_x \alpha(x>\delta x) = 0 \\
        \partial_x W (x<0) &= \partial_x W(x>\delta x) = 0 \\
    \end{aligned}
\end{equation}
that is, $\alpha$ and $W$ are piecewise constant.

A constant $j$ acts just as a Galilean frame shift. Working in the frame $j=0$, we have:
\begin{equation}
    \begin{aligned}
        v_{\text{eig}-} = \Q'(\alpha_-) W_- \\
        v_{\text{eig}+} = \Q'(\alpha_+) W_+
    \end{aligned}
\end{equation}
Assuming $W_- = W_+ > 0$ (the generalization to $W<0$ is not difficult), we therefore find:
\begin{equation}
    v_{\text{eig}-}>0>v_{\text{eig}+}
\end{equation}
so the eigenvelocity flows into the black-box region from both sides.

This bunching up of the ``characteristics" has asymmetric effects on the flow variables. We have constructed this example so that the mean void fraction within the region:
\begin{equation}
    \bar{\alpha} \equiv \frac{1}{\delta x}\int_0^{\delta x} \alpha \,dx
\end{equation}
remains constant in time:
\begin{equation}
    \frac{d \bar{\alpha}}{dt} = \frac{1}{\delta x}\left[(\Q W)_- \, - (\Q W)_+ \right]= 0
\end{equation}

The same may not be said for the mean relative momentum:
\begin{equation}
    \overline{W} \equiv \frac{1}{\delta x}\int_0^{\delta x} W \,dx
\end{equation}
Considering only the effect of the advective term (Eq.\ \ref{eq:relmotion}), this quantity is set to grow as long as $\Delta\!\Q'$ is negative:
\begin{equation}
    \frac{d \overline{W}}{dt} = \frac{1}{\delta x}\left[(\Q' W^2)_-\! - (\Q' W^2)_+ \right]= -\frac{\Delta\!\Q' W^2}{\delta x}
\end{equation}
where we have dropped the $+/-$ subscripts on the background $W$. Without any conservative forces, the mean energy density is given by (cf.\ Eq.\ \ref{eq:kinetic}):
\begin{equation}
    \overline{T} \equiv \frac{1}{\delta x}\int_0^{\delta x} \half\!\Q W^2 \,dx
\end{equation}
Using Eq.\ \ref{eq:localadvpower}, we find that this quantity also grows:
\begin{equation}
    \frac{d \overline{T}}{dt} = \frac{1}{\delta x}\left[(\Q'\!\Q W^3)_{\smallminus} \scalebox{0.8}[1.0]{\( - \)} \, (\Q'\!\Q W^3)_+\right] = \scalebox{0.9}[1.0]{\( - \)} \frac{\Delta\!\Q'\!\Q W^3}{\delta x}
\end{equation}

This accumulation of momentum and energy around a localized change in $\alpha$ explains the growth of momentum ``spikes" as seen in Figure \ref{fig:FiniteTimeST}(b)/(c). We may also predict a narrowing of the spikes over time because of the inward flow of the eigenvelocity.


The energy and momentum growth rates diverge as $\delta x \rightarrow 0$, an effect somewhat analogous to the divergence of $\text{Im}(\omega)$ as $k \rightarrow \infty$ in the linear regime under pure advection. However, the behavior of the force terms in this highly nonlinear test case is not at all similar.

Surface tension, for instance, has no limiting effect on the growth of $\overline{W}$ as long as the flow conditions around the black box remain piecewise constant:
\begin{equation}
    \frac{d \overline{W}}{dt} = \; ... + \left. \frac{\partial_{x_{}}^2 \alpha}{[1 + (H \partial_x \alpha)^2]^{3/2}} \right|_0^{\delta x} = \: ... + 0
\end{equation}
nor does it contribute to the energy flux (see Eq.\ \ref{eq:surfEflux}):
\begin{equation}
    \mathcal{F}_{\text{local},\sigma}(x=0) = \mathcal{F}_{\text{local},\sigma}(\delta x) = 0
\end{equation}
despite its effectiveness as a linear limiter.

We now consider a typical viscosity term (Eq. \ref{eq:simplevisc}):
\begin{equation}
    \partial_t W = ... + \partial_x \! \left(\nu \partial_x W \right)
\end{equation}
Regardless of the type of viscosity (kinematic or turbulent) we find that the dissipative effect on the average relative momentum is 0:
\begin{equation}
    \int_0^{\delta x} \! \partial_x \! \left(\nu \partial_x W \right) \, dx= \nu \left. \partial_x W^{^{}}_{_{}} \! \right|_0^{\delta x} = 0
\end{equation}
although the average energy is dissipated (Eq.\ \ref{eq:dlocalwaveenergydt}):
\begin{equation}
    \int_0^{\delta x} \! \Q W \, \partial_x \! \left(\nu \partial_x W \right) dx \approx - \Q \! \int_0^{\delta x} \! \nu (\partial_x W)^2 \, dx
\end{equation}
neglecting the effect of the variation of $\Q$ over the region.

It seems that viscosity would tend to ``spread" $W$ out within the black box, reducing energy but conserving total momentum. 

Granted, the piecewise constant boundary conditions are not likely to remain untouched. Viscosity and surface tension would eventually carry significant gradients to the boundaries, breaking the simplicity of the thought experiment.

Nevertheless, the dissipation of relative flow energy via viscosity gives some hope that runaway growth may be stopped. In order to make a complete analysis of this effect, assumptions about the behavior of the flow variables within the black-box region are necessary.

\subsection{Shock-Spikes and Turbulent Limiting in the Nonlinear Limit} \label{spikeheuristics}

\begin{figure}
    \includegraphics[width=0.45\textwidth]{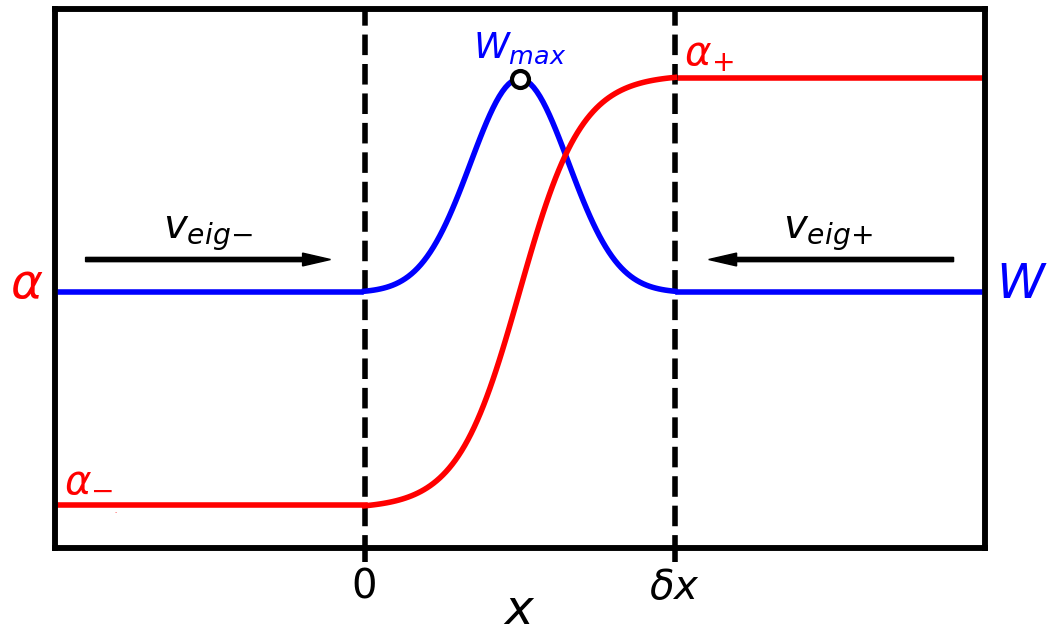}
    \caption{Rough, presumptive forms of the flow variables within the black-box region.}
    \label{fig:filledinflowregion}
\end{figure}

The numerical results in the inviscid case (Fig.\ \ref{fig:FiniteTimeST}) give support to the straightforward assumption that as runaway growth starts within the black box, $\alpha$ takes the form of a narrow shock, and $W$ the form of a narrow spike. See Figure \ref{fig:filledinflowregion} for an illustration. 

We call the peak value of $W$, assumed to occur roughly at the center of the region, $W_{max}$. The difference between the peak and the background momentum density is labeled $\Delta W$:
\begin{equation}
    \Delta W \equiv W_{max} - W_\pm
\end{equation}
As an additional shorthand we define the difference in $\alpha$ across the region:
\begin{equation}
    \Delta \alpha \equiv \alpha_+ - \alpha_-
\end{equation}

We may estimate spatial derivatives of the flow variables at the peak using these deltas and the length scale $\delta x$. For example, the gradient of $\alpha$ is roughly:
\begin{equation} \label{eq:gradalphaestimation}
    \left. \frac{\partial \alpha}{\partial x} \right|_\text{peak} \!\approx \frac{\Delta \alpha}{\delta x}
\end{equation}
Its second derivative is roughly 0, but the third derivative comes out to:
\begin{equation}
    \left. \frac{\partial^3 \alpha}{{\partial x}^3} \right|_\text{peak} \!\approx - \frac{\Delta \alpha}{\delta x^3}
\end{equation}
The gradient of $W$ is assumed to be 0, but its second derivative is significant:
\begin{equation} \label{eq:grad2Westimation}
    \left. \frac{\partial^2 W}{{\partial x}^2} \right|_\text{peak} \!\approx -\frac{\Delta W}{\delta x^2}
\end{equation}


Plugging Eqs.\ \ref{eq:gradalphaestimation}-\ref{eq:grad2Westimation} into the $W$-equation (Eq.\ \ref{eq:relmotion}) yields an approximation for the growth rate of the spike. We handle the key terms one by one.

The advective term comes out to:
\begin{equation} \label{eq:roughadv}
    \partial_x \mathcal{A}_W = \partial_x \! \left( \half \Q' W^2\right) 
    \approx \half W_{max}^2 \Delta\!\Q' \, \delta x^{-1}
\end{equation}
using:
\begin{equation}
    \partial_x \Q' = \Q'' \partial_x \alpha \approx \Q'' \Delta \alpha \ \delta x^{-1} \approx \Delta\!\Q' \,\delta x^{-1}
\end{equation}
To approximate the surface tension term, we first expand:
\begin{equation}
    \begin{aligned}
    &F_{\sigma,W} = \sigma\!H \frac{\partial}{\partial x} \! \left( \! \frac{\partial_x^2 \alpha}{[1 + (H \partial_x \alpha)^2]^{3/2}} \! \right) = \\
    &\sigma\!H\!\left( \! \frac{\partial_x^3 \alpha}{[1 + (H \partial_x \alpha)^2]^{3/2}} - \frac{3H^2(\partial_x^2 \alpha)^2\partial_x\alpha}{[1 + (H \partial_x \alpha)^2]^{5/2}}\! \right)
    \end{aligned}
\end{equation}
Since $\left. \partial_x^2 \alpha \right|_\text{peak} \approx 0$, the second part vanishes. Additionally, keeping $\delta x$ sufficiently small:
\begin{equation}
    H \partial_x \alpha \approx H \Delta\alpha \,\delta x^{-1} \gg 1
\end{equation}
Thus, the denominator of the first part simplifies:
\begin{equation} \label{eq:STdenomscaling}
    [1 + (H \partial_x \alpha)^2]^{\frac{3}{2}} \approx (H \partial_x \alpha)^3
\end{equation}
leaving:
\begin{equation}
    F_{\sigma,W} \approx \sigma\!H \! \frac{\partial_x^3 \alpha}{(H \partial_x \alpha)^3} \approx - \sigma\!H \! \frac{\Delta\alpha \,\delta x^{-3}}{(H \Delta\alpha \, \delta x^{-1})^3}
\end{equation}
Performing the final cancellations reveals a quantity independent of $\delta x$:
\begin{equation} \label{eq:roughSTforce}
    F_{\sigma,W} \approx -\sigma H^{-2} \Delta\alpha^{-2}
\end{equation}
Finally, we have the viscosity term, for which we take the IC-free form (Eq.\ \ref{eq:simplevisc}) as representative:
\begin{equation}
    F_{\text{visc},W} = \partial_x(\nu\partial_xW) = \nu \partial_x^2 W + \partial_x\nu \cdot \partial_x W
\end{equation}
The second term vanishes assuming $\left. \partial_x W \right|_\text{peak} = 0$, so:
\begin{equation}
    F_{\text{visc},W} = (\nu_k + \nu_t) \, \partial_x^2 W
\end{equation}
where $\nu_k$ is the kinematic viscosity and $\nu_t$ is the turbulent viscosity. From Eq.\ \ref{eq:turbvisc}:
\begin{equation} \label{eq:turbviscform}
    \nu_t = l_m |u_r| = l_m \Q \alpha^{-1}(1-\alpha)^{-1} W
\end{equation}
where we have dropped the absolute value bars just for the case $W>0$. Then:
\begin{equation}
    F_{\text{visc},W} \approx - \!\left(\nu_k + \frac{l_m \Q W_{max}}{\alpha(1-\alpha)}\right) \frac{\Delta\!W}{\delta x^2}
\end{equation}
where $\alpha$ and $\Q$ are taken at the location of the $W$-peak.

We now combine the above terms as follows:
\begin{equation}
    d_tW_{max} = -\partial_x \mathcal{A}_W + F_{\sigma,W} + F_{\text{visc},W}
\end{equation}
The advective term $-\partial_x \mathcal{A}_W$ drives growth. In taking $\delta x$ closer and closer to zero (a narrower and narrower shock), it diverges as $\delta x^{-1}$.\ $F_{\sigma,W}$, having no dependence on $\delta x$, becomes irrelevant for narrow shocks. This fact may be surprising given the effectiveness of surface tension as a linear well-poser, but it explains its breakdown in the nonlinear regime (Sec.\ \ref{LinBreakdown}). It is the viscous force, $F_{\text{visc},W}$, that contributes effective nonlinear limiting scaling with $\delta x^{-2}$.


We may consider momentum spikes to be well-behaved if, above some bounding $W_{max}$, they always tend down:
\begin{equation}  \label{eq:boundedcond}
    \frac{d_tW_{max}}{W_{max}} \le 0
\end{equation}
For a narrow shock the surface tension is negligible, so Eq.\ \ref{eq:boundedcond} is equivalent to:
\begin{equation}
    \frac{\partial_x \mathcal{A}_W}{W_{max}} \le \frac{ F_{\text{visc},W}}{W_{max}}
\end{equation}
or roughly:
\begin{equation}
    \half W_{max} | \Delta\!\Q' | \, \delta x^{-1}\!\, \le (\nu_k + \nu_t) \Delta\!W \, W_{max}^{-1} \, \delta x^{-2}
\end{equation}
In the simple case $\Delta\!W \approx W_{max}$ we arrive at:
\begin{equation} \label{eq:spikeinequality1}
    W_{max} \le 2(\nu_k + \nu_t) | \Delta\!\Q' |^{-1} \delta x^{-1}
\end{equation}

The limiting behavior depends on the type of viscosity. With kinematic (constant) viscosity only, the inequality reduces to:
\begin{equation} \label{eq:spikeinequality2}
    W_{max} \le 2\nu_k | \Delta\!\Q' |^{-1} \delta x^{-1}
\end{equation}
showing the wrong kind of threshold.\ Above a certain bound ${W_{max} > 2 \nu_k |\Delta\!\Q'|^{-1} \delta x^{-1}}$\; one obtains super-exponential growth instead of stable limiting. 

The difficulty presented by Eq.\ \ref{eq:spikeinequality2} appears even worse when considering the integrated momentum of the spike:
\begin{equation}
    p_{spike} \propto W_{max} \, \delta x
\end{equation}
which must only exceed a value proportional to $\nu_k | \Delta\!\Q' |^{-1}$ to cause runaway growth.

With the viscosity outcompeted in this way, the dynamics quickly become dominated by the advective term:
\begin{equation} \label{eq:superexponential}
    d_t W_{max} \approx \half |\Delta\!\Q'| \, \delta x^{-1} W^2_{max} 
\end{equation}
This simple ODE is solved by:
\begin{equation} \label{eq:divergentW}
    W_{max}(t) \approx \frac{W_{max}(0)}{1- \half |\Delta\!\Q'| W_{max}(0) \, \delta x^{-1} \, t}
\end{equation}
exhibiting divergence at time:
\begin{equation} \label{eq:divergencetime}
    t_{blowup} \approx  \frac{2\delta x}{|\Delta\!\Q'|W_{max}(0)}
\end{equation}

Finite divergence times in the ultra-nonlinear regime explain the blowup behavior observed in Section\,\ref{LinBreakdown}. We may verify those results against Eqs.\ \ref{eq:divergentW}-\ref{eq:divergencetime} quantitatively. Instead of following $W_{max}$ to infinity numerically, we check the time it takes to go from one finite value to another. Selecting $W_{max}(0) = 2 \frac{\text{m}}{\text{s}}$, and with:
\begin{equation}
    |\Delta\!\Q'| = 2\Delta\alpha \approx 0.16  \qquad \delta x = 1/4000\,\text{m}
\end{equation}
Eq.\ \ref{eq:divergentW} predicts that the time for $W_{max}$ to reach $3 \frac{\text{m}}{\text{s}}$ should be $5.2 \cdot 10^{-4}$\,s. The code with $\Delta x = 1/4000\,$m gets $5.9 \cdot 10^{-4}$\,s (see Fig.\ \ref{fig:FiniteTimeST}c for rough confirmation). These findings are in approximate agreement.


Turbulent viscosity, as we have formulated it, appears to avert this blowup issue. Leaving out the kinematic viscosity, we expand the stability condition Eq.\ \ref{eq:spikeinequality1}:
\begin{equation}
    W_{max} \le \frac{2l_m \Q_{max} W_{max}}{\alpha_{max}(1-\alpha_{max})} | \Delta\!\Q' |^{-1} \delta x^{-1}
\end{equation}
or, equivalently:
\begin{equation}
    \delta x \le \frac{2l_m \Q_{max}}{ \alpha_{max}(1-\alpha_{max}) | \Delta\!\Q' |}
\end{equation}

In the turbulent model the peak height disappears from the inequality entirely. We see that spikes are unconditionally stable below a certain \textit{width} that varies directly with the turbulent mixing length $l_m$.

This very different stability behavior is due to the proportionality between viscosity and relative momentum:
\begin{equation} \label{eq:TVproportional}
    \nu_t \propto |u_r| \propto |W|
\end{equation}
which the concept of turbulent viscosity has allowed us to introduce.

It may then be that all shock-spikes grow and narrow until they reach the turbulent $\delta x$ threshold, after which they dissipate. What remains is to test this hypothesis numerically.

\subsection{Numerical Results from the Turbulence-Limited Model} \label{cascade}

We now analyze the effects of turbulent viscosity (Eq.~\ref{eq:turbvisc}) numerically. We modify the case investigated in Sec.\ \ref{LinBreakdown} to include turbulent viscosity with mixing length $l_m=1$\,mm (Eqs.\ \ref{eq:turbviscappx}-\ref{eq:viscousforce}). $\Delta x = 1/4$\,mm as before. The addition of turbulent viscosity prevents blowups altogether. The simulation results can be viewed in video form at \cite{Video1}.


\begin{figure} 
    \begin{subfigure}[b]{0.45\textwidth}
        \includegraphics[width=\textwidth]{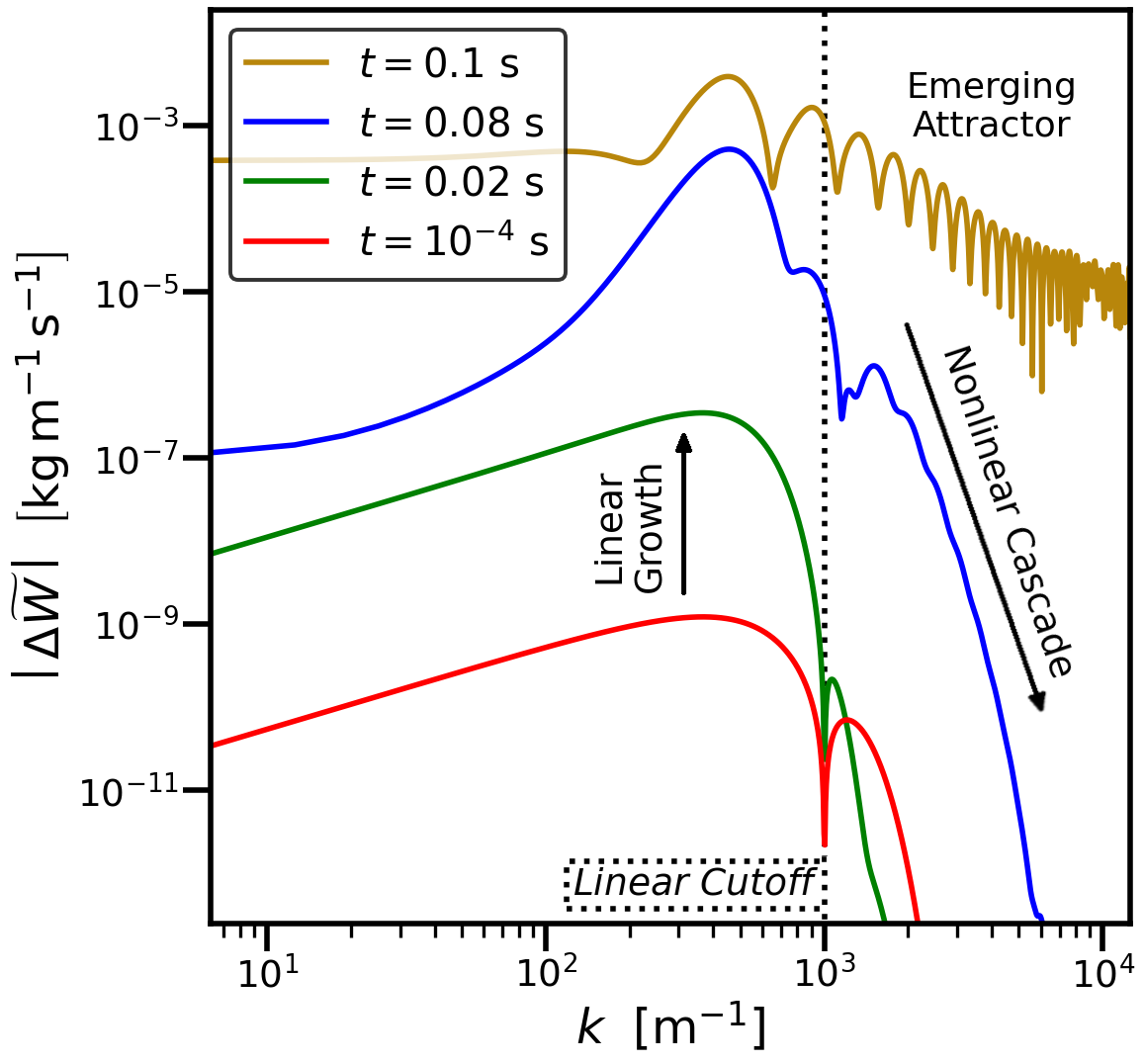}
        \caption{Time evolution of relative momentum Fourier transform. Logarithmic scale. \newline} 
    \end{subfigure}
    \begin{subfigure}[b]{0.45\textwidth}
        \includegraphics[width=\textwidth]{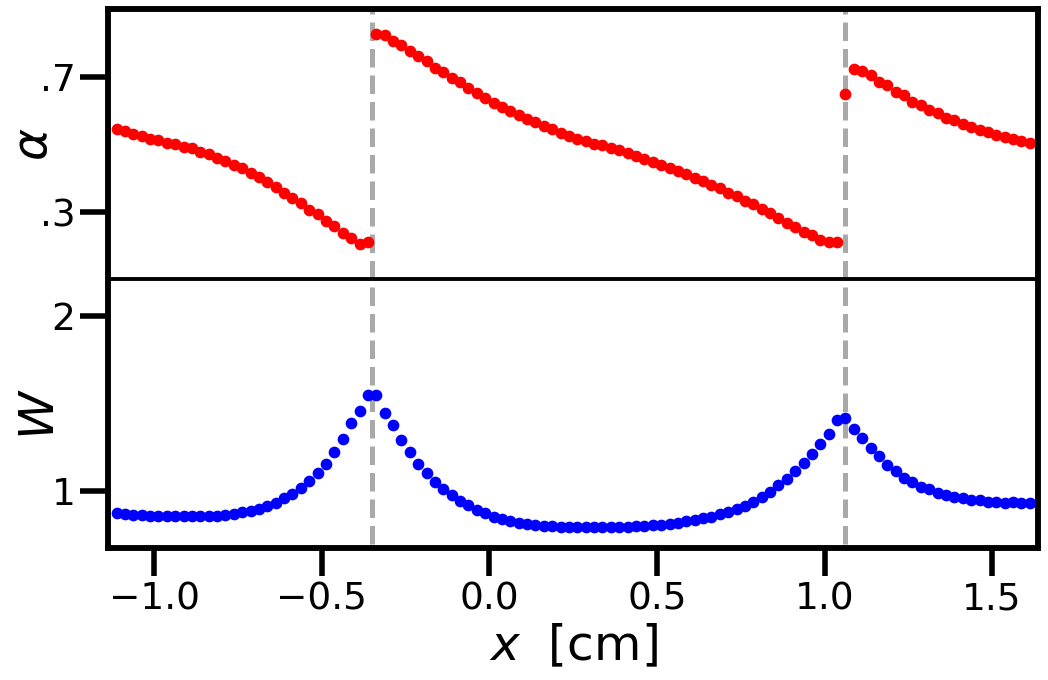}
        \caption{Interface shocks at $t = 0.1$, lined up with comparatively gentle peaks in relative momentum. \newline}
    \end{subfigure}
    \caption{The case of Fig.\ \ref{fig:FiniteTimeST} with turbulent viscosity. The nonlinear cascade approaches a chaotic attractor (a). There continue to be $W$-spikes aligned with interface shocks, but now without single-node outlier peaks (b).} \label{fig:FiniteTimeTV}
\end{figure}

Figure \ref{fig:FiniteTimeTV}a shows that the linear response in $W$ to the Gaussian interface perturbation (Eqs.\ \ref{eq:gaussianpert}-\ref{eq:gaussianpertvalues}) differs slightly versus the inviscid case. Wave growth progresses at roughly half speed due to the damping effect of viscosity, and instead of exhibiting oscillatory behavior, the post-cutoff wavenumbers undergo linear decay (see Appx.\ \ref{appx:linear}). A small signature of nonlinear growth is visible in the post-cutoff modes as early as 0.02\,s.

As before, a nonlinear cascade emerges at some intermediate time; see Fig.\ \ref{fig:FiniteTimeTV}(a), 0.08\,s. However, whereas the inviscid cascade approaches a flat line (Fig.\ \ref{fig:FiniteTimeST}a), the turbulent cascade holds to a downward sloping curve; see Fig.\ \ref{fig:FiniteTimeTV}(a), 0.1\,s. The former (flat) Fourier transform corresponds to a delta function in real space, signaling imminent blowup, whereas this new Fourier transform corresponds to resolvable cusp peaks in the real-space distribution of the relative momentum. These peaks are displayed in Figure \ref{fig:FiniteTimeTV}b.

\begin{figure}  
    \includegraphics[width=0.45\textwidth]{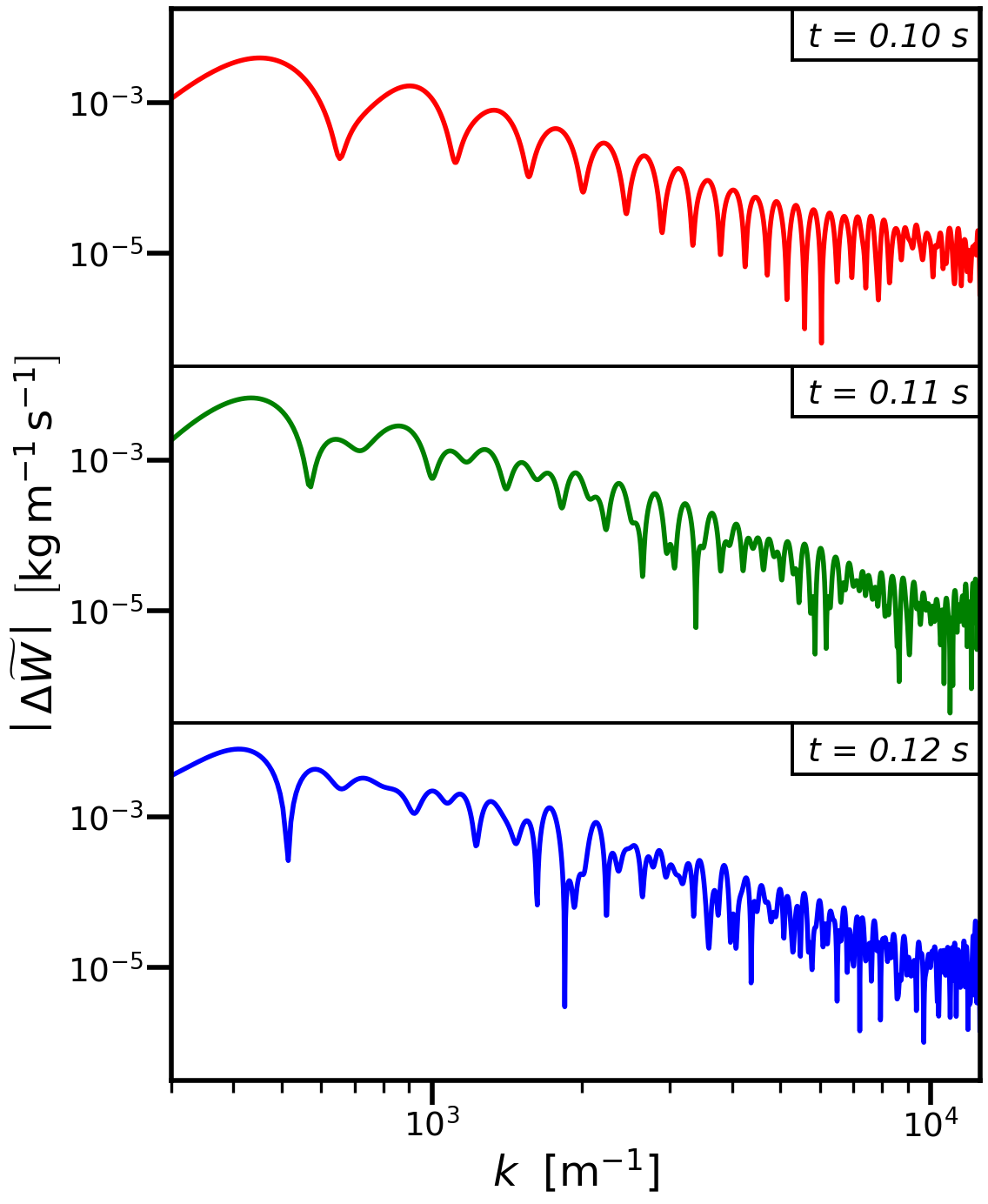}
    \caption{Transition of $W$'s Fourier transform from a periodic structure towards chaos. Logarithmic scale.}  \label{fig:ChaosTransitionTV}
\end{figure}

Figure \ref{fig:ChaosTransitionTV} shows the initially oscillatory superstructure of the Fourier transform, overlaying the overall downwards slope, as it becomes gradually more chaotic. After power cascades down the spectrum, ``noise" originating at high wavenumber travels back up. 
This Fourier-space noise is probably seeded by tiny numerical errors early on, but it provides a good analogue to the noise one might expect from any number of sources in a real case. It does not manifest itself strongly in the real-space variables. 
It also tends to remain within a narrow range around the downward slope, indicating a kind of chaotic attractor. However, in this case, the viscosity continually dissipates energy, so trying to follow this dynamic over longer times yields null results.

We therefore add a driving mechanism, that being channel-parallel gravity ($g_x$). We also include interfacial drag ($C_D$) to balance it in totally smooth flow.

\begin{figure}  
    \includegraphics[width=0.45\textwidth]{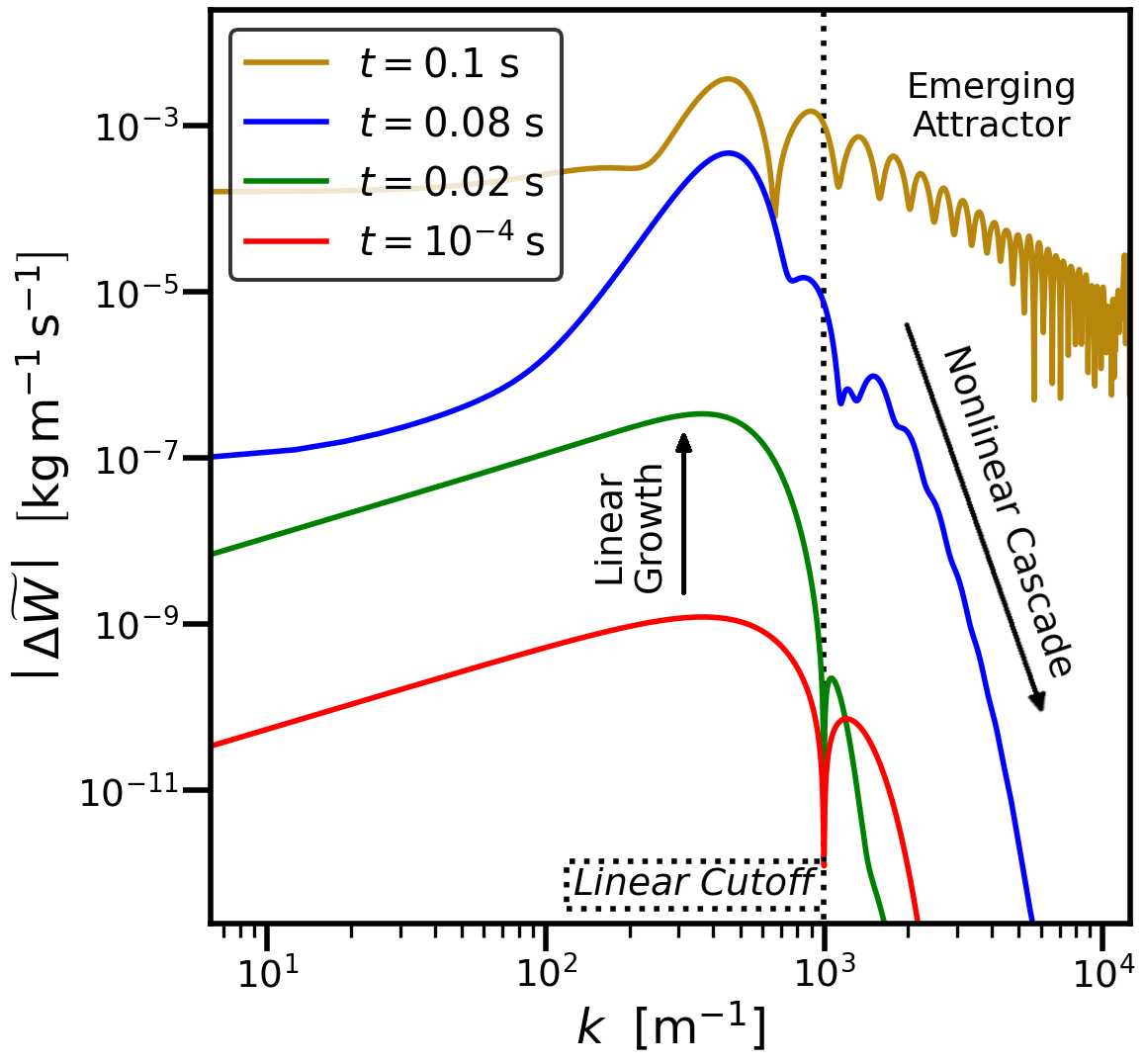}
    \caption{Time evolution of relative momentum Fourier transform with parallel gravity (driver) and interfacial drag (dampener). Logarithmic scale.}  \label{fig:FiniteTimeD+D}
\end{figure}

Since the parallel gravity term in the $W$-equation depends on $\delrho$, we may no longer hold the densities equal. Instead we use:
\begin{equation}
    \rho_1 = 1.1 \qquad \rho_2 = 0.9
\end{equation}
We then set $g_x = -1 \frac{\text{m}}{\text{s}^2_{}}$, $f_\text{int} = 9.1\cdot10^{-4}$, and $\rho_D = \rho_1$ (see Eqs.\ \ref{eq:dragforce} and \ref{eq:limdraglaw}). These values would maintain a steady state of $\alpha(x) = 0.5$ and  $W(x) = 1 \frac{\text{m}}{\text{s}}$, i.e.\ our initial conditions sans perturbation.

Adding these new parameters leaves the short-time dynamics essentially unmodified (Figure \ref{fig:FiniteTimeD+D}). Small differences may owe as much to the slight changes in density as the addition of the new forces.

\begin{figure}
    \includegraphics[width=0.45\textwidth]{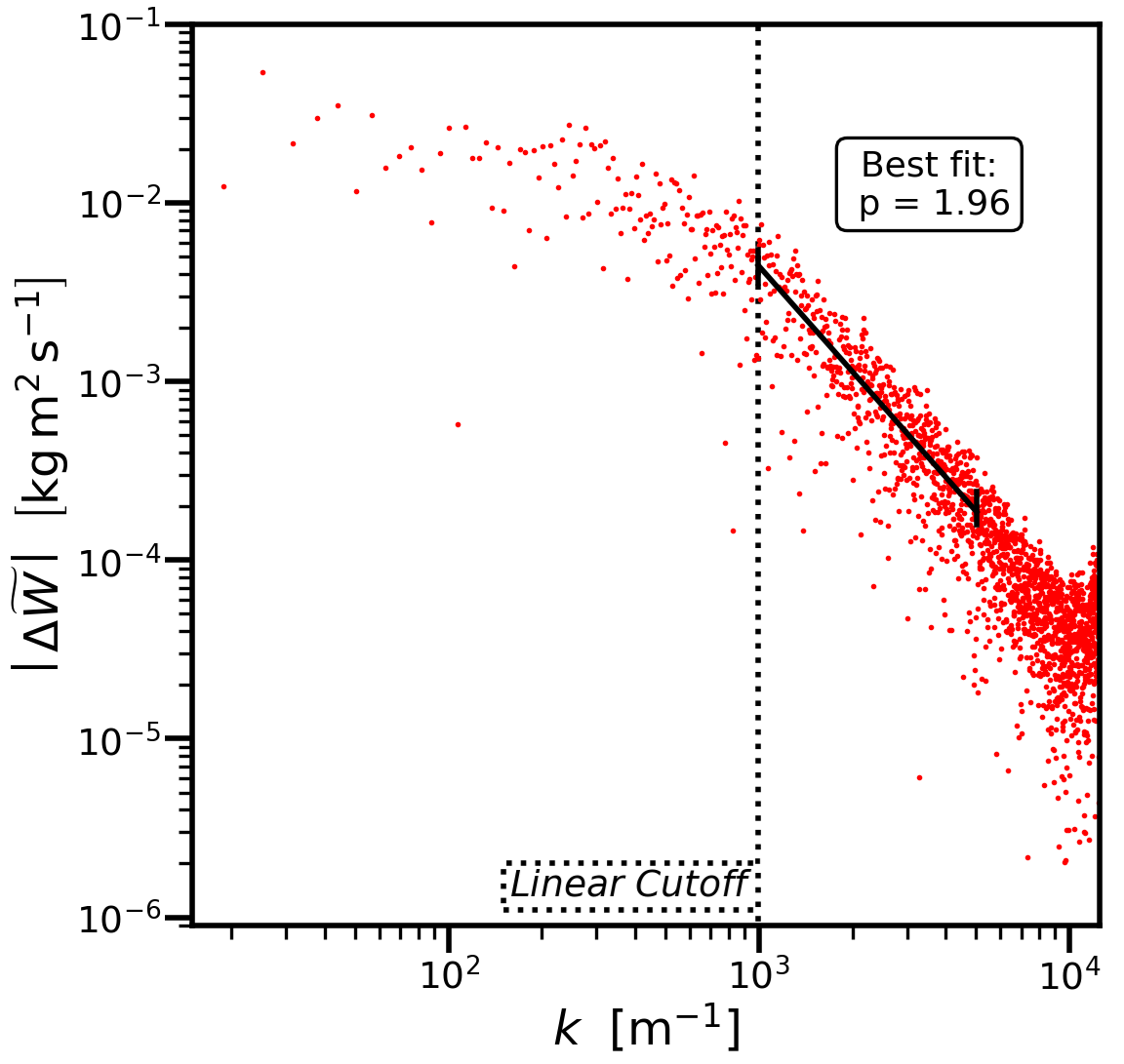}
    \caption{Relative momentum Fourier spectrum at 0.5 s with power-law fit and corresponding exponent. Logarithmic scale.}  \label{fig:FFTfit}
\end{figure}

The Fourier spectrum of $W$ transitions to chaos as before, resulting in magnitudes scattered around a decaying power distribution:
\begin{equation} \label{eq:attractor}
    |\Delta \widetilde{W}(k>k_\text{cut})| \propto k^{-p} + \text{noise}
\end{equation}
The spectrum at 0.5 s is shown in Figure \ref{fig:FFTfit}, along with an approximate fit of the decay exponent $p$ as determined from that instant alone. The fitting procedure is designed to capture an overall trend without privileging fluctuations or high $k$, where there are more data and low-level numerical noise plays an outsized role. We therefore prune the data to $k$'s between the linear cutoff and 70 \% of the mesh maximum and take a centered moving average. We fit the resulting ``clean" dataset to Eq.\ \ref{eq:attractor} with a residual given by squared deviations of the log's, weighted by $k^{-1}$.

\begin{figure}
    \includegraphics[width=0.45\textwidth]{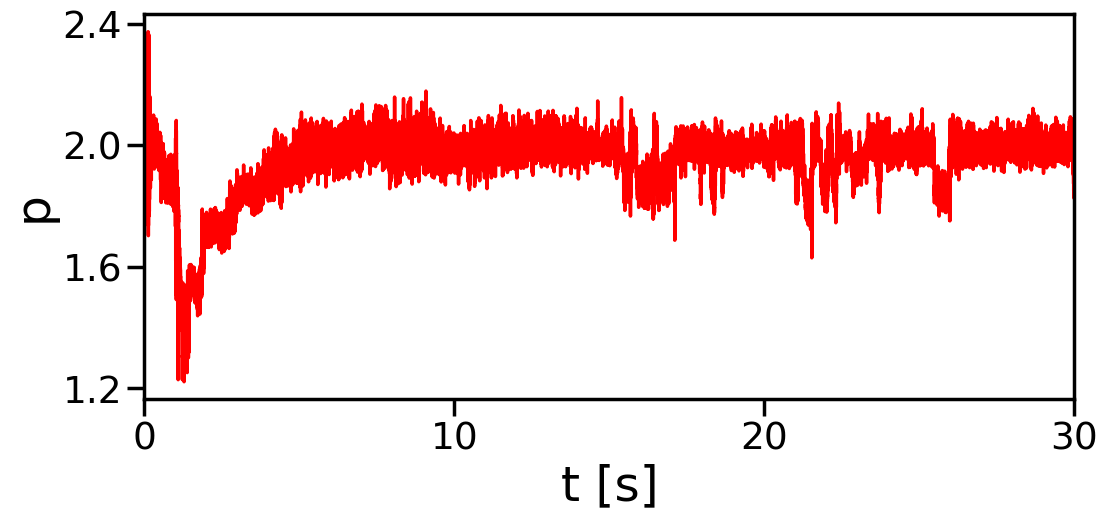}
    \caption{Time evolution of Fourier spectrum exponent.}  \label{fig:FFTexpovertime}
\end{figure}

The evolution of the fitted exponent $p$ from the time when the power law becomes descriptive ($\sim\,$0.1 s) is shown in Figure \ref{fig:FFTexpovertime}. Though fluctuations occur over various timescales, $p$ remains near 2 with some consistency. $p>1$ is required for boundedness of the real-space distribution when accounting for $k\to\infty$. Meaningful deviations occur during the initial development of the flow (up to 5 s) and later as a result of merging or splitting of cap bubbles. An overlay of high frequency noise arises from the propagation of small changes and/or numerical errors through the Fourier transform and the fitting algorithm. Even still, the results always exhibit $p>1$.

\begin{figure}
    \includegraphics[width=0.45\textwidth]{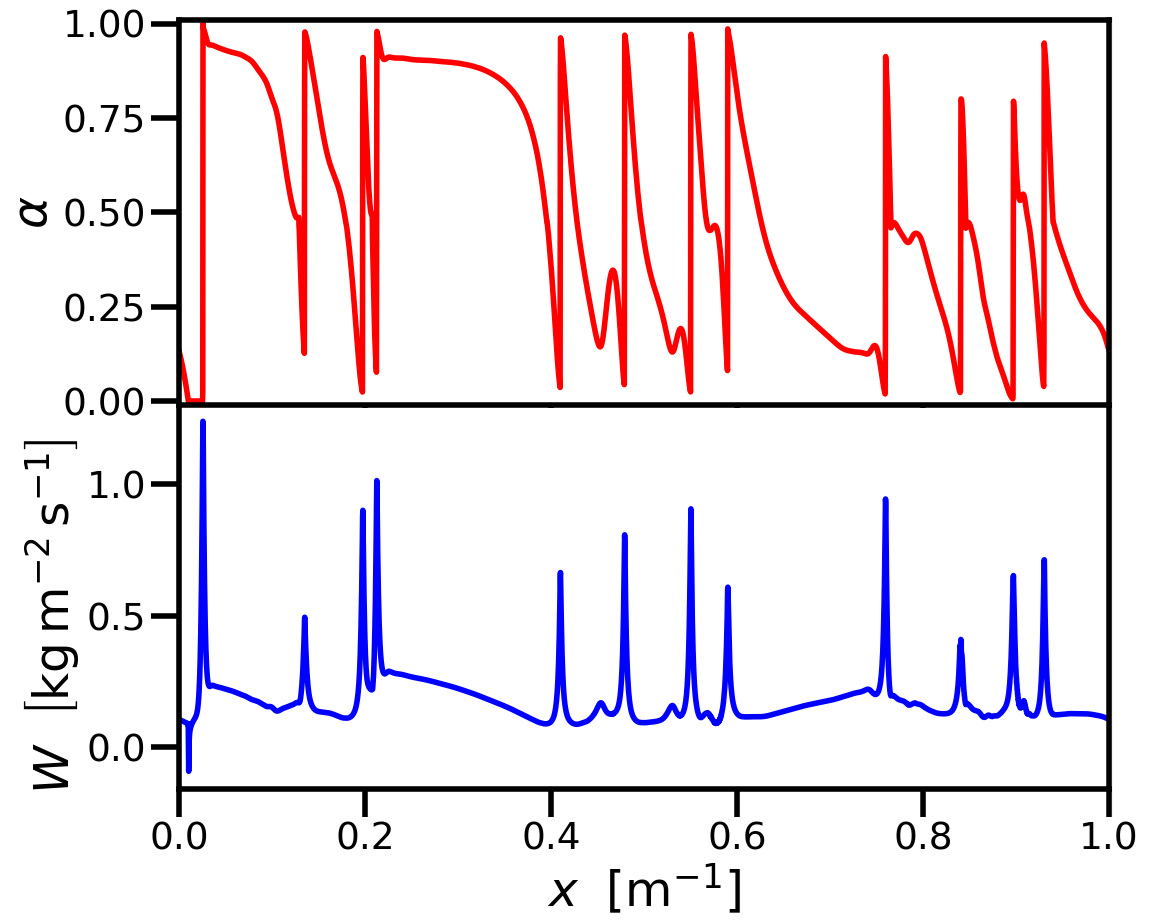}
    \caption{Void fraction and relative momentum profiles during transient churn flow, t = 5\;s.}  \label{fig:ChurnFlow}
\end{figure}

\begin{figure}
    \includegraphics[width=0.45\textwidth]{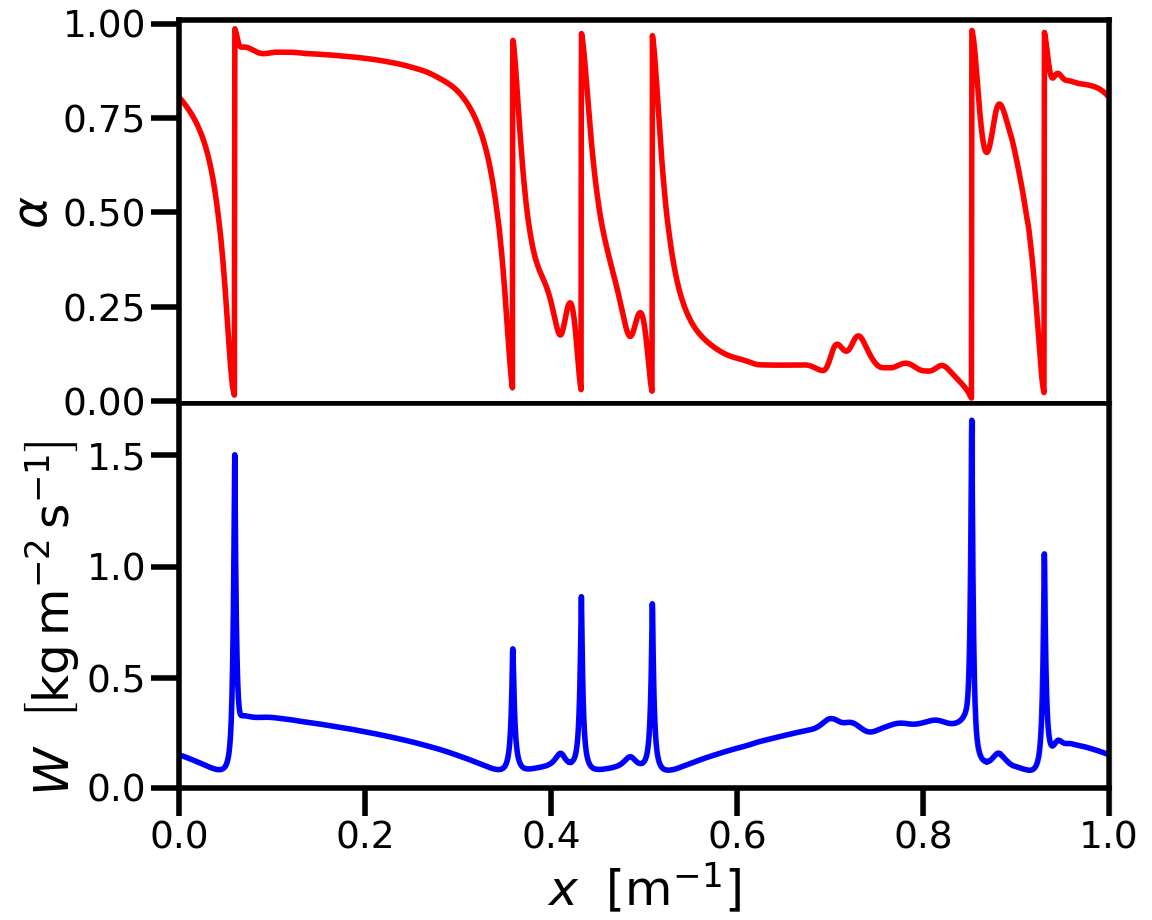}
    \caption{Void fraction and relative momentum profiles in pseudo-slug flow, t = 20\;s.}  \label{fig:PseudoSlugFlow}
\end{figure}

We now discuss the corresponding real-space behavior in detail. As the effect of the initial perturbation spreads across the channel, the flow evolves into a collection of ``bubbles" in either phase, divided by shock-spikes. A kind of churn occurs as these bubbles split and recombine chaotically (Figure \ref{fig:ChurnFlow}). By 20 s, this churn has mostly settled down into a finalized series of Taylor (cap) bubbles, (Figure \ref{fig:PseudoSlugFlow}). This case does not exhibit full-cross-section ``slugs" following the cap bubbles, but the flow is largely static. We dub it a ``pseudo-slug" flow. 

Judging by the repeated splitting of the of the large heavy-phase bubble (viewable at \cite{Video1}), there appears to be a preference for cap bubbles composed of the lighter phase, but the similar densities make this difference relatively subtle.

\begin{figure}
    \begin{subfigure}[b]{0.45\textwidth}
        \includegraphics[width=\textwidth]{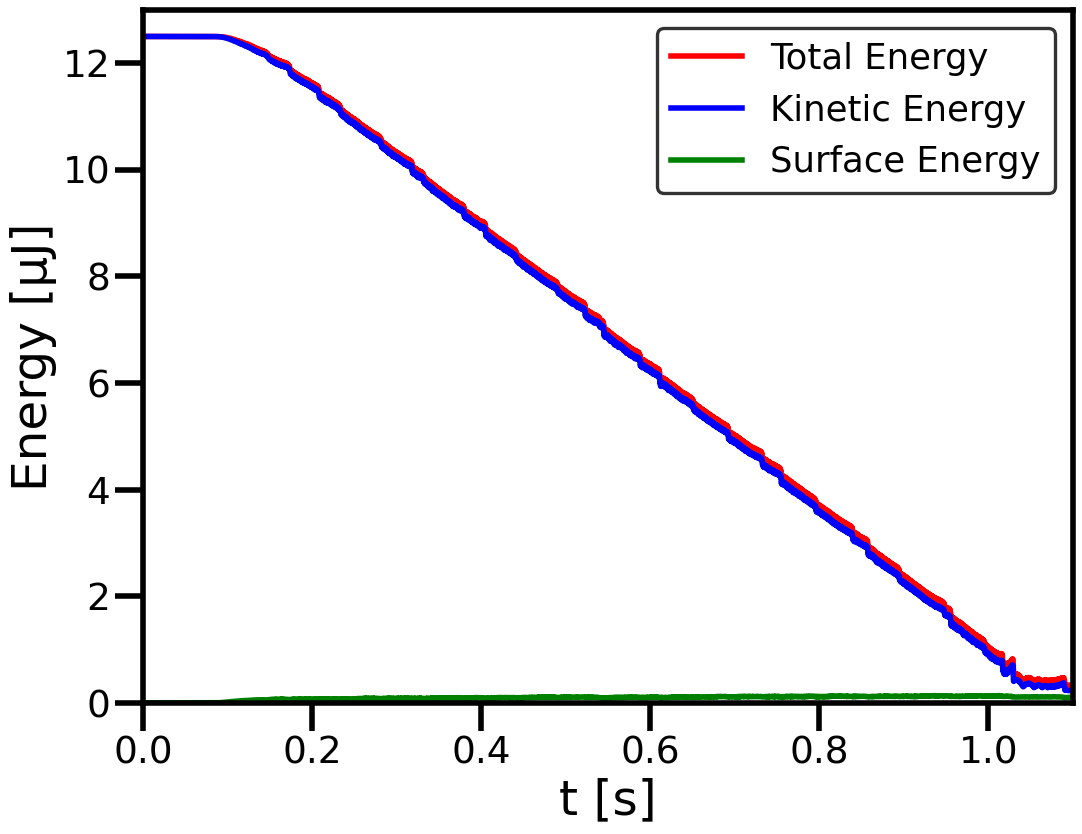}
        \caption{Flow energy from $t=0$ s to 1.1 s.} 
    \end{subfigure}
    \begin{subfigure}[b]{0.45\textwidth}
        \includegraphics[width=\textwidth]{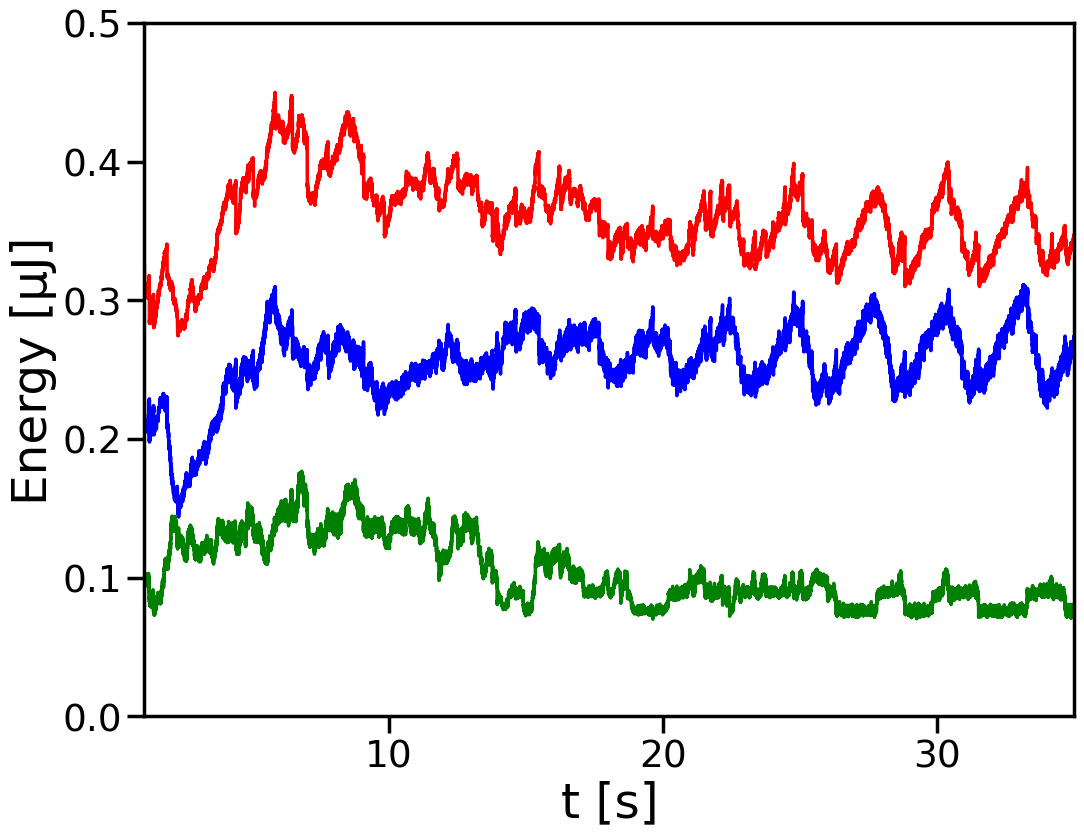}
        \caption{Flow energy from $t=1.1$ s to 35 s.}
    \end{subfigure}
    \caption{Integrated flow energy over two timescales. Cross-sectional area $A =H^2=1 \,\text{cm}^2$.} 
    \label{fig:EnergyOverTime}
\end{figure}

The behavior of the flow energy, shown in Figure \ref{fig:EnergyOverTime}, also tracks what is expected during slugging. Initially ($t<0.1$ s) the perturbation remains small, so the energy stays near its smooth-stratified equilibrium, and the surface energy (Eq.\ \ref{eq:surfenergy} minus its base value $\sigma/H$) is insignificant. The end of this period is marked by the formation of full-height shock-spike waves. They spread through the channel at a constant speed ($t<1$ s), reducing the total energy linearly. An intervening period of churn ensues ($t<20$ s), followed by semi-regular fluctuations of the flow energy near a new ``equilibrium".

The overall reduction in energy after the formation of slugs ultimately owes to the increasing strength of interfacial drag as $\alpha$ tends to extreme values (Eq.\ \ref{eq:limdraglaw}).

We make one further refinement to the present model: local averaging of the turbulent viscosity (Eq. \ref{eq:localavgvisc}). Taking into account the fact that eddies should be affected by flow conditions across their length is an upgrade in terms of physicality. Furthermore, local averaging has advantages for nonlinear stability.

We run another simulation using the same parameters and initial conditions as previously, plus the new averaging length parameter $\delta_m\equiv2l_m \equiv 2$mm. A video of the results can be viewed at \cite{Video2}.


\begin{figure}
    \includegraphics[width=0.45\textwidth]{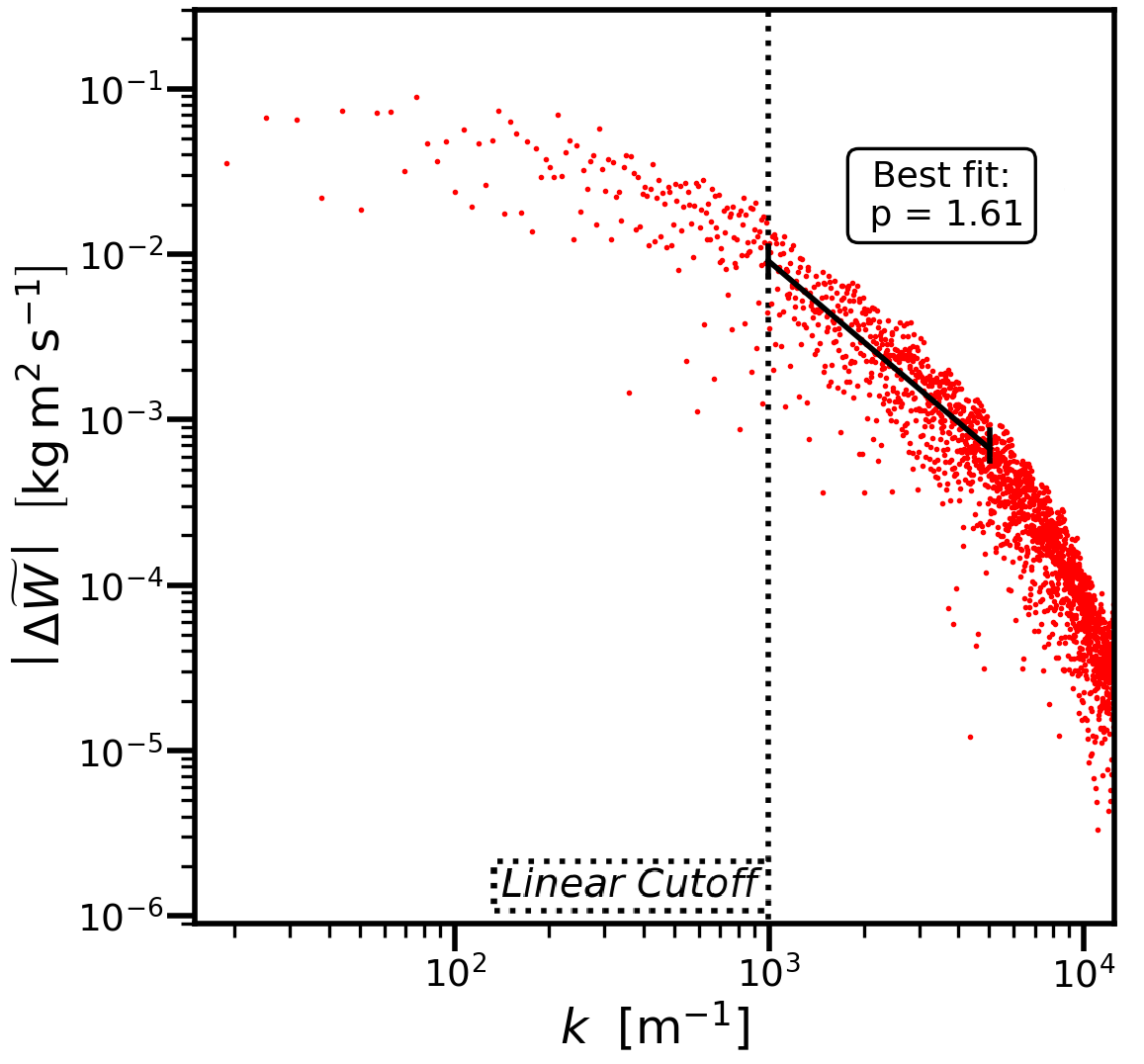}
    \caption{Relative momentum Fourier spectrum at 0.5 s with power-law fit and corresponding exponent, using local averaging of the turbulent viscosity. Log scale.} 
    \label{fig:FFTfitlocalavg}
\end{figure}

\begin{figure}
    \includegraphics[width=0.45\textwidth]{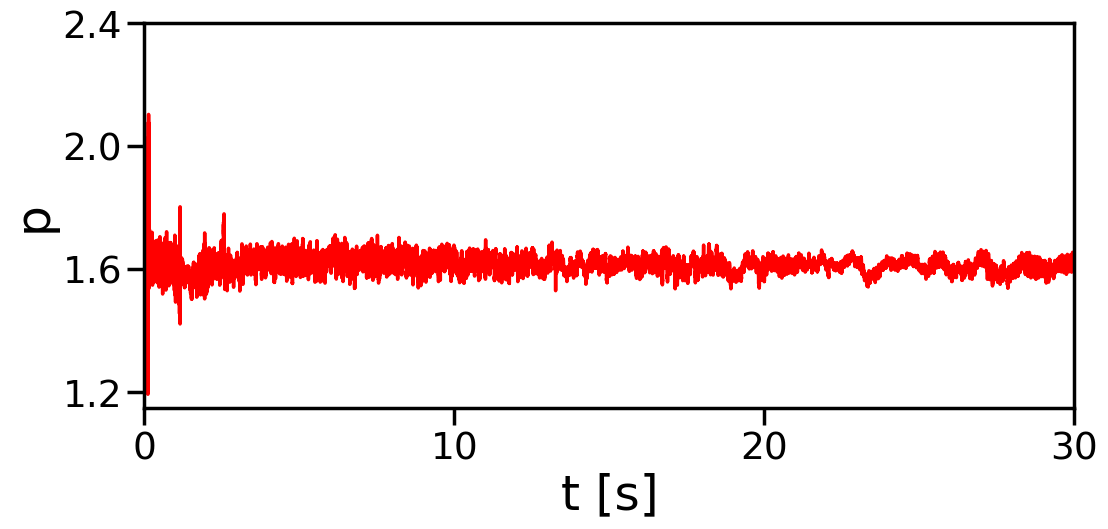}
    \caption{Time evolution of Fourier spectrum exponent, using local averaging of the turbulent viscosity.}  \label{fig:FFTexpovertimelocalavg}
\end{figure}

The early-time development of the relative momentum Fourier spectrum is nearly indistinguishable from Figure \ref{fig:FiniteTimeD+D}, so we do not show it. More substantial differences in the $W$ spectrum emerge within 0.5 s, see Figure \ref{fig:FFTfitlocalavg}. The downward slope through intermediate $k$ is more subdued than with non-averaged turbulent viscosity; however, instead of demonstrating a sudden increase in magnitude at the highest $k$, this new spectrum bends downward more sharply above what we call a ``dissipation threshold", in this case $k_\text{dsp}\approx6\cdot10^{-3}$. This structure maintains itself for the duration of the run.

 This new spectrum is similar to a Kolmogorov cascade \cite{Kolmogorov1941}. Over an inertial range -- where dissipation is comparatively inactive -- a power-law decrease is observed. Beyond it, the action of viscosity causes an increasingly abrupt decline. This resemblance is partly coincidental. Kolmogorov analyzed isotropic single-phase flows in three dimensions, tracking the energy of turbulent fluctuations. We work with two fluids in 1-D, tracking a bulk flow momentum. The tendency to slope downwards more sharply at high $k$ nevertheless shares a single cause, that being increased dissipation. 


The fit exponent $p$ hovers at 1.6 instead of 2 over the duration of the run, with significantly smaller deviations from the mean overall, see Figure \ref{fig:FFTexpovertimelocalavg}. The condition $p > 1$ is still met.

\begin{figure}
    \includegraphics[width=0.45\textwidth]{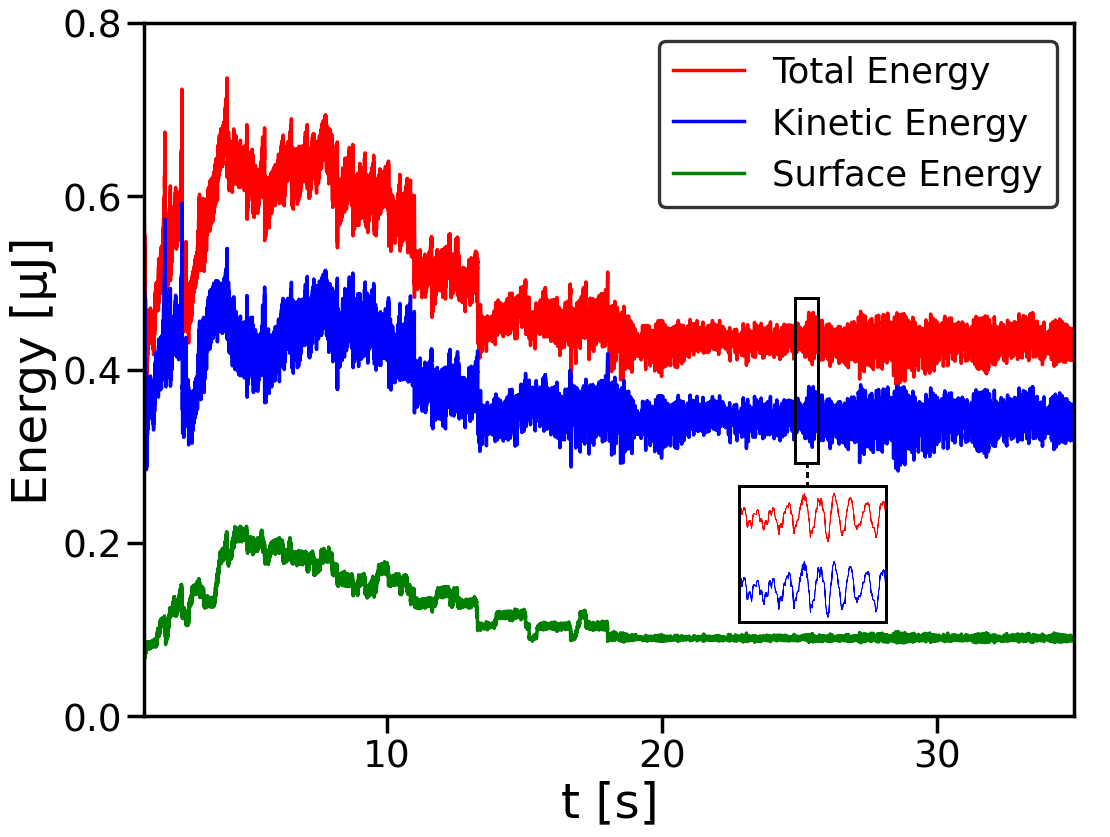}
    \caption{Integrated flow energy from $t=1.2$ s to 35 s, using local averaging of the turbulent viscosity.}  \label{fig:EnergyOverTimelocalavg}
\end{figure}

The real-space behavior is marginally different. The period of churn between 1 and 20 seconds is slightly rougher, and the final series of cap bubbles is more stable. These findings may be confirmed by viewing \cite{Video2}.

As for the energy, we can remark that the final pseudo-slug state with local averaging of the turbulent viscosity is comparatively more energetic, settling at about $0.44\,\mu$J vs. $0.34\,\mu$J, see Fig.\ \ref{fig:EnergyOverTimelocalavg} vs.\ Fig.\ \ref{fig:EnergyOverTime}. This difference lies mostly in the kinetic component, and it may owe to decreased turbulent dissipation or a complex interaction with the interfacial drag. The shorter, smaller oscillations observed in the final state energy are in line with increased cap bubble stability.

\begin{figure}
    \includegraphics[width=0.45\textwidth]{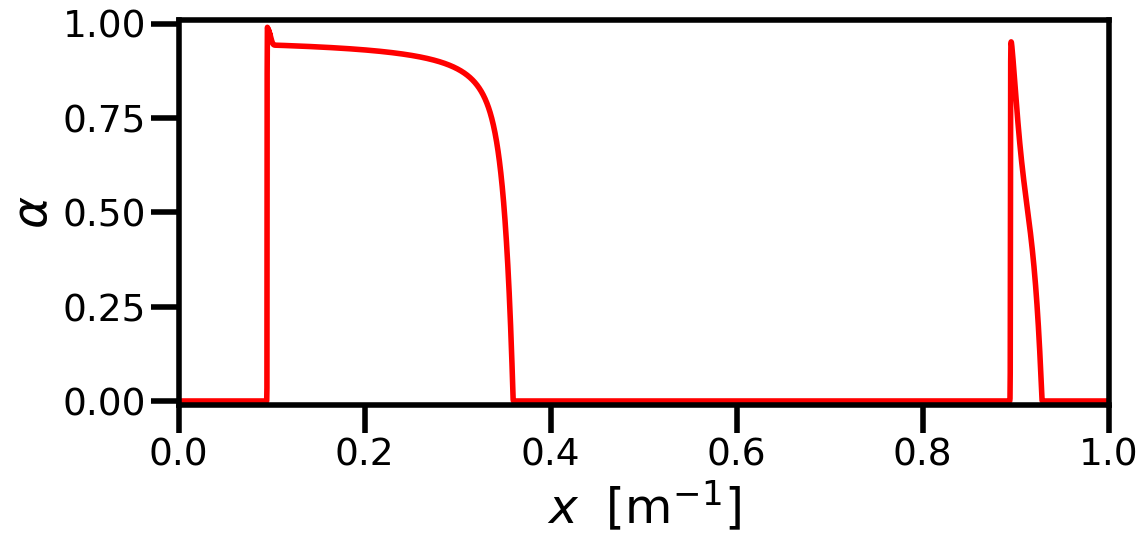}
    \caption{Void fraction profile in a case resulting in slug flow, t = 15\;s.}
    \label{fig:SlugFlow}
\end{figure}

So far we have found a kind of pseudo-slug flow, exhibiting the beginnings of Taylor bubbles in both phases, but without the bubbles detaching from one another and forming fully defined slugs. We now decrease the mean void fraction to 0.25, add wall drag with $f_1 = f_2 = f_\text{int}$ (see Eq.\ \ref{eq:walldrag}), and set a total volumetric flux $j=0.5$. Every other parameter is kept the same, and we continue with local averaging of $\nu_t$. This new case achieves full slug flow, see Figure \ref{fig:SlugFlow}.

A video of the full simulation may be viewed at \cite{Video3}. One key ingredient in these results is lower energy in the relative flow (Eq.\ \ref{eq:localwaveenergy}) of about 0.11 $\mu$J in the final state, owing to dissipation from the wall drag. Higher settled energy, on the other hand, tends to cause stronger and more persistent churn.

Starting from a Kelvin-Helmholtz instability, bounded linearly by surface tension and nonlinearly by turbulent viscosity, we have demonstrated the emergence of slug flow in the TFM without any mechanisms that assume or impose such a structure. 

Clear Taylor bubbles, with caps correctly oriented in the direction of relative motion, are one possible result of the previously described shock-spike waves. They can be thought of as a kind of ``soliton" produced by an inertially unstable, but \textit{well-posed} TFM.



\subsection{Perturbative Series Analysis} \label{series}

We now embark on a rigorous analytical treatment of the Fourier mode interactions in an inertially unstable TFM. This approach provides a degree of insight into the preceding results, but we leave it for last because it is outside our present scope to pursue it to the point of decisive conclusions about nonlinear stability. For that purpose, the heuristic analysis of Sec.\ \ref{spikeheuristics} is much more economical.

Our numerical results show that low-$k$ perturbations in the TFM excite higher-$k$ harmonics. Eq.\ \ref{eq:2ndorderconvol} illustrates the general mathematical structure underlying these cross-wavelength interactions, but to analyze them in more detail, we test a perturbative series solution:
\begin{equation} \label{eq:pertseries}
    \begin{aligned}
        &\Delta \alpha = \eps \! \left( \widetilde{\alpha}_1 e^{i\theta} + \widetilde{\alpha}_{\smallminus 1} e^{\smallminus i\theta^*} \right) +\\
        &\qquad + \eps^2 \! \left( \widetilde{\alpha}_2 e^{2i\theta} + \widetilde{\alpha}_{\smallminus 2} e^{\smallminus 2i\theta^*} \right) + \mathcal{O}(\eps^3) \\
        &\Delta W = \eps \! \left( \widetilde{W}_1 e^{i\theta} + \widetilde{W}_{\smallminus 1} e^{\smallminus i\theta^*}\right) +\\
        &\qquad + \eps^2 \! \left( \widetilde{W}_2 e^{2i\theta} + \widetilde{W}_{\smallminus 2} e^{\smallminus 2i\theta^*} \right) + \mathcal{O}(\eps^3)
    \end{aligned}
\end{equation}
where $\eps$ is a perturbative parameter such that $0<\eps\ll 1$. $\widetilde{\alpha}_{\pm j}$ and $\widetilde{W}_{\pm j}$ denote magnitudes of order-$j$ perturbations. Real solutions require:
\begin{equation} \label{eq:realnesscond}
    \begin{aligned}
    \widetilde{\alpha}_j &= \widetilde{\alpha}_{\smallminus j}^{\, *} \\
    \widetilde{W}_j &= \widetilde{W}_{\smallminus j}^{\, *} \\
    \end{aligned}
\end{equation}
The phase $\theta$ is defined as:
\begin{equation}
    \theta \equiv k x - \omega t
\end{equation}
where $\omega$ may be complex:
\begin{equation}
    \omega = \text{Re}(\omega) + i \, \text{Im}(\omega)
\end{equation}
meaning that the phase also has an imaginary part:
\begin{equation}
    \text{Im}(\theta) = - \text{Im}(\omega) \cdot t
\end{equation}
hence the necessity of the complex conjugate of $\theta$ in minus-mode exponents (Eq.\ \ref{eq:pertseries}) to construct real-valued solutions. We could equivalently write:
\begin{equation} \label{eq:pertseriesRe}
    \begin{aligned}
        \Delta \alpha &= \eps \,\text{Re} \!\left( \widetilde{\alpha}_1 e^{i\theta} \right) + \eps^2 \,\text{Re}\! \left( \widetilde{\alpha}_2 e^{2i\theta}\right) + \mathcal{O}(\eps^3) \\
        \Delta W &= \eps \,\text{Re} \!\left( \widetilde{W}_1 e^{i\theta} \right) + \eps^2 \,\text{Re} \!\left( \widetilde{W}_2 e^{2i\theta}\right) + \mathcal{O}(\eps^3)
    \end{aligned}
\end{equation}
but the expanded form is more useful in understanding the emergence of nonlinear corrections to $\omega$.

Perturbing the PDEs around some equilibrium and substituting in this hypothetical solution (Eq.\ \ref{eq:pertseries}) allows for the grouping of perturbations by order (power of $\eps$) and wavelength (power of $e^{i \theta}$). Grouping together the first order ($\eps^1$) terms results in linearization; supplying $\widetilde{\alpha}_1$ yields two solutions for the linear frequency $\omega_0$ and the coefficient $\widetilde{W}_1$.

At second order one obtains secondary coefficients $\widetilde{\alpha}_{2}$ and $\widetilde{W}_{2}$. At third order, the terms multiplying $e^{3i\theta}$ yield the tertiary coefficients, but interestingly there are also 3rd-order terms multiplying $\eps^3 e^{i\theta + 2 \text{Im}(\omega) t}$: nonlinear products of perturbations in $\eps^2 e^{2i\theta}$ and $\eps e^{\smallminus i\theta^*}$.

In Stokes' original application of the series method \cite{Stokes1847}, $\omega$ is purely real, so these cross terms are fully accounted for by a correction to $\omega$:
\begin{equation}
    \omega = \omega_0 + \eps^2 \omega_2 + ...
\end{equation}
such that the time derivatives on the left-hand side of the PDEs yield:
\begin{equation}
    \partial_t \Delta \phi = \eps \widetilde{\phi}_1 e^{ikx} \partial_t e^{\smallminus i\omega t} = (\eps \omega_0 + \eps^3 \omega_2 + ...) \widetilde{\phi}_1 e^{i\theta}
\end{equation}
for $\phi =$ either field variable, $\alpha \text{ or } W$.

In Stokes' case, the above frequency correction of order $\eps^3$ perfectly balances contributions in $\eps^3 e^{i\theta + 2 \text{Im}(\omega) t}$ from the nonlinear RHS terms because $\text{Im}(\omega) = 0$. In our case, however, a constant $\omega_2$ cannot compensate the additional factor $e^{2 \text{Im}(\omega) t}$. This factor is meaningful; it describes the time-dependence of the perturbation amplitudes given mode growth or decay:
\begin{equation}
    \eps(t) \rightarrow \eps e^{\text{Im}(\omega) t}
\end{equation}
The frequency correction depends on $\eps^2$, therefore matching with a factor of $e^{2 \text{Im}(\omega) t}$. Luckily, for small enough perturbations we may approximate:
\begin{equation} 
    e^{\text{Im}(\omega) t} \approx e^{\text{Im}(\omega_0) t}
\end{equation}
at least eliminating the difficult circular dependence of $\omega_2$ on itself. In the stricter limit $t \ll \text{Im}(\omega_0)^{\smallminus 1}$ we may ignore the time-dependent factor altogether:
\begin{equation} \label{eq:smalltimeapprox}
    e^{\text{Im}(\omega) t} \approx 1
\end{equation}
We work under this assumption first.

We construct an unstable model ($\Q''_0 <0$) with surface tension and turbulent viscosity, exactly like the first model in Section \ref{cascade}. We use the turbulent viscosity (Eq.\ \ref{eq:turbvisc}) sans local averaging to ease the analysis. We do not account for kinematic viscosity or other forces. 

The nonlinear corrections have no dependence on $j$, which acts merely as a Galilean frame shift. We take $\Q'_0 = 0$ for simplicity, and we assume the effect of higher derivatives of $\Q$ ($\Q'''_0$, $\Q^{(4)}_0$, etc.)\ is negligible. We assume that the simple form of the viscous force (Eq.\ \ref{eq:simplevisc}) is valid. We furthermore take $\alpha_0$, appearing in the turbulent viscosity (Eq.\ \ref{eq:turbviscform}), equal to $1/2$. Again, we do not consider the effect of local averaging of the turbulent viscosity.  

This system is governed by three dimensionless groups; the inertial instability strength:
\begin{equation} \label{eq:dimgroup1}
    N_\text{in} \equiv \sqrt{\frac{|\Q''_0|}{\Q_0}}
\end{equation}
the Weber number:
\begin{equation}
    \text{We} \equiv \frac{\Q_0 W_0^2 H}{\sigma} 
\end{equation}
and the turbulent Ohnesorge number:
\begin{equation}
    \Oh \equiv \frac{\nu_T}{\sqrt{\Q_0 \sigma \! H}} = \frac{\Q_0}{\alpha_0(1-\alpha_0)} \cdot \frac{\ell_{vis} |W_0|}{\sqrt{\Q_0 \sigma \! H}}
\end{equation}
Finally, the frequencies (growth rates) can be stated in terms of a scaling parameter:
\begin{equation} \label{eq:dimgroup-1}
   \omega_\text{norm} \equiv \frac{|\Q''_0|\Q_0 W_0^2}{\sqrt{Q_0 \sigma \! H}}
\end{equation}

We select some $k = \kappa k_\text{cut}$ as our principal perturbation wavenumber, where $k_\text{cut}$ is the wavenumber of the linear growth cutoff (see Eq.\ \ref{eq:kcutoff}):
\begin{equation}
    k_\text{cut} = |W_0| \sqrt{\frac{|\Q''_0|}{2 \sigma \! H}} 
\end{equation}
and $0 < \kappa < 1$. 

For this $k$ we compute $\omega_0$ and $\widetilde{W}_1$ in the growth and decay modes:
\begin{equation} \label{eq:linearlizationwithgroups}
    \begin{aligned}
        \omega_0 - jk &= \frac{i \kappa \omega_\text{norm}}{4} (-\Oh \, \kappa \pm \bar{\Delta}^{\frac{1}{2}}) \\
        \widetilde{W}_1 &= \frac{\sqrt{2} \, W_0 N_\text{in} (\omega_0 - jk)}{ \kappa \omega_\text{norm}} \widetilde{\alpha}_1
    \end{aligned}
\end{equation}
where:
\begin{equation}
    \bar{\Delta} \equiv \Oh^2  \kappa^2 - 4 \kappa^2 + 4
\end{equation}

We then use the $+$ (growth) mode values to compute secondary amplitudes:
\begin{equation} \label{eq:2ndamps}
    \begin{aligned}
        \widetilde{\alpha}_2 &= \frac{i \sqrt{2} N_\text{in}}{2 \kappa} \frac{F_1(\Oh, \kappa)} {(-\Oh^2 + 6) \kappa + \Oh \bar{\Delta}^{\frac{1}{2}}} \, \widetilde{\alpha}_1^2 \\
        \widetilde{W}_2 &= - \frac{W_0 N_\text{in}^2} {2 \kappa} \frac{F_2(\Oh, \kappa)} {(-\Oh^2 + 6) \kappa + \Oh \bar{\Delta}^{\frac{1}{2}}} \, \widetilde{\alpha}_1^2
    \end{aligned}
\end{equation}

There are multiple ways to satisfy the 3rd-order conditions in $e^{i\theta}$. We choose two frequency corrections, one for each variable: 
\begin{equation} \label{eq:freqcorr}
    \begin{aligned}
    \omega_2^\alpha =& - \frac{i \omega_\text{norm} N_\text{in}^2}{4} \frac{F_3(\Oh, \kappa)}{(-\Oh^2 + 6) \kappa + \Oh \bar{\Delta}^{\frac{1}{2}}} \, \widetilde{\alpha}_1^2 \\
    \omega_2^W =& \ i \omega_\text{norm} \left( \frac{N_\text{in}^2}{4} \frac{\We \cdot F_4 + F_5}{F_6} - 6 \, \Oh \, \kappa^2 \right) \widetilde{\alpha}_1^2
    \end{aligned}
\end{equation}
The functions $F_n(\Oh,\kappa)$ are listed in Appendix \ref{appx:pertfuncs}.

From third order onward, the symbolic algebra becomes very cumbersome. However, at second order we may already reap insights on three separate lines:

\begin{enumerate}
    \item Elaboration of the nonlinear cascade mechanism identified in Subsection A.
    \item Nonlinear waveforms, given by the higher-order amplitudes. 
    \item Nonlinear time evolution, and potential limiting, governed by the amplitude-dependent frequency corrections. 
\end{enumerate}

Concerning the first point: we find that components of order $n$ grow as $e^{n\text{Im}(\omega)t}$. This means that growth from a low-$k$ ``source" mode not only feeds into higher-$k$ modes, but the resulting nonlinear growth can also be far faster.

On the second point: we now draw the waveforms resulting from the superposition of first- and higher-order excitations. We begin with the linear waveforms. 

The choice of $\widetilde{\alpha}_1$'s phase is a choice of overall phase, therefore arbitrary. Take real $\widetilde{\alpha}_1 > 0$ and subsume $\eps$ into the series coefficients so that:
\begin{equation}
    \Delta \alpha = 2 \widetilde{\alpha}_1 \cos{\theta} + ...
\end{equation}
Then $\widetilde{W}_1$ is purely imaginary (Eq.\ \ref{eq:linearlizationwithgroups}), resulting in a 90$^o$ phase shift:
\begin{equation}
    \Delta W = - 2 |\widetilde{W}_1| \sin{\theta} + ...
\end{equation}

\begin{figure}
    \begin{subfigure}[b]{0.45\textwidth}
        \includegraphics[width=\textwidth]{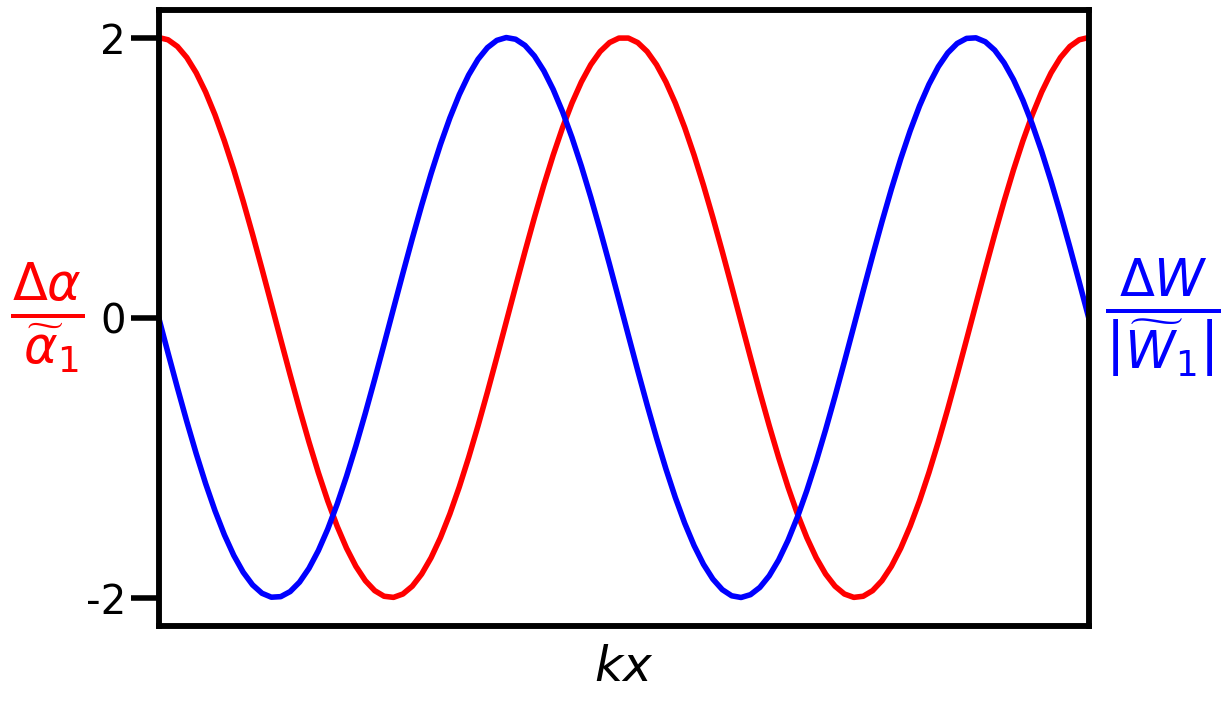}
        \caption{Linear waveforms, scaled by linear amplitudes. \newline} 
    \end{subfigure}
    \begin{subfigure}[b]{0.45\textwidth}
        \includegraphics[width=\textwidth]{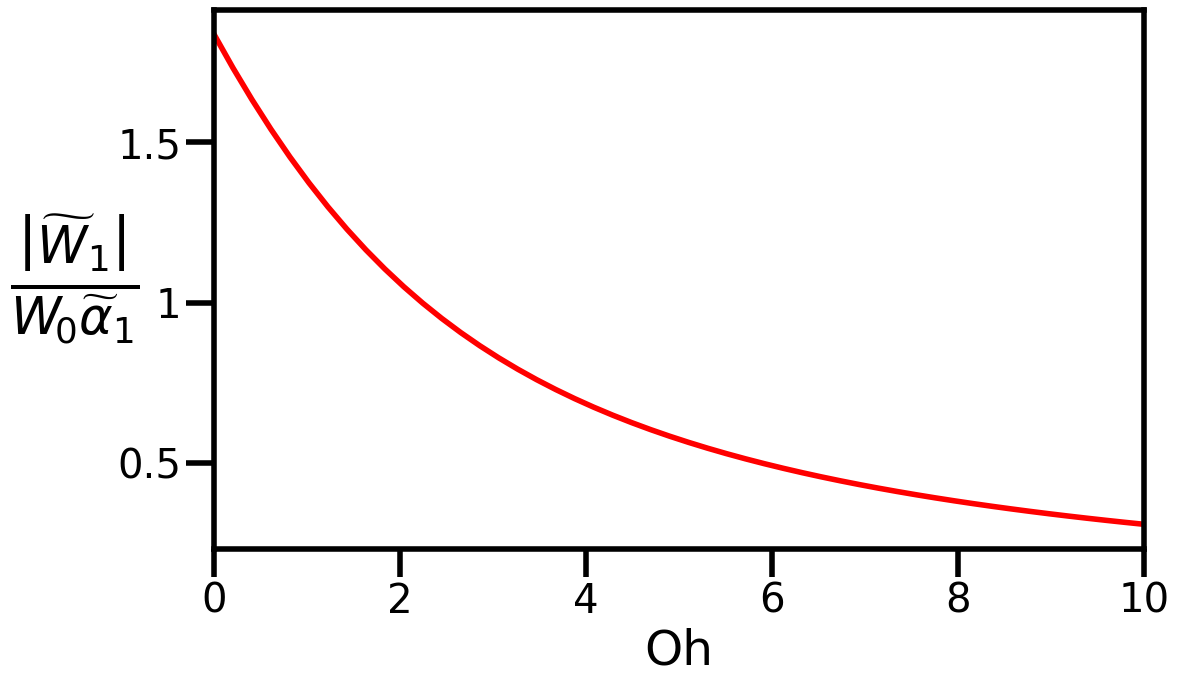}
        \caption{1st-order $W$-wave magnitude relative to background ($W_0$) and $\alpha$-wave magnitude ($\widetilde{\alpha}_1$) vs.\ Ohnesorge number.}
    \end{subfigure}
    \caption{Linear waveforms and relative magnitudes. $\kappa = \half$, $N_\text{in} = 3$.}  \label{fig:linearwaveform}
\end{figure}

The linear waveforms are shown in Figure \ref{fig:linearwaveform}a. The key feature here is the 90$^\circ$ phase shift between the two variables due to the ratio $\widetilde{W}_1/\widetilde{\alpha}_1$ being purely imaginary, which goes hand in hand with the dispersive part of $\omega$ also being imaginary. Note too the inverse relationship between the magnitude of the $W$-wave and Oh (Fig.\ \ref{fig:linearwaveform}b).

\begin{figure}
    \begin{subfigure}[b]{0.45\textwidth}
        \includegraphics[width=\textwidth]{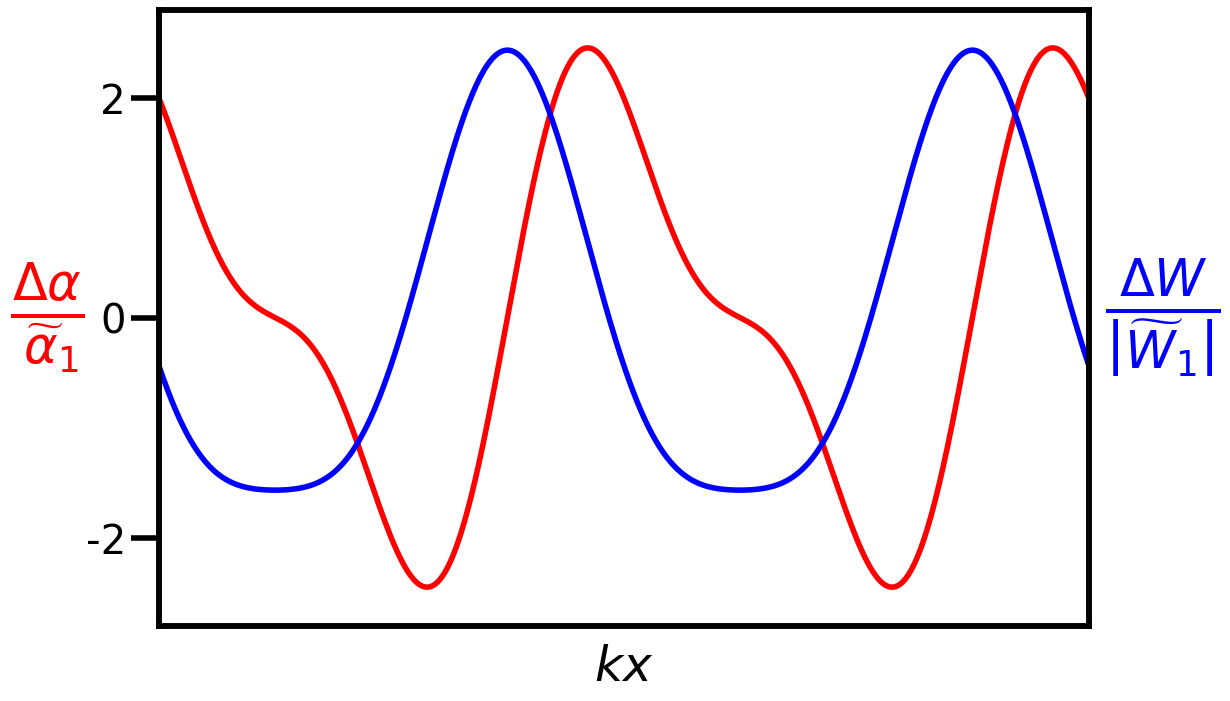}
        \caption{2nd-order waveforms, scaled by linear amplitudes. $\widetilde{\alpha}_1 = 0.07$, $\text{Oh} = 0.5$. \newline} 
    \end{subfigure}
    \begin{subfigure}[b]{0.45\textwidth}
        \includegraphics[width=\textwidth]{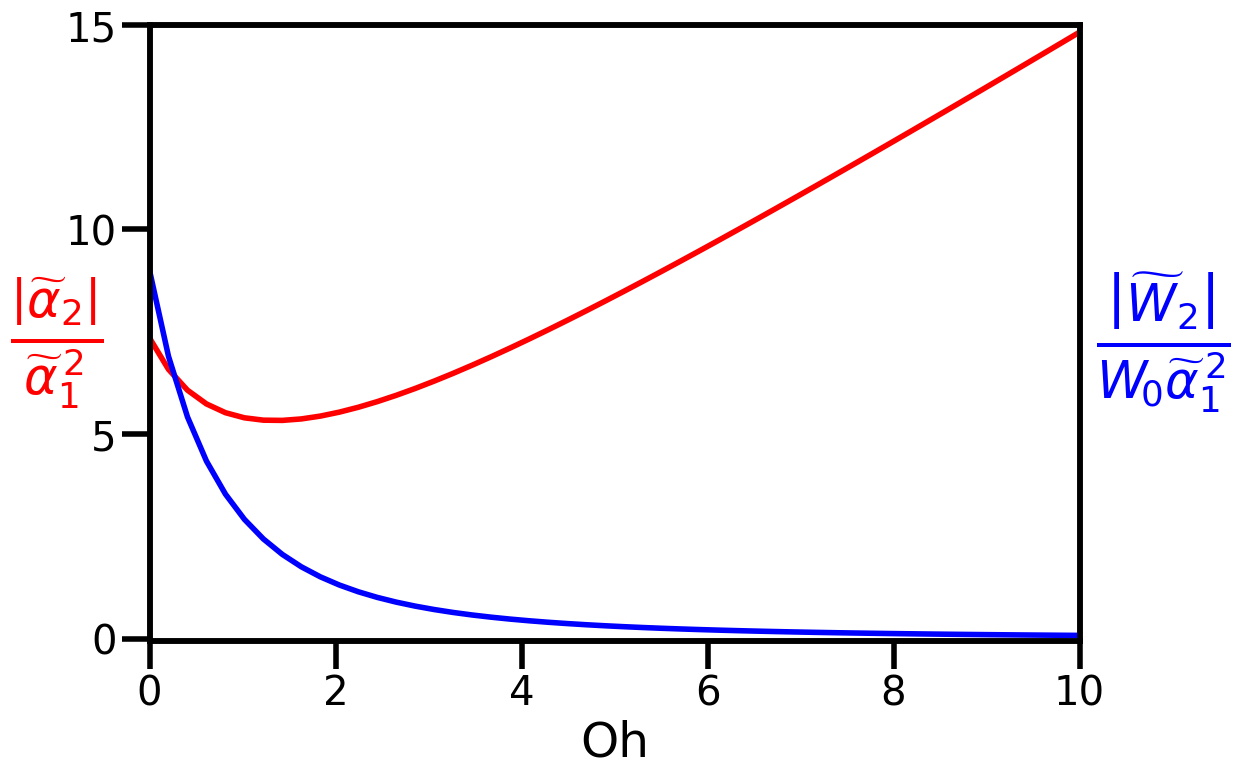}
        \caption{Normalized 2nd-order void and W magnitudes vs.\ Ohnesorge number.}
    \end{subfigure}
    \caption{Nonlinear waveforms and 2nd-order magnitudes. $\kappa = \half$, $N_\text{in} = 3$.}  \label{fig:2ndorderwaveform}
\end{figure}

The inclusion of 2nd-order terms affects the waveforms asymmetrically. $\widetilde{\alpha}_2$ is imaginary with positive sign while $\widetilde{W}_2$ is real and negative (Eq.\ \ref{eq:2ndamps}), contributing corrections:
\begin{equation}
    \begin{aligned}
        \Delta \alpha &= 2 \widetilde{\alpha}_1 \cos{\theta} - 2 |\widetilde{\alpha}_2| \sin{2\theta} + ... \\ 
        \Delta W &= -2 |\widetilde{W}_1| \sin{\theta} - 2 |\widetilde{W}_2| \cos{2\theta} ...
    \end{aligned}
\end{equation}

The effect of these added perturbations is shown in Figure \ref{fig:2ndorderwaveform}. The relative momentum ($W$) begins to take the form of a Stokes wave, with high and narrow peaks. By contrast, the crests and troughs of the void ($\alpha$) wave are shifted closer together around the peak in $W$. We also note the continued limiting effect of Oh on $W$-wave magnitudes, the opposite being applicable to $\alpha$.

Although computing the results of the nonlinear series to high order is difficult, we can extend the pattern seen in the terms so far in the following fashion:
\begin{equation}
    \begin{aligned}
    &\widetilde{\alpha}_{n+1} = i c_{\alpha} \left| \frac{\widetilde{\alpha}_2
    }{\widetilde{\alpha}_1} \right| \widetilde{\alpha}_n ; &\widetilde{W}_{n+1} = i c_{W} \left| \frac{\widetilde{W}_2
    }{\widetilde{W}_1} \right| \widetilde{W}_n
    \end{aligned}
\end{equation}
for $n \geq 3$, where $c_{\alpha}$ and $c_{W}$ are adjustable parameters. In this way we may gain an understanding of the qualitative (if not quantitative) features of fully nonlinear solutions to an inertially unstable TFM.

\begin{figure}
    \includegraphics[width=0.45\textwidth]{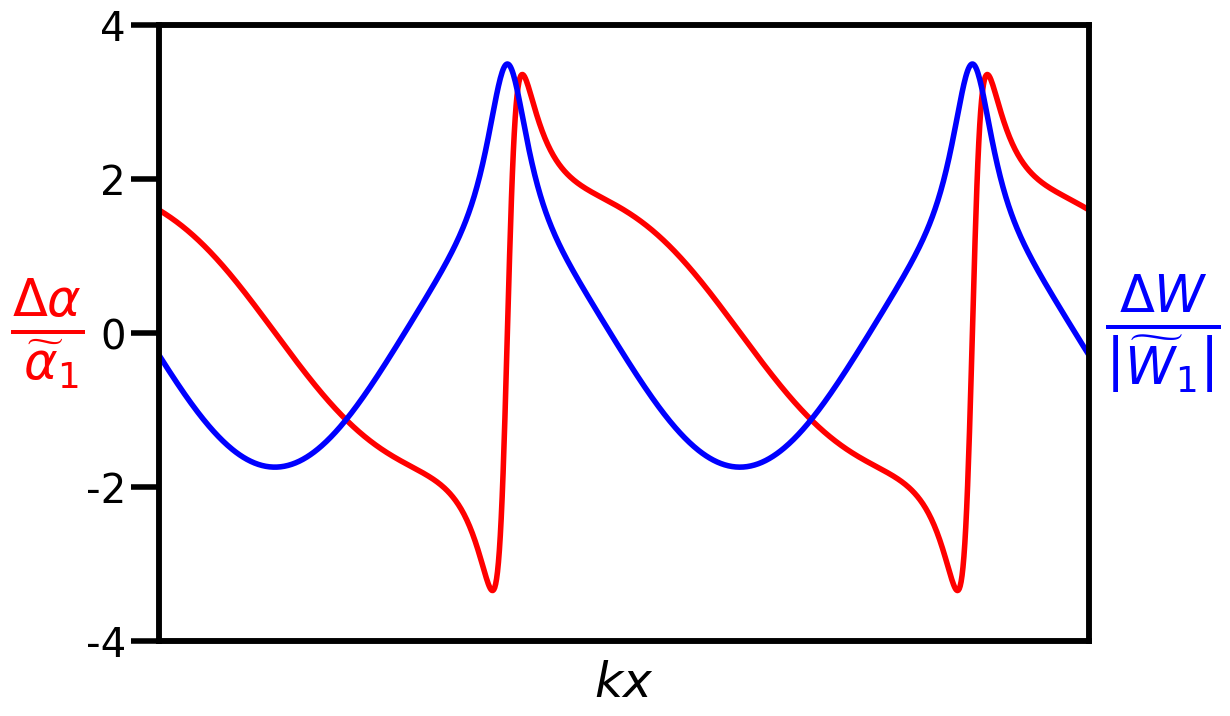}
    \caption{Extended nonlinear void and relative momentum waveforms, $c_{\alpha} = 2,\ c_W = 3.3$.}  \label{fig:nonlinearwaveform}
\end{figure}

The result of this procedure, illustrated in Figure \ref{fig:nonlinearwaveform}, is a reproduction of the shock-spike waves identified initially in Subsection A. They emerge naturally from this analytical method, confirming the assumptions of our heuristic approach to nonlinear stability (Sec.\ \ref{spikeheuristics}) and the results of the numerical analysis (Sec.\ \ref{cascade}). 



On the other hand, if one reintroduces perpendicular gravity ($g_y$) at a sufficient strength to make the model linearly stable, i.e.\ $\delrho H \! g_y > \Q'' W_0^2$, one can recover the key features of Stokes waves. By accounting for the Froude number:
\begin{equation} 
    \text{Fr} \equiv \sqrt{\frac{|\Q''|}{\delrho}}\frac{W_0}{\sqrt{Hg_y}}
\end{equation}
one can compute higher-order amplitudes and frequency corrections as before. 

There is one caveat; including gravity as the only force requires apparently infinite nonlinear corrections due to a constant velocity $\omega/k$ among all modes. This breakdown occurs because differences in $\omega/k$ are required to compensate nonlinear effects from the right-hand sides of the PDEs. Otherwise, waves of ``permanent type" cannot be solved for.

\begin{figure}
    \includegraphics[width=0.45\textwidth]{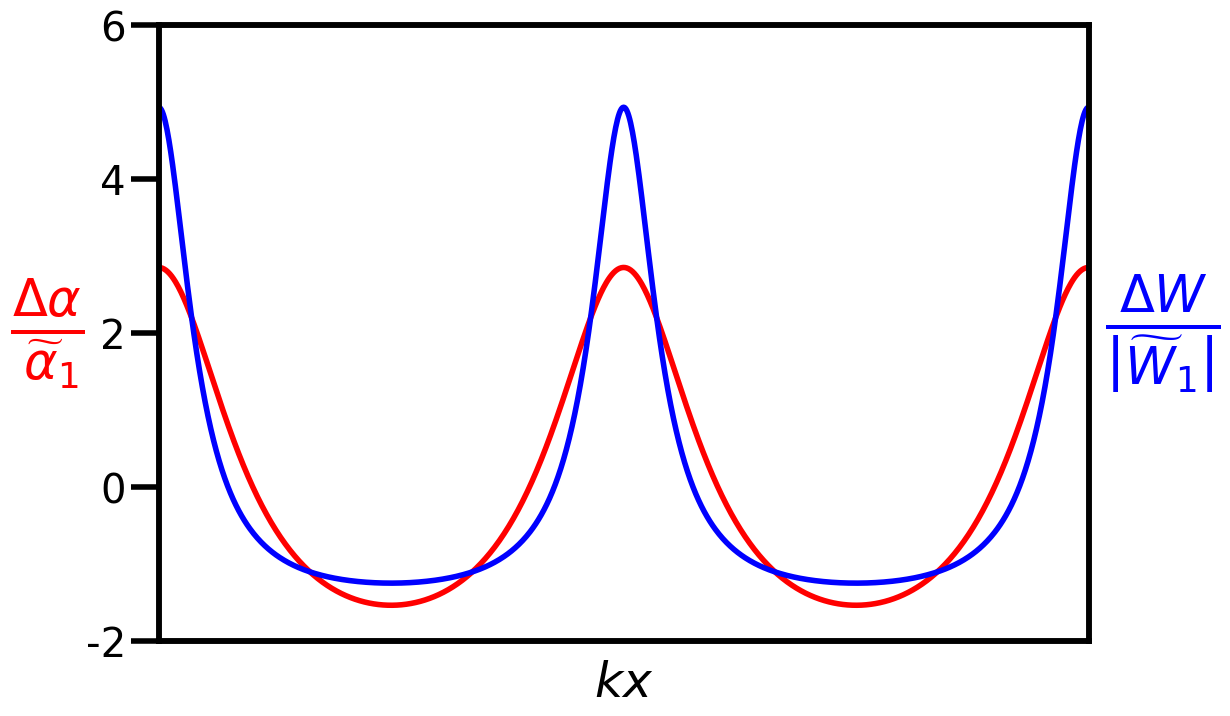}
    \caption{Extended nonlinear void and relative momentum waveforms in a gravity-stabilized TFM, $\kappa = \half$, $N_\text{in} = 3$, $\text{Fr}=\sqrt{2}$, $\widetilde{\alpha}_1 = 0.07$, $c_{\alpha} = c_W = 1$.}  \label{fig:nonlinearwaveformStokes}
\end{figure}

With both surface tension and gravity (no dissipation), variation in $\omega/k$ reappears. $\omega$ is purely real, leading to waveforms composed of only cosines:
\begin{equation}
    \Delta\phi= 2 \sum_n \widetilde{\phi}_n \cos(n\theta)
\end{equation}
with $\phi = $ either field variable, $\alpha$ or $W$. Consequently, a Stokes-esque waveform manifests in both variables (Figure \ref{fig:nonlinearwaveformStokes}). 

We now return to the unstable case, and the third point: the amplitude-dependent frequency corrections (Eq.\ \ref{eq:freqcorr}). If a given model is to be nonlinearly (but not linearly) stable, the imaginary part of some higher-order corrections to $\omega$ must be negative. Then there may be some amplitude $\widetilde{\alpha}_0$ at which:
\begin{equation}
    \text{Im}(\omega)(\widetilde{\alpha}_0) = \text{Im}(\omega_0) + \eps^2 \, \text{Im}(\omega_2) + \dots = 0
\end{equation}
indicating that wave growth plateaus at said amplitude. Carrying the perturbation analysis to just the first frequency corrections $\omega_2^\alpha$ and $\omega_2^W$ reveals a range of nonlinearly limited conditions.

\begin{figure}
    \includegraphics[width=0.45\textwidth]{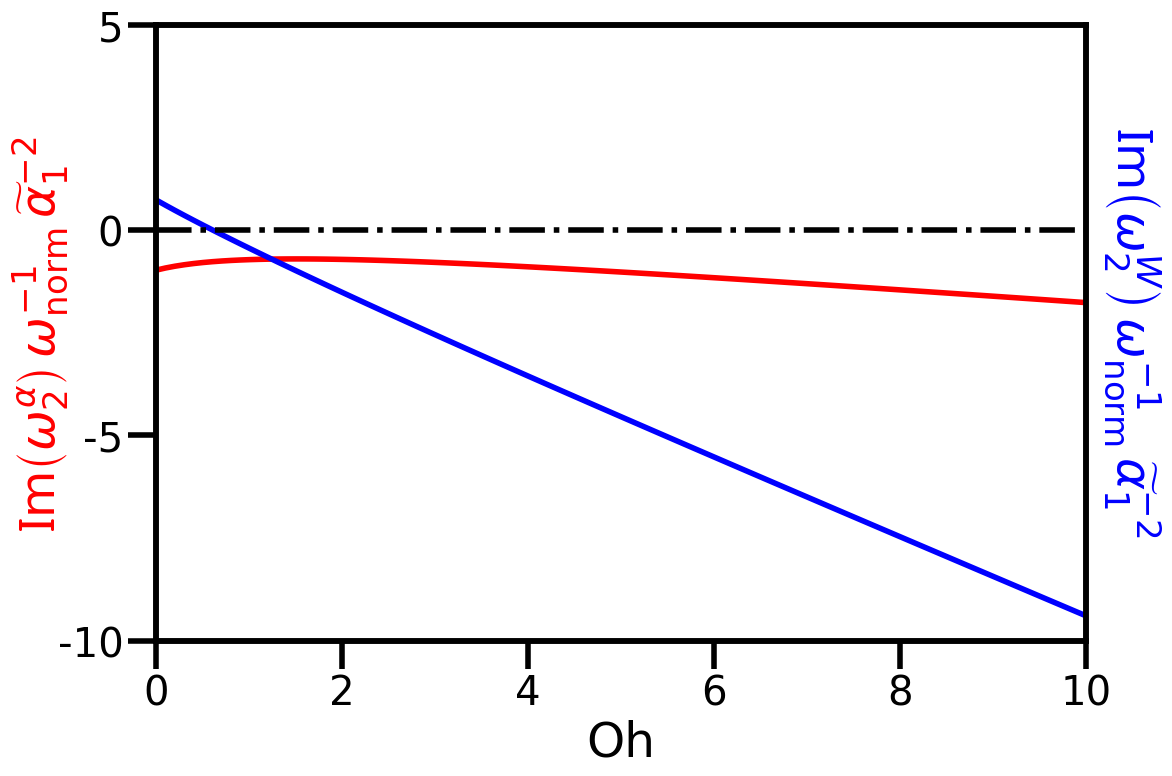}
    \caption{Normalized frequency corrections plotted vs.\ Ohnesorge number in a stable case, $\kappa = \half$, $N_\text{in} = 1$, $\text{We} = 25$.}  \label{fig:freqcorrsstable}
\end{figure}

Figure \ref{fig:freqcorrsstable} shows one such case, with relatively low $N_\text{in}$ and We. A reasonable amount of viscosity ($\text{Oh}\gtrsim0.61$) is enough to ensure both $\omega_2^\alpha,\ \omega_2^W<0$. As long as $N_\text{in}$ and $\text{We}$ are low enough, a similar result is obtained.

\begin{figure}
    \includegraphics[width=0.45\textwidth]{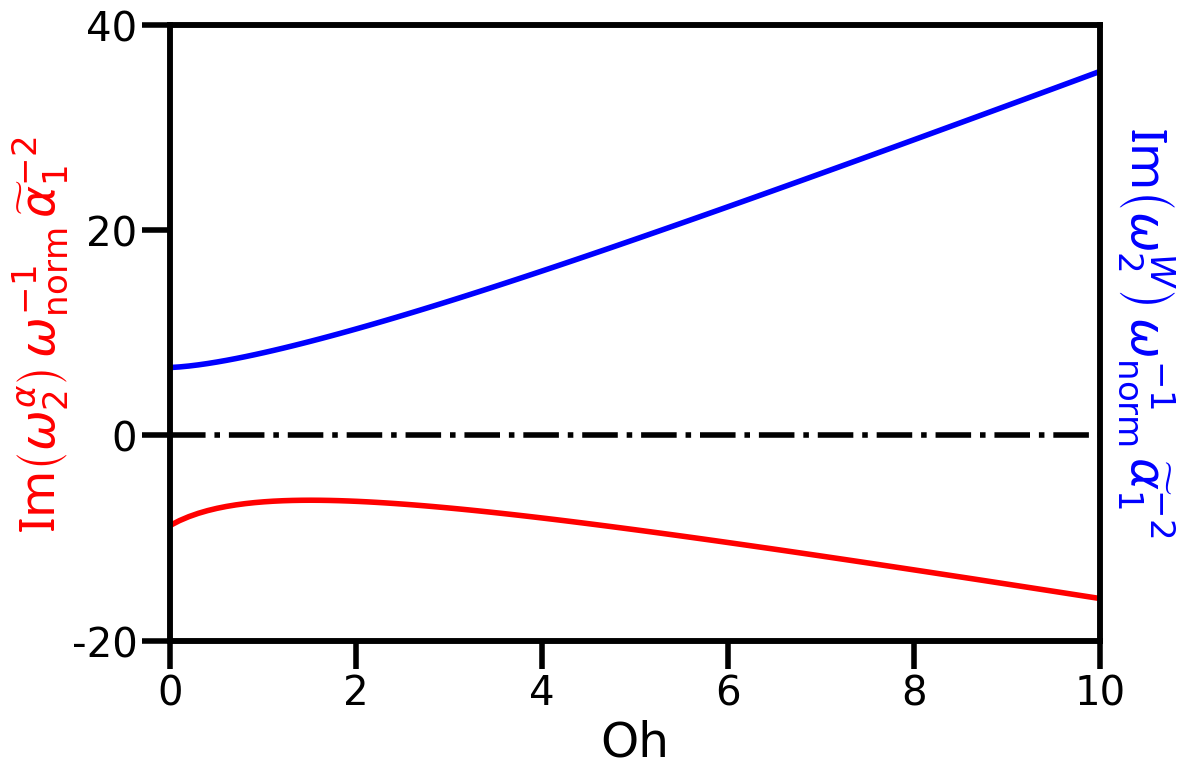}
    \caption{Normalized frequency corrections plotted vs.\ Ohnesorge number in an unstable case, $\kappa = \half$, $N_\text{in} = 3$, $\text{We} = 25$.}  \label{fig:freqcorrsunstable}
\end{figure}

However, not all parameter sets can achieve nonlinear stability at 2nd order. Increasing the strength of the inertial instability ($N_\text{in}$) eventually results in a failure to limit $W$-waves regardless of the viscosity, see Figure \ref{fig:freqcorrsunstable}. The growth rate gets larger, not smaller, as Oh increases. High Weber numbers result in a similar failure. These results indicate super-exponential growth of $W$ as the influence of $\omega_2^W$ increases.

This increased growth is unlikely to be the end of the story. Higher order corrections to $\omega^W$ may be stabilizing. We have also assumed that the feedback of amplitude growth or decay over time is negligible (Eq.~\ref{eq:smalltimeapprox}), making the present results applicable only at very short times. 

In effect, we have so far solved for amplitudes of the form:
\begin{equation}
    \begin{aligned}
        \widetilde{\alpha}_{j}(t) &= \widetilde{\alpha}_j(0) \, e^{|j| \text{Im}\left(\omega^\alpha\right)t} \\
        \widetilde{W}_j(t) &= \widetilde{W}_j(0) \,e^{|j| \text{Im}(\omega^W\!)t}
    \end{aligned}
\end{equation}

However, evolving amplitudes mean evolving $\omega$'s, making simple exponentials à la $e^{\text{Im}(\omega)t}$ an insufficient description. One can get around this problem by seeking to state only the time derivatives of the amplitudes:
\begin{equation}
    \begin{aligned}
    \frac{d \widetilde{\alpha}_j}{dt}\!(t) &= |j| \, \widetilde{\alpha}_j \, \text{Im}\!\left[ \omega^\alpha(\widetilde{\alpha}_1) \right] \\
     \frac{d \widetilde{W}_j}{dt}\!(t) &= |j| \, \widetilde{W}_j  \, \text{Im}\!\left[ \omega^W\!(\widetilde{\alpha}_1) \right] \\
    \end{aligned}
\end{equation}
where the $\omega$'s are allowed to evolve flexibly. Most of the foregoing analysis stays intact up to this point. 

Now, however, parameterizing the $\omega$'s in terms of just $\widetilde{\alpha}_1$ becomes unjustified, as the multiple growth rate corrections cause previously ``constant" amplitude ratios to evolve. Once a sufficient degree of nonlinearity enters in, variability of $\widetilde{W}_1/\widetilde{\alpha}_1$ will invalidate even the linear step of the analysis.

One is left with the prospect of considering the time dependence of each amplitude in terms of every other amplitude individually, for example:
\begin{equation} \label{eq:generalseries}
    \begin{aligned}
    d_t\widetilde{\alpha}_1 &= C^{\alpha_1}_{\alpha_1} \widetilde{\alpha}_1 + C^{\alpha_1}_{W_1}\widetilde{W}_1 + C^{\alpha_1}_{\alpha_2\alpha_1^*} \widetilde{\alpha}_2 \widetilde{\alpha}_1^* + C^{\alpha_1}_{\alpha_2W_1^*} \widetilde{\alpha}_2 \widetilde{W}_1^* \\
    &+C^{\alpha_1}_{W_2\alpha_1^*} \widetilde{W}_2 \widetilde{\alpha}_1^*
    +C^{\alpha_1}_{W_2W_1^*} \widetilde{W}_2 \widetilde{W}_1^* + \dots
    \end{aligned}
\end{equation}
The $C$'s are obtained by plugging Eq. \ref{eq:pertseries} into the PDEs and collecting terms in $e^{i\theta}$ only (not $e^{2i\theta}$ etc.) to match the wavelength that goes with $\widetilde{\alpha}_1$. Now that the time dependence resides in the amplitudes, $\theta \equiv kx$.

Constructing an equation like Eq. \ref{eq:generalseries} for each amplitude $\widetilde{\alpha}_j$ or $\widetilde{W}_j$ up to an order $j_\text{max}$ results in a system of nonlinear ODEs.

What this approach gains in generality, it loses in usefulness. Even at low order, the ODEs may only be approachable numerically. One could think of them as a finite element discretization of the PDEs on a Fourier basis set. With anything like sufficient resolution, they would become unwieldy.

The results obtained with the simpler perturbation method, assuming constant $\omega$'s and ratios between amplitudes, still support the rest of the theoretical and numerical analysis overall. Disagreements occur where the perturbation method is weakest, i.e.\ in predicting limiting behaviors over long times.



\begin{acknowledgements}
Thank you to Martín López-de-Bertodano, who originally introduced me (A. López-de-Bertodano) to the Two-Fluid Model.
\end{acknowledgements}

\appendix
\section{Derivation of Force Terms} \label{appx:momeqs}
The derivation of the flux-variable TFM equations using the variational method is carried out in \cite{Clausse2021}, from which we take our advective terms. This method may also be used to derive the surface tension and gravity forces. We include these conservative forces in our present mechanistic derivation, but both approaches give identical results.

We constitute the forces on each fluid, $i$, as follows:
\begin{equation}
    \begin{aligned}
        F_i = &\, \alpha_i \rho_i g_x - D_i - (-1)^iD_\text{int} + V_i  \\
        & - \partial_x (\alpha_i P_i) + P_{\text{int},i}\,\partial_x \alpha_i
    \end{aligned}
\end{equation}

Gravitational acceleration parallel to the channel, $g_x$, supplies a body force proportional to the mass of each fluid. For the wall drag on each fluid, $D_i$, we use a simple drag law:
\begin{equation} \label{eq:walldrag}
    D_i \equiv \frac{f_i \mathcal{P}_i}{2A} \rho_i |u_i| u_i
\end{equation}
where $f_i$ is the Fanning friction factor for fluid $i$, $A$ is the cross-sectional area of the channel, and $\mathcal{P}_i$ is the wall perimeter in contact with fluid $i$. For a rectangular channel in stratified flow:
\begin{equation}
    \mathcal{P}_i = w + 2\alpha_iH
\end{equation}
where $w$ is the channel width. However, under a threshold $\alpha_D$, we supply limiting:
\begin{equation}
    \mathcal{P}_i(\alpha_i<\alpha_D) = \alpha_i(w+2H)
\end{equation} 
so that $\mathcal{P}_i$ goes to zero, avoiding infinite rigidity. We take $\alpha_D = 0.02$ where not otherwise specified.

Fluid-fluid interfacial drag is given by:
\begin{equation}
    D_\text{int} \equiv \frac{f_\text{int} \mathcal{P}_\text{int}}{2A} \rho_D |u_r|u_r
\end{equation}
The interfacial Fanning friction factor $f_\text{int}$ will be reconstituted into the interfacial drag coefficient $C_D$ in the final equations. The drag density $\rho_D$ should be chosen based on the extent to which cross-sectional velocity variations are expected within each fluid. Similarly to the $\mathcal{P}_i$'s, the interfacial perimeter $\mathcal{P}_\text{int}$ will be (effectively) set to 0 at extreme $\alpha$'s -- see Eq.\ \ref{eq:limdraglaw}.

The viscous force on each fluid: 
\begin{equation} \label{eq:viscousforcewithstress}
    V_i = \alpha_i \, \partial_x \tau_{\text{visc},i}
\end{equation}
is driven by the gradient of the viscous stress $\tau_{\text{visc},i}$, which is given by:
\begin{equation}
    \tau_{\text{visc},i} = \rho_i \nu \, \partial_x u_i
\end{equation}
The viscosity $\nu$ (taken equal between the two fluids for simplicity) is split into two components:
\begin{equation}
    \nu = \nu_k + \nu_t
\end{equation}
where $\nu_k$ is the kinematic viscosity and $\nu_t$ is the turbulent eddy viscosity.

We estimate the eddy viscosity with a simple mixing length model. The equation for time-averaged turbulent momentum flux near a boundary:
\begin{equation}
    \tau_{t} = \beta \,l_m \Delta u \frac{du}{dy}
\end{equation}
is due to Prandtl \cite{Prandtl1925}, who was loyal and sympathetic to the Nazi regime \cite{PrandtlIdeo}.\ $y$ is the coordinate normal to a boundary, $l_m$ is the mixing length, $\Delta u$ is the transverse eddy speed (assumed roughly equal to the difference in bulk velocity across a $y$-distance of $l_m$) and $\beta$ is the area fraction occupied by eddies. 

This flux is defined in the $y$ direction, but assuming that eddy structure around boundaries is isotropic in $x$ (the coordinate aligned with bulk flow) and $y$ (the boundary-normal coordinate), we may generalize a local eddy viscosity:
\begin{equation}
    \nu_t = \beta \,l_m \Delta u
\end{equation}
describing momentum diffusion in both the $x$ and $y$ directions. Since $l_m$ is an undetermined parameter, we simply condense $\beta$ into it:
\begin{equation}
    \nu_t \rightarrow l_m \Delta u
\end{equation}

Consider fluid-fluid interfaces to act like boundaries, at least insofar as they generate substantial transverse velocity gradients. The scale of $\Delta u$ near an interface is then given by $|u_r|$ (the difference between the average velocities of the fluids) along with dimensionless quantities that may again be folded into $l_m$. Taking the cross-sectional average yields:
\begin{equation} \label{eq:turbviscappx}
    \bar{\nu}_t = \bar{l}_m |u_r|
\end{equation}
We drop the averaging bars from here on out and treat $l_m$ as a constant, yielding Eq.\ \ref{eq:turbvisc}:
\begin{equation} \label{eq:turbvisc_apdx}
    \nu_t = l_m |u_r|
\end{equation}
This is one of the turbulent viscosity models proposed by Fullmer \cite{Fullmerthesis}. 

With these considerations, the viscous force on each phase becomes:
\begin{equation} \label{eq:viscousforce}
    V_i = \rho_i \alpha_i \partial_x\!\left[ \left(\nu_k + l_m|u_r|\right) \partial_x u_i \right]
\end{equation}

Under certain unstable inertial couplings, a TFM with the above turbulent viscosity can exhibit very slow convergence, or even momentum spike blowups like those observed in an inviscid model (Sec.\ \ref{LinBreakdown}). In other words, the model can be ill-posed.

Unfavorable behavior occurs when destabilizing effects from the second part of the viscous stress, seen by expanding Eq.\ \ref{eq:viscousforcewithstress}:
\begin{equation} \label{eq:viscforceexp}
 V_i = \alpha_i \rho_i \! \left[ \nu \partial_x^2 u_i + \partial_x \nu \cdot \partial_xu_i\right]
\end{equation}
hinder or outweigh the stabilization provided by the first part. The dependence of $\nu$ on the relative velocity (Eq. \ref{eq:turbvisc_apdx}), as much as it is crucial to nonlinear well-posing, provides the gateway. If the inertia $\Gamma$ leads to overly large variations of $u_r$ across a shock-spike (Sec.\ \ref{nonlinear}), good behavior is not guaranteed.

For this reason, we introduce ``local averaging" of Eq.\ \ref{eq:turbvisc_apdx} over a preset length $\delta_m$:
\begin{equation}
    \nu_t(x) \equiv l_m \int_{x-\delta_m}^{x+\delta_m} |u_r|\, \frac{dx}{2\delta_m}
\end{equation}
The physical reasoning is that, given a turbulent length scale $l_m$, the momentum diffusion caused by eddies must be influenced by the flow conditions across a similar length $\delta_m$, seeing as the eddies have a finite length.

This averaging tamps down on destabilizing effects coming from the second term in Eq.\ \ref{eq:viscforceexp} by smoothing out overly sharp gradients in $u_r$ as they contribute to $\partial_x \nu$. It somewhat weakens the nonlinear stabilization from the first part, but still basically maintains the all-important proportionality $\nu_t \propto |W|$ (Eq.\ \ref{eq:TVproportional}). Linear damping is unaffected.

The argument is not that these turbulent viscosity models are particularly exact, but that they are simple, physical, and effective.

The average bulk pressure on each phase, $P_i$, acts on an area proportional to $\alpha_i$; the resulting force density is the gradient $\partial_x (\alpha_i P_i)$. In stratified flow, on a finite volume of length $\delta x$, the pressure at the interface of phase $i$ ($P_{\text{int},i}$) acts over an area:
\begin{equation}
    \delta a = w \delta x \sqrt{1 + (H \partial_x \alpha)^2}
\end{equation}
The component of the interfacial pressure acting in the $x$ direction is:
\begin{equation}
    (P_{\text{int},i})_x = P_{\text{int},i} \cdot \frac{H \partial_x \alpha}{\sqrt{1 + (H \partial_x \alpha)^2}}
\end{equation}
The resulting force density is shown to be:
\begin{equation}
    \frac{\delta F}{A\delta x} = (P_{\text{int},i})_x \, \frac{\delta a}{H\!w\delta x} =  P_{\text{int},i}\,\partial_x \alpha_i
\end{equation}

A hydrostatic approximation for the pressures across each phase gives:
\begin{equation} \label{eq:phasepressurediffs}
    \begin{aligned}
        P_1 &= P_{\text{int},1} + \half\rho_1g_yH (1-\alpha) \\
        P_2 &= P_{\text{int},2} - \half\rho_2g_y\alpha
    \end{aligned}
\end{equation}
The surface tension results in a pressure jump across the interface. Call the height of the interface $h_\text{int} = H\alpha$. Then its curvature is:
\begin{equation}
    \kappa_c = \partial_x^2 h_\text{int} \left(1+h_\text{int}^2\right)^{-3/2} 
\end{equation}
and the pressure jump:
\begin{equation}
    \Delta P_\text{int} = \sigma \kappa_c
\end{equation}
ergo:
\begin{equation} \label{eq:intpressurediff}
    P_{\text{int},1} = P_{\text{int},2} + \sigma_{_{}}\! H \partial_x^2\alpha\left[1+(H\partial_x\alpha)^2\right]^{-3/2}   
\end{equation}

These relations are valid only for stratified flow, and substantial modifications may be necessary when extending to other flow regimes. For example, in bubbly flow of uniform radius $r_b$, the pressure jump becomes constant:
\begin{equation}
    \Delta P_\text{int} = \sigma \kappa_c = \frac{2\sigma}{r_b}
\end{equation}
leading the surface tension terms in both the mixture and relative momentum equations to vanish. The absence of a high-order ST term necessitates linear stabilization via the inertial coupling.

Any of 4 pressures ($P_i$, $P_{\text{int},i}$) may be chosen as the reference pressure $P$, although this choice must be respected when supplying boundary conditions. We choose:
\begin{equation} \label{eq:refpressure}
    P = P_{\text{int},1}    
\end{equation}
This choice has no effect on the $g_y$ and $\sigma$ terms in the relative momentum equation, but it does affect their form in the mixture momentum equation. 

We now demonstrate combining the forces into the mixture and relative momentum equations. We use the fact that the mixture momentum may be stated:
\begin{equation}
    \rho_m j - \delrho J = (1-\alpha) \rho_1u_1 + \alpha \rho_2 u_2
\end{equation}
which is just the sum of the phase momenta. With no IC, the term under the time derivative in the relative momentum equation simplifies to:
\begin{equation} \label{eq:noICW}
    W - \delrho j = \Gamma_0 J - \delrho j = \rho_2u_2 - \rho_1u_1
\end{equation}
which is the difference of the phase momenta divided by their respective void fractions.

Thus the force terms must be combined as follows:
\begin{equation} \label{eq:forcecombos}
    \begin{aligned}
        F_W &= \frac{F_2}{\alpha} - \frac{F_1}{1-\alpha} \\
        F_m &= F_1 + F_2 \\
    \end{aligned}
\end{equation}
where $F_W$ is the relative momentum force, and $F_m$ is the mixture momentum force. We assume that the forces derived without inertial coupling remain applicable generally. Some assumption of this type must be made to include non-conservative forces, which cannot be derived by the variational method.

Carrying out Eq.\ \ref{eq:forcecombos} yields the RHS's of Eqs.\ \ref{eq:relmotion} and \ref{eq:mixmotion}. We draw attention to two terms; first, the interfacial drag:
\begin{equation}
    \begin{aligned}
        F_{\text{int},m} &= 0 \\
        F_{\text{int},W} &= \frac{f_\text{int} \mathcal{P}_\text{int}}{2\alpha (1-\alpha)A} \rho_D |u_r|u_r
    \end{aligned}
\end{equation}
The ``equal and opposite" actions cancel in the mixture equation. For the relative momentum equation, we define a simplified drag coefficient:
\begin{equation} \label{eq:dragforce}
    C_D \equiv \frac{f_\text{int} \mathcal{P}_\text{int} H}{2\alpha (1-\alpha)A} \Rightarrow F_{\text{int},W} = \frac{C_D\rho_D}{H} |u_r|u_r
\end{equation}
For stratified flow in a rectangular channel at intermediate void fractions:
\begin{equation} \label{eq:simpleperimeter}
    \mathcal{P}_\text{int} \approx w \Rightarrow \frac{\mathcal{P}_\text{int}}{A} \approx \frac{1}{H}
\end{equation}
so we may simplify:
\begin{equation} \label{eq:limdraglaw}
    C_D = \frac{f_\text{int}}{2 \max\left[\alpha (1-\alpha), \alpha_D\right]} 
\end{equation}
where the $\max$ condition avoids infinite rigidity. We set $\alpha_D = 0.02$ where not otherwise specified. 

Secondly, we examine the effect of viscosity:
\begin{equation}
    \begin{aligned}
        F_{\text{visc},m} &= \sum_i V_i \\
        F_{\text{visc},W} &= \sum_i \tfrac{(-1)^i}{\alpha_i}V_i
    \end{aligned}
\end{equation}
Based on Eq.\ \ref{eq:viscousforce}, we may rewrite $F_{\text{visc},W}$:
\begin{equation} \label{eq:fullvisc}
    F_{\text{visc},W} = \partial_x \! \left[ \nu \partial_x (\rho_2u_2 - \rho_1u_1) \right]
\end{equation}
Without inertial coupling (Eq.\ \ref{eq:noICW}), this term simplifies:
\begin{equation} \label{eq:simplevisc_apdx1}
    F_{\text{visc},W} = \partial_x \! \left[ \nu \partial_x (W - \delrho j) \right]
\end{equation}
and assuming incompressibility (Eq.\ \ref{eq:incomp}):
\begin{equation} \label{eq:simplevisc_apdx2}
    F_{\text{visc},W} = \partial_x \! \left( \nu \partial_x W \right)
\end{equation}

In practice, we extend the use of Eqs.\ \ref{eq:simplevisc_apdx1}-\ref{eq:simplevisc_apdx2} to general inertial couplings to avoid destabilizing effects caused by the appearance of other gradients.

\section{Detailed Linear Analysis of the 2-Equation System} \label{appx:linear}

A full dispersion relation, including all forces from Eq.\ \ref{eq:relmotion} except the wall drag, may be written:
\begin{equation} \label{eq:fulldisp}
    \begin{aligned}
        \bar{\omega} =&  \,\,v_\text{eig}k - i\!\left(\Omega + c W + \frac{\nu k^2}{2}\right) \\
        \Delta =&  \Q\!\left(-i c' W^2 k + \half \Q'' W^2 k^2 + g_y\delrho H k^2 + \sigma H k^4 \right) \\
        & - \left(c W + \frac{\nu k^2}{2} \right)^2
    \end{aligned}
\end{equation}
where $\omega = \bar{\omega} \pm \Delta^\frac{1}{2}$. The simplified form of the viscosity (Eq.\ \ref{eq:simplevisc}) has been used. We have also included the effect of constant boiling with the convention:
\begin{equation}
    \partial_x j = \frac{\delrho}{\rho_1 \rho_2}\Gamma_g \equiv \Omega
\end{equation}
where $\Gamma_g$ is the mass transfer rate from the liquid to the gaseous phase, not to be confused with the inertial function $\Gamma(\alpha)$.

To further simplify the above expressions we have used:
\begin{equation}
    c \equiv \frac{\rho_D C_D}{H\alpha^2(1-\alpha)^2}
\end{equation}
following \cite{Clausse2023}. $c'$ denotes the derivative of $c$ w.r.t.\ $\alpha$. We have used the simplified form of the viscosity (Eq.\ \ref{eq:simplevisc}).


Restricting our focus to the higher-order forces, surface tension and viscosity, we obtain:
\begin{equation}
    \omega =\! v_\text{eig}k -\! \frac{i\nu k^2}{2} \!\pm\! \left[\Q k^2\!\left(\sigma\! H k^2 + \half\Q''W^2 \right) \!-\! \frac{\nu^2  k^4}{4}\right]^{\frac{1}{2}}_{}
\end{equation}
For simplicity, substitute in values of $\Q = 1/4$ and \mbox{$\Q'' = -2$} corresponding to the simple inertia of Eq.\ \ref{eq:simpleinertia} with $\alpha_0 = 1/2$:
\begin{equation}
    \omega = v_\text{eig}k + \frac{i}{2}\left[-\nu k^2 \pm ( W^2k^2 - \sigma H k^4 + \nu^2  k^4 )^{\frac{1}{2}}\right]
\end{equation}
This form of the linear frequency is valid in the unstable range $k < W/\sqrt{\sigma H}$. It is representative, to constant scalings, of any unstable inertia.

The eigenvectors corresponding to the above eigenvalues are, respectively:
\begin{equation}
    \psi_{\pm}
    \!=\!
    \begin{pmatrix}
        1\\ 2i \!\left[-\nu k \pm k^{-1}( W^2k^2 - \sigma H k^4 + \nu^2  k^4 )^{\frac{1}{2}}\right]_{_{_{}}}
    \end{pmatrix}
\end{equation}

Taking the initial perturbation used in Section \ref{linstab}:
\begin{equation}
    \begin{pmatrix}
        \Delta \alpha\\ \Delta W
    \end{pmatrix}
    _{\!0} =
     \begin{pmatrix}
        \eps\cos(kx)\\ 0
    \end{pmatrix}
\end{equation}
we decompose it into eigenvector components:
\begin{equation}
    \begin{aligned}
        & \begin{pmatrix}
            \eps \cos(kx)\\ 0
        \end{pmatrix} = \frac{\eps}{2}
        \begin{pmatrix}
            1 \\ 0
        \end{pmatrix} e^{ikx} +
        \frac{\eps}{2}
        \begin{pmatrix}
            1 \\ 0
        \end{pmatrix} e^{-ikx} \\
        & = \frac{\eps}{4}\left( \frac{2 \bar{\Delta}^{\frac{1}{2}} + \nu k^2}{2 \bar{\Delta}^{\frac{1}{2}}} \psi_+(k)+ \frac{2 \bar{\Delta}^{\frac{1}{2}} - \nu k^2}{2 \bar{\Delta}^{\frac{1}{2}}} \psi_-(k) \right) e^{ikx} \\
        & +\, \frac{\eps}{4}\left( \frac{2 \bar{\Delta}^{\frac{1}{2}} + \nu k^2}{2 \bar{\Delta}^{\frac{1}{2}}} \psi_+(\smallminus k) + \frac{2 \bar{\Delta}^{\frac{1}{2}} - \nu k^2}{2 \bar{\Delta}^{\frac{1}{2}}} \psi_-(\smallminus k) \right) e^{-ikx} \\
    \end{aligned}
\end{equation}
where:
\begin{equation}
    \bar{\Delta} = \frac{1}{4} \left( W^2k^2 - \sigma H k^4 + \nu^2  k^4 \right)
\end{equation}
which just is the negative of $\Delta$, see Eq.\ \ref{eq:fulldisp}. Inserting the time dependence terms $e^{-i\omega t}$ per eigenvector and recombining obtains:
\begin{equation} \label{eq:generallinresponse}
    \begin{aligned}
    &\begin{pmatrix}
        \Delta \alpha\\ \Delta W
    \end{pmatrix}\!(t) \!=\! \eps\!
     \begin{pmatrix}
        \!\left[ \cosh{\!\bar{\Delta}^{\frac{1}{2}}t} + \frac{\nu k^2}{2 \bar{\Delta}^{\frac{1}{2}}} \sinh{\!\bar{\Delta}^{\frac{1}{2}}t} \right]^{^{^{}}}_{_{_{_{}}}} \! \cos \theta \!
        \\
        -\frac{k(W^2-\sigma Hk^2)}{\bar{\Delta}^{\frac{1}{2}^{^{^{}}}}} \sinh{\!\bar{\Delta}^{\frac{1}{2}}t} \ \sin \theta
    \end{pmatrix}_{_{}} \!e^{\smallminus
    \frac{\nu k^2}{2}t}
    \end{aligned}
\end{equation}
where the overall phase $\theta$ is defined as:
\begin{equation}
    \theta \equiv k(x-v_\text{eig}t)
\end{equation}
With this general expression one may explain the linear behavior observed in various limits. Firstly, in the purely inertial case (Sec.\ \ref{pureadvect}):
\begin{equation}
        \nu = 0 \qquad \bar{\Delta} = \frac{W^2k^2}{4}
\end{equation}
we find that the response of the $W$-wave:
\begin{equation} \label{eq:Wlinresponse}
    \Delta W(t) = -2\eps W \sinh{\!\frac{Wk}{2}}t \,\sin{\theta}
\end{equation}
grows exponentially with a rate that diverges as $k\to\infty$. (So does the perturbation in $\alpha$.)

Restoring the surface tension, we see that:
\begin{enumerate}
    \item The response in $W$ goes to 0 specifically when:
    \begin{equation}
        W^2 -\sigma H k^2 = 0
    \end{equation}
    i.e. $k = k_\text{cut}$. This effect is not modified by  viscosity. It manifests itself as a spectrum ``dip" seen in Figures \ref{fig:FiniteTimeST}, \ref{fig:FiniteTimeTV}, and \ref{fig:FiniteTimeD+D}. 
    
    \item It can be verified that for $k > k_\text{cut}$, the \mbox{$W$-response} becomes oscillatory rather than exponential. The math can be short-cut as follows:
    \begin{equation}
    \begin{aligned}
        \bar{\Delta} < 0 \Rightarrow -\bar{\Delta} = \Delta >0 \Rightarrow  \\
        \bar{\Delta}^{-\frac{1}{2}}\sinh \bar{\Delta}t \sim \Delta^{-\frac{1}{2}}\sin \Delta t
    \end{aligned}
    \end{equation}
    Differences in oscillatory frequency between modes cause the post-cutoff peaks seen in Figure \ref{fig:FiniteTimeST}. 
\end{enumerate}

The latter effect is severely dampened by the viscosity. Due to the added $\nu^2 k^4$ term, $\bar{\Delta}$ may either change sign beyond the linear cutoff, or, if $\nu^2 \geq 4Q \sigma H$, never change sign at all. Additionally, the overall decay factor $e^{\smallminus \frac{\nu k^2}{2}t}$ (Eq. \ref{eq:generallinresponse}) may overshadow an oscillatory signature even if it is hypothetically present. 

In the case of Figures \ref{fig:FiniteTimeTV} and \ref{fig:FiniteTimeD+D}, $\nu^2 = 4Q \sigma H$ exactly, so there is no linear oscillatory behavior.

\section{Numerical Methods} \label{appx:numerics}

For the purpose of this analysis, the Fourier transform of a function $f(x)$ is defined as:
\begin{equation} \label{eq:contFT}
    \mathcal{F}[f](k) \equiv \frac{1}{\ell}  \int_0^\ell e^{-ikx} f(x) \, dx
\end{equation}
assuming a finite periodic domain. The inverse transform is then defined:
\begin{equation}
    \mathcal{F}^{-1}[\widetilde{f}](x) \, = \! \sum_{k=2 \pi / \ell}^\infty e^{ikx} \widetilde{f}(k)
\end{equation}
On a finite set of mesh points with separation $\Delta x$, we can approximate Eq.\ \ref{eq:contFT} via:
\begin{equation}
    \mathcal{F}[f](k) \approx \frac{1}{\ell} \sum_{i=1}^N e^{-ikx_i} f(x_i) \, \Delta x
\end{equation}

Numerical solutions are carried out via explicit 2nd-order time-marching on a finite difference method. All variables are regarded as being located at the center of evenly sized mesh-cells of length $\Delta x$.

Advective flux terms are discretized on a per-equation basis according to the upwinded flux-limiter (FL) method, a very succinct description of which may be found in \cite{Roe1986}. This method is conservative and up to second-order accurate in space, mixing in first-order upwind differencing in the vicinity of sharp gradients to avoid spurious oscillations.

A variety of 2nd-order differencing methods may be used as a base for the FL. Though Lax-Wendroff is common in problems of pure advection, it is not easily applicable to problems with additional force terms. We use simple center differencing instead.

Taking as an example the $\alpha$-equation (Eq.\ \ref{eq:voidprop}), one supplies to the flux limiter the advected variable $\alpha$, the flux $\alpha u_2 =\alpha j + \Q W$, and the upwinding velocity, given by the partial derivative of the flux w.r.t.\ the advected variable, in this case $j+\Q'W$. This velocity, shared by the void and relative momentum equations, is just the eigenvelocity $v_\text{eig}$ discussed in Section \ref{characteristics}.

The flux limiter takes a pre-specified limiter function ($\phi$) of the ratio ($r$) of node-to-node deltas of the advected variable to determine the levels of 1st-order upwind and 2nd-order central differencing applied to each face flux. The face upwinding velocity is determined by averaging the velocities at the two adjoining nodes. Though many limiters exist, we find that the simple minmod limiter:
\begin{equation}
    \phi(r) = \max \! \left[0,\min(1,r)\right]
\end{equation}
works best in our case. The combined flux is given by:
\begin{equation}
    F = \phi(r)F_\text{2nd-order}+\left(1-\phi(r)\right)F_\text{1st-order-upwind}
\end{equation}

We discovered it was necessary to include the surface tension within the same FL framework for numerical stability. We view the $W$-equation ST term (Eq.\ \ref{eq:relmotion}) as supplying an added momentum ``flux" (stress) equal to:
\begin{equation}
    \Phi_\sigma = \sigma H \frac{\partial_x^2 \alpha}{[1 + (H \partial_x \alpha)^2]^{3/2}}
\end{equation}
Obtaining the variation of this flux w.r.t.\ the local flow variables is complicated by the presence of $\alpha$'s derivatives; some notion of a functional derivative may be used to do so in the future. We ignore its potential contribution to the upwinding velocity in the present study.

An open end in the present numerical approach is that it is not precisely suited to the unstable two-equation system. A canonical multi-variable FL method linearizes and diagonalizes a system of conservation laws \cite{LeVequebook} along the lines of Sec.\ \ref{linstab}/Appx.\ \ref{appx:linear}. The flux limiter is then implemented on each ``eigen-equation". This methodology is intended for purely hyperbolic systems where the resulting characteristic velocities are always real. No bespoke methods for advective terms resulting in complex eigenvalues were found.

It is also true that the present numerical approach is not Galilean invariant; adding a total velocity to the system generally results in additional numerical dissipation. A moving-mesh scheme along the lines of \cite{EPursiMuove2010} may be applied to resolve this issue in the future.

We implement a method to ensure that entropy-violating shocks do not propagate. This failure case occurs when the characteristic (upwinding) velocity crosses 0 across a single-node shock, and there are roughly equal fluxes either side of it \cite{Harten1976}. If the characteristic velocities point inward, one has correctly rendered a stable shock. If they lead outward, the shock should be smoothed out over time, producing a ``rarefaction wave". The numerical method leaves the shock untouched because it lacks information about the intermediate region.

To address this issue, we make a targeted fix to the differencing of the advective term in the $\alpha$ equation. Whenever $v_\text{eig}$ crosses from negative to positive between two nodes $j$ and $j+1$, instead of taking the flux at the adjoining face to be some combination of the node-centered fluxes:
\begin{equation}
    F_{j+\frac{1}{2}} = \beta F_j + (1-\beta)F_{j+1}
\end{equation}
with $\beta$ dictated by the flux limiter, and the flux at each node given by:
\begin{equation}
    F_j= \alpha_ju_{2_j}
\end{equation}
we instead average the sub-components of the flux individually:
\begin{equation}
    F_{j+\frac{1}{2}} = \frac{\alpha_j + \alpha_{j+1}}{2} \frac{u_{2_j} + u_{2_{j+1}}}{2}
\end{equation}
We find that this ``entropy fix" both dissipates spurious shocks and avoids their formation.

We observe light spurious noise in the present method when the characteristic velocity sits close to 0 over a significant part of the domain. Such cases usually only occur when the model is extremely dissipative. 

It is not necessary to create a special modeling regime for regions where $\alpha$ is near 0 or 1. We instead make a simple change to the basic flux limiter method that ensures $\alpha$ remain between $0$ and $1$ very closely.

We find that, as long as $\lim_{\alpha \to 0 \,\text{or}\, 1}\!\Q=0$, overshoots of $\alpha$ beyond 0 or 1 tend to be isolated to single nodes at the top (or bottom) of shocks. They are also self-limiting or even self- correcting. This self-correction can be enhanced if the sign of the upwinding velocity at the overshoot node is imposed on both of the adjacent face fluxes. This simple approach makes no change to the underlying equations, keeps the numerical method completely conservative, and brings overshoots down from $10^{-2}$ or more to about $10^{-4}$.


Second derivatives appearing in the viscous term and ST flux are evaluated according to the simple 2nd-order discretization:
\begin{equation}
    (\partial_x^2 \phi)_i \equiv \frac{\phi_{i+1}-2\phi_i + \phi_{i-1}}{\Delta x^2}
\end{equation}

where $\phi$ stands in for $\alpha$ or $W$. The square of the first derivative in the denominator of the ST flux is discretized as the average of the squares of the forward and backward differences:
\begin{equation}
    (\partial_x \alpha)^2_i \approx \frac{1}{2} \left[ \left( \frac{\alpha_{i+1}-\alpha_i}{\Delta x} \right)^2 + \left( \frac{\alpha_i-\alpha_{i-1}}{\Delta x} \right)^2 \right]
\end{equation}
This averaging scheme is second-order accurate, and preferable to center-differencing insofar as it does not return 0 if the forward and backward differences cancel, better limiting the action of the surface tension in the presence of large gradients.

Remaining first derivatives in the buoyant ($g_y$) term:
\begin{equation}
    g_y\delrho H \partial_x \alpha
\end{equation}
and the viscous cross term:
\begin{equation}
    \partial_x \nu \cdot \partial_x W 
\end{equation}
are center-differenced. 

We totally separate the discretization of the time and space derivatives. We opt for an explicit Runge Kutta method to iterate forward in time. Of the wide selection available, we choose the second-order midpoint method for its simplicity and good performance per computation. 

We make this choice despite the findings of Gottlieb and Shu \cite{Gottlieb2001}. They showed that for purely advective problems, Heun's method is the optimal 2nd-order Strong Stability Preserving (SSP) method. On TFM problems involving surface tension and turbulent viscosity, Heun's method does tolerate higher timesteps than the midpoint method without solutions numerically ``blowing up", but they instead become noisy and differ significantly from converged solutions. Excepting such results, we find that Heun's method only increases the ``allowable" timestep length by $\sim1\%$. We judge that the value of this slight speed-up does not outweigh the cost of filtering bogus solutions. 

Gottlieb and Shu's optimal 3rd-order method increases the allowable timestep by $\sim20\%$, nonetheless failing to justify its more than $50\%$ higher computational cost.


\section{Energy Conservation and Flux} \label{appx:energy}

To demonstrate energy conservation w.r.t.\ local transport we take the time derivative of the ``local energy" (Eq.\ \ref{eq:localwaveenergy}):
\begin{equation} \label{eq:dlocalwaveenergydt}
    \dot{\boldsymbol{\mathcal{E}}} = 
    \int_0^1 \half \partial_t (\Q W^2) \, dx =
    \int_0^1 \half \Q' W^2 \partial_t \alpha + \Q W \partial_t W \, dx
\end{equation}
then using the force-free equations (Eq.\ \ref{eq:pureadvectsys}) we replace:
\begin{equation}
    \dot{\boldsymbol{\mathcal{E}}} \! = \! - \! \int_0^1 \half \Q' W^2 \partial_x(\Q W + \alpha j) + \Q W \partial_x (\half \Q' W^2 + W j) \, dx
\end{equation}
If $\partial_x j = 0$, combining terms reveals the conservation law:
\begin{equation} \label{eq:localadvpower}
    \dot{\boldsymbol{\mathcal{E}}} = \left. - (\Q' W + j) \cdot \half \Q W^2 \right|^1_0
\end{equation}
We may therefore write the energy flux:
\begin{equation} \label{eq:localEflux}
    \mathcal{F}_{\text{local}} = (\Q' W + j) \, T_{\text{local}}
\end{equation}

Including the surface tension force means one must consider the surface energy (Eq.\ \ref{eq:surfenergy}). The time derivative of this energy, along with the force term in $\partial_t W$, contribute an additional power:
\begin{equation}
    \begin{aligned}
        \dot{\boldsymbol{\mathcal{E}}}_{\sigma} = \int_0^1 \sigma \! H \! \left( \Q W  \partial_x \tfrac{\partial_{x_{}}^2 \alpha}{[1 + (H \partial_x \alpha)^2]^{3/2}} \right. \\ 
        - \left. \partial_x^2 (\Q W) \tfrac{\partial_{x_{}} \alpha}{[1 + (H \partial_x \alpha)^2]^{1/2}} \right) dx
    \end{aligned}
\end{equation}
Application of derivative rules reveals:
\begin{equation}
    \partial_x \tfrac{\partial_{x_{}}^2 \alpha}{[1 + (H \partial_x \alpha)^2]^{3/2}} = \partial_x^2 \tfrac{\partial_{x_{}} \alpha}{[1 + (H \partial_x \alpha)^2]^{1/2}}
\end{equation}
so the power simplifies:
\begin{equation} \label{eq:localpower}
    \begin{aligned}
        \dot{\boldsymbol{\mathcal{E}}}&_{\sigma} = \int_0^1 \Q W \partial_x^2 \mathcal{S} - \mathcal{S} \, \partial_x^2 (\Q W) \, dx \\
        &\mathcal{S} \equiv \sigma H \partial_x \alpha \left[1 + (H \partial_{x} \alpha)^2 \right]^{-1/2}_{_{}}
    \end{aligned}
\end{equation}
Integrating by parts on the above leaves only the boundary terms, yielding an energy flux:
\begin{equation} \label{eq:surfEflux}
    \mathcal{F}_{\text{local},\sigma} = \mathcal{S} \, \partial_x (\Q W) - \Q W \partial_x \mathcal{S}
\end{equation}
We may also begin with the system energy (Eqs.\ \ref{eq:kinetic}, \ref{eq:surfenergy}):
\begin{equation}
    \begin{aligned}
    \mathbf{E} \equiv \int_0^1 T + U_{\sigma} \, dx  = \int_0^1 \half \Q W^2 - \delrho \Q W \! j \\ 
    + \half \rho_m j^2 \, dx + \frac{\sigma}{H_{}} [1 + (H \partial_x \alpha)^2]^{1/2} \, dx
    \end{aligned}
\end{equation}
Applying Eqs.\ \ref{eq:voidprop}-\ref{eq:mixmotion} yields a total energy flux:
\begin{equation}
    \begin{aligned}
    \mathcal{F}_{total} = \, & \half \Q W (\Q' W^2 + j W + \delrho j^2) + j (\mathcal{A}_m - \half \rho_m j^2) \\
    & + \, \mathcal{S} \, \partial_x (\Q W + \alpha j) - (\Q W + \alpha j) \partial_x \mathcal{S}
    \end{aligned}
\end{equation}
with $\mathcal{S}$ defined per Eq.\ \ref{eq:localpower}. \newline


\section{Functions Appearing in the Perturbative Series Expansion} \label{appx:pertfuncs}
Below are the dimensionless functions appearing in Eqs.\ \ref{eq:2ndamps}-\ref{eq:freqcorr}:
\begin{equation}
    \begin{aligned} 
        F_1 & \equiv (2 \Oh - \Oh^3) \kappa^3 + \Oh^2 \bar{\Delta}^\frac{1}{2} \kappa^2 - \Oh \kappa + 3 \bar{\Delta}^\frac{1}{2} \\
        F_2 & \equiv (\Oh^4-3\Oh^2) \kappa^4 + (\Oh -\Oh^3) \bar{\Delta}^\frac{1}{2} \kappa^3 \\
        & + (5 \Oh^2 - 12) \kappa^2 - 3 \Oh \bar{\Delta}^\frac{1}{2} \kappa + 6 \\
        F_3 & \equiv - (\Oh + \Oh^3) \kappa^3 + (\Oh^2 + 3) \bar{\Delta}^\frac{1}{2} \kappa^2 + 6 \bar{\Delta}^\frac{1}{2} \\
        F_4 & \equiv 3 \kappa^5 \! \left[ (\Oh^3 - 5\Oh) \kappa^2  \!+\! (3 - \Oh^2) \bar{\Delta}^\frac{1}{2} \kappa \!+\! 2 \Oh \right] \\
        F_5 & \equiv (- \Oh^{7} + 4 \Oh^{5} - 2 \Oh^{3}) \kappa^{6} \\
        & + (2 \Oh^{6} - 6 \Oh^{4}+ 2 \Oh^{2}) \bar{\Delta}^\frac{1}{2} \kappa^{5} \\
        & + \left[- (8 + \bar{\Delta}) \Oh^{5} + (22 + 2 \bar{\Delta}) \Oh^{3} + 10 \Oh \right] \kappa^{4} \\ 
        & + (14 \Oh^{4} - 34 \Oh^{2} - 6 ) \bar{\Delta}^\frac{1}{2} \kappa^{3} \\
        & + \left[- (9 + 6 \bar{\Delta}) \Oh^3 - (26 - 12 \bar{\Delta} ) \Oh\right] \kappa^{2}  \\
        & + (10 \Oh^{2} + 18) \bar{\Delta}^\frac{1}{2} \kappa  + (4 - \bar{\Delta}) \Oh \\
        F_6 & \equiv (5 \Oh^2 - \Oh^4) \kappa^3 + (2 \Oh^3 - 8 \Oh) \bar{\Delta}^\frac{1}{2} \kappa^2 \\
        & + (3 \bar{\Delta} - \Oh^2 \bar{\Delta} - 2 \Oh^2) \kappa + 2 \Oh \bar{\Delta}^\frac{1}{2}
    \end{aligned}
\end{equation}
\newline
\newline
\newline
\newline
\newline
\newline
\newline



\bibliography{references.bib}

\end{document}